\documentclass[12pt]{article}
\usepackage{lscape}
\usepackage{geometry}
\usepackage[onehalfspacing]{setspace}
\usepackage{amssymb}
\usepackage{amsfonts}
\usepackage{graphicx}
\usepackage{amsmath}
\usepackage{epstopdf}
\usepackage{natbib}
\usepackage{bibunits}
\usepackage{morefloats}
\usepackage{float}
\usepackage{comment}
\usepackage{pdflscape}
\usepackage{adjustbox}
\usepackage{subcaption}
\usepackage{longtable}
\usepackage{apalike}
\usepackage{xfrac}
\usepackage[dvipsnames,table]{xcolor}
\usepackage{etoolbox,siunitx}
\robustify\bfseries

\usepackage{hhline,multirow}
\usepackage{yfonts}

\usepackage{longtable}
\usepackage{threeparttablex}

\newcolumntype{d}[1]{D{.}{.}{#1}}

\usepackage{qtree}
\usepackage{multicol}
\usepackage{booktabs}
\usepackage{threeparttable}
\usepackage{tabularx}
\usepackage{dcolumn}
\newcolumntype{d}[1]{D..{#1}} 

\usepackage{siunitx}
\usepackage{mathpazo} 
\usepackage[scaled]{helvet} 
\usepackage{courier} 
\normalfont
\usepackage[T1]{fontenc}
\usepackage{listings}
\usepackage{epigraph}
\def\sym#1{\ifmmode^{#1}\else\(^{#1}\)\fi}

\usepackage[pdftex,bookmarks,colorlinks]{hyperref}
\hypersetup{
  colorlinks,
  citecolor=DarkerPineGreen,
  linkcolor=red,
  urlcolor=blue}
\usepackage{cleveref}

\newtheorem{theorem}{Theorem}

\newtheorem{algorithm}[theorem]{Algorithm}

\DeclareMathOperator*{\argmin}{arg\,min}

\usepackage{etoolbox}

\definecolor{dukeblue}{rgb}{0.0, 0.0, 0.61}
\definecolor{darkred}{rgb}{0.8,0,0}
\definecolor{DarkerPineGreen}{RGB}{0, 90, 80} 

\makeatletter
\patchcmd{\epigraph}{\@epitext{#1}}{\itshape\@epitext{#1}}{}{}
\makeatother

\makeatletter
\def\munderbar#1{\underline{\sbox\tw@{$#1$}\dp\tw@\z@\box\tw@}}
\makeatother

\renewenvironment{abstract}
 {\small
  \begin{center}
  \bfseries \abstractname\vspace{-.5em}\vspace{0pt}
  \end{center}
  \list{}{
    \setlength{\leftmargin}{1.44cm}    \setlength{\rightmargin}{\leftmargin}  }  \item\relax}
 {\endlist}
\usepackage{graphicx}
\usepackage{wrapfig}

\usepackage{algorithm,algorithmic}

\usepackage{soul}

\begin{document}
\setstcolor{orange}
\sloppy
\title{\vspace*{-0cm} \huge {Maximally Machine-Learnable Portfolios}}
\small
\small 
\author{\hspace*{-0.8cm} Philippe Goulet Coulombe\thanks{%
Departement des Sciences Économiques, \href{mailto:p.gouletcoulombe@gmail.com}{\texttt{goulet\_coulombe.philippe@uqam.ca}}.   For helpful discussions and comments,  we would like to thank Frank Diebold,  Dave Rapach,  Erik Christian Montes Sch\"utte,  Hugo Subtil,  Dalibor Stevanovic,  and Boyuan Zhang.  For research assistance,  we are grateful to Mikael Frenette and Félix-Antoine Gaudreault.}\\ \hspace*{-0.8cm} Université du Québec à Montréal  
	\and 
	\hspace*{0.4cm} Maximilian G\"obel\\ \hspace*{0.5cm} Bocconi University
}
\date{\vspace{-0.4cm}
\small
\small
\bigskip
\bigskip
First Draft: December 5, 2022 \\ 
This Draft: April 4,  2024 
}
  
\maketitle
 
\begin{abstract}

\noindent When it comes to stock returns,  \textit{any form} of predictability can bolster risk-adjusted profitability.  We develop a collaborative machine learning algorithm that optimizes portfolio weights so that the resulting synthetic security is maximally predictable.  Precisely,  we introduce MACE,  a multivariate extension of Alternating Conditional Expectations that achieves the aforementioned goal by wielding a Random Forest on one side of the equation,  and a constrained Ridge Regression on the other.  There are two key improvements with respect to Lo and MacKinlay’s original maximally predictable portfolio approach.  First,  it accommodates for any (nonlinear) forecasting algorithm and predictor set.  Second,  it handles large portfolios.   We conduct exercises at the daily and monthly frequency and report significant increases in predictability and profitability using very little conditioning information.  Interestingly,  predictability is found in bad as well as good times,  and MACE successfully navigates the debacle of 2022.

\end{abstract}

\thispagestyle{empty}

\clearpage

\clearpage 
\setcounter{page}{1}

\newgeometry{left=2 cm, right= 2 cm, top=2.3 cm, bottom=2.3 cm}

\section{Introduction}

A natural trading strategy is to buy (\textcolor{black}{sell}) assets \textcolor{black}{whose price} one expects to appreciate (depreciate) with \textit{higher certainty}.  That is,  out of a basket of securities,  it may be preferable to focus active trading efforts on the most predictable assets given one's information set.  It is well known that marginal predictive accuracy improvements can translate into substantial profits without equally substantial risks for investors.  However,  such desirable assets are in very short supply,  if they can be identified at all.  Consequently,  a natural question is whether we can reach to the ideal by constructing more predictable synthetic securities as linear combinations of (mostly) unpredictable existing ones.  This paper devises a data mining technique that drills out those -- Maximally Machine-Learnable Portfolios (MMLP) -- by directly optimizing portfolio weights so to maximize forecasting accuracy,  and thereby risk-adjusted returns. 

\vskip 0.1cm

The origins of such ideas lie within \cite{loMacKinlay1997}'s maximally predictable portfolios (MPP),  where a set of weights $\boldsymbol{w}$ are chosen so to maximize the $R^2$ of $\boldsymbol{w}'\boldsymbol{r}_{t}$ with a tightly specified linear regression based on a few factors.  In this convenient sparse (both in returns and predictors) linear framework,  obtaining the MPP reduces to solving an eigenvalue problem (akin to canonical correlation analysis) subject to a constraint.\footnote{In practice,  solving the non-convex fractional programming optimization problem this quickly becomes a daunting task  -- particularly when considering many assets, factors,  and constraints on the portfolio's composition \citep{GotohFujisawa2014}.  For this reason,  many alternative numerical methods have been proposed to solve more successfully the MPP problem  \citep{Yamamotoetal2007,Konnoetal2010,Konnoetal2010b, TakayaetKonno2010,GotohFujisawa2014}.   Recent applications include  \cite{harris2022} and \cite{BaoetAl2022}.}  Fundamental limitations are apparent.  First,  the predictive function is of rather limited sophistication.  Namely,  it is linear, low-dimensional,  and most often unregularized.  This limits the obtained MPPs to lie within a narrow space of maximally \textit{linearly}-predictable portfolios.  Clearly,  additional patterns of predictability are likely to be found when allowing for complex nonlinear relationships while keeping an eye on the characteristically hostile signal-to-noise ratio (SNR) of asset pricing applications.  Second,  the portfolio side of the equation is similarly unregularized,  opening two evenly unpleasant routes.  One can either limit the number of assets to be included and severely bound the space of MPP candidates,  or run into the well-documented estimation problems of (large) covariance matrices \citep{ledoit2004honey}.  MMLPs are designed to avoid all of the above by including a powerful nonlinear nonparametric  function approximator on the right hand side,  and regularization schemes for both the learnable portfolio weights and the corresponding predictive function.  Those qualities,  flexible nonlinearities and thoughtful regularization,  have both been instrumental to the ML renaissance in empirical asset pricing \citep{gu2020empirical,ChenPelgerZhu2021,Nagel2021},  and will be so again crossing the bridge from MPP to MMLP. 

\vskip 0.2cm

{\sc \noindent \textbf{MACE}.} We introduce MACE,  which stands for Multivariate Alternating Conditional Expectations,  and is a multifaceted generalization of  \cite{BreimanFriedman1985}'s ACE algorithm.  The latter was originally designed for nonlinearly transforming a univariate regression target to maximize association.  As the name suggests,  ACE achieves its aim by alternating the estimation of two functions (one for \textcolor{black}{the right-hand side (RHS)},  and another for \textcolor{black}{the left-hand side (LHS)}) taking the other as fixed at each iteration,  very much in the spirit of EM algorithms.   Adapting ACE to the MMLP problem,  MACE modifies it in two key aspects.  First,  the LHS is multivariate (an extension) and linear (a restriction).   Second,  it replaces ACE's rudimentary additive polynomial models by a Ridge Regression (RR) on the portfolio side and a Random Forest (RF) on the prediction side.  RR provides a linear and regularized fit for the LHS, avoiding overfitting and non-plausible allocations.  RF is a powerful off-the-shelf predictive algorithm that (i) handles high-dimensional data,  (ii) can approximate a wide range of unspecified nonlinearities,  (iii) requires little tuning,  (iv) very rarely overfits.  Then comes a  panoply of algorithmic details that make the cohabitation of aforementioned elements possible: a learning rate,  block out-of-bag sampling,   decreasingly random optimization,  and bagging strategies rather than predictions.   Those are all extensively discussed in the paper. 

\vskip 0.1cm

 \textcolor{black}{Maximizing a portfolio's predictability given the available information set makes economic sense.  In fact,  for a mean-variance investor,  the utility improvement from a predictable portfolio (vs.  an unpredictable one) is a monotonically increasing function of its $R^2$ \citep{campbell2008predicting}.  Thereby,  constructing the portfolio with the highest out-of-sample $R^2$ is equivalent to finding that \textcolor{black}{portfolio which} will generate the highest utility when \textit{traded} (as a whole) according to a basic \textit{univariate} mean-variance rule.   MACE maximizes the $R^2$, or put differently minimizes the "costly" portfolio variance -- \textcolor{black}{that part of the variation in our portfolio's return stream that} cannot be neutralized by trading based on our information set.  Thus,  MACE's objective is as economically general as the mean-variance utility maximization framework,  which, despite limitations such as neglecting higher-order moments or not directly including transaction costs, is still the dominant approach to portfolio construction. }

\vskip 0.1cm

Regarding portfolio construction, the majority of the literature relies on a two-step procedure,  prediction coming first and portfolio construction, following various fixed rules,  coming next.  MACE, in contrast, fits into a  stream of studies that optimizes portfolio weights directly -- explicitly or implicitly encompassing the prediction step.   An early contribution is \cite{BrandtSantaClaraValkanov2009},  who exploit linearity in characteristics  to do so.  \cite{FiroozyeEtAll2022} also deploy linearity to rewrite their simultaneous mean-variance/prediction problem with vector autoregressive predictions as canonical regression analysis.  \cite{CongEtAl2021} \textcolor{black}{achieve this direct portfolio optimization} in a nonlinear nonparametric setup with a reinforcement learning algorithm targeting Sharpe Ratios,  and fueled with a large database à  la \cite{gu2020empirical}.  \cite{GOPZ2024} also propose a deep learning pipeline (for statistical arbitrage) which end goal is to optimize the Sharpe ratio.  Closer to our approach,  \cite{KellyMalamudPedersen2023} exploits the potential of cross-correlation among asset-specific signals to find \textit{predictability} within stock returns.  They decompose the ``prediction matrix'' into its principal components,  which delivers multiple investment portfolios that are  in order from most to least predictable -- conditional \textit{linearly} on the chosen signal.  While bearing some resemblance to canonical regression analysis and thus \cite{loMacKinlay1997}'s MPP,  it differs by the use of asset-specific signals and its scalability.  MACE's advantages with respect to those alternatives are hereby visible.  It is nonlinear and nonparametric,  yet remains simple,  transparent and rather traditional in its trading decisions.  It works with or without terabytes of asset-specific data which has been the usual backbone of key contributions of the ML-finance literature.  While most of the literature approaches the prediction problem by first fixing the target and then collecting relevant predictors for it,  MACE rather explores the space of possible forecasting targets and selects what works best based on the available information set.  In the era of abundant alternative data,  it offers an algorithmic solution to an increasingly common question: what is this particular dataset useful for?  Statistically, the resulting optimization problem is some kind of semi-parametric canonical correlation analysis  \citep{michaeli2016nonparametric} supplied with various desirable features for financial forecasting.  Economically,  it is aligned with the mean-variance principle.

\vskip 0.2cm

{\sc \noindent \textbf{Statistical Arbitrage at the Daily Frequency}.} We consider two applications.  The first is creating portfolios of the 20, 50,  and 100 most capitalized firms on the NASDAQ for trading at the daily frequency.  We evaluate those from January 2017 to December 2022,  an era for which gains of ML-based statistical arbitrage are expected to be low,  if they exist at all \citep{krauss2017deep}.  The information set is lagged returns of the portfolio itself.  Thus,  in this setup MACE is looking for maximally \textit{nonlinearly} mean-reverting portfolios.   And indeed it does find some,  scoring enviable returns and risk-reward ratios.  Nonlinearities prove instrumental to such results as the degree \textcolor{black}{of} mean-reversion (when approximated linearly) is shown  to be highly  state-dependent.  Out-of-sample $R^2$ testifies to that, ranging from a moderate 0.5-0.9\% in calmer periods,  to 12\% during the first wave of Covid-19, and a staggering 20\% when zooming in on  March 2020.  MACE is shown to heavily rely on day-to-day oscillations to achieve swift returns  in tumultuous months -- a behavior learned in part from the financial crisis.  Most importantly,  it also outperforms benchmarks outside of high volatility episodes,  both in bull and bear markets.   In particular, all MACEs deliver positive returns in 2022, ranging from 5\% to 23\%.  This and other features lead MACE, sometimes with only 20 highly liquid stocks,  to nearly double the market's Sharpe Ratio.

\vskip 0.1cm

This application expands on various strands of the statistical arbitrage literature,  where many tactics (ranging from heuristics to cointegration tests) have been proposed to identify mean-reverting portfolios or pairs with predictable spreads (see \cite{krauss2017} for an extensive survey).  Typically,  the candidate securities are fixed ex-ante rather than ``discovered'',  and mean reversion is linear.  Nonetheless,  when a good discovery is made -- like that of \cite{MedhatSchmeling2022} exploiting very different time series behaviors for low- vs. high-turnover stocks -- gains can be huge.   Our approach mines for such discoveries.  Hence,  more closely related are the works of \cite{Aspremont2011},  \cite{CuturiAspremont2013},  and \cite{FogarasiLevendovszky2013} who also focus on constructing maximally mean reverting portfolios.  Linearity is inherent, and allows for such problems to be reformulated as extensions of canonical correlation analysis with a varying degree of elaboration.   An important focus of the literature,  as it is the case for MPPs,  has been on improving computations and placing reasonable constraints (like sparsity) on the allocation \citep{ZhaoPalomar2016,ZhaoPalomar2018,GriveauCalderhead2021}.  Nonetheless,  linearity remains pervasive, with some directly targeting linear autocorrelation statistics  \citep{ZhaoPalomar2016},  and others embedding directly linear forecasting models,  like Vector Autoregressions,  within the optimization framework \citep{GriveauCalderhead2021,FiroozyeEtAll2022}.  Therefore,  MACE, through its use of RF,  evidently widens the space of exploitable time series dependence for statistical arbitrage.  Additionally, the portfolio side of the equation is not constrained computationally nor statistically from including a myriad of stocks and a long daily sample,  a case of interest from a conceptual standpoint (testing market efficiency) and eventually practical in an era of shrinking transaction costs.

\vskip 0.2cm

{\sc \noindent \textbf{Monthly Trading Based on Macroeconomic Indicators}.} The second application is at the monthly frequency and utilizes the canonical \cite{WelchGoyal2008} data set to construct an MMLP with a subset of large\textcolor{black}{-cap} individual stock returns from CRSP.  Thus,  it uses no firm characteristics and is rather looking for aggregate predictability based on trivially available macroeconomic indicators.  All RFs (MACE or not) struggle to deliver positive $R^2$'s from the late 1980s up to the mid-2000s.  However,  MACE hits an $R^2$ of above 4\% in the last 15 years of our sample ending in 2019 -- an era for which predictability and associated economic gains (ML-based or not)  are often reported to have waned \citep{chordia2014have,han2018firm,gu2020empirical,CongEtAl2021}.  This is achieved in part during the  financial crisis  and the \textcolor{black}{years thereafter}, where MACE limits losses considerably,  and catches up with the pre-crisis trend as early as mid-2009.  Other non-MACE portfolios using RF as the predictive function also manage to somehow mitigate losses during the meltdown,  but then fail to find and leverage predictability \textcolor{black}{when exiting} the Great Recession.  Moreover,  during the slowdown of 2018,  only MACE continues with mostly unabated upward-trending returns.   We find,  using interpretable ML tools,  that this success is attributable to MACE uncovering a portfolio with a subtle response to elevated volatility,  in the form of a nonlinearly time-varying risk premium. 

\vskip 0.1cm

\textcolor{black}{Our approach thus differs from} \cite{gu2020empirical} and the vast body of (monthly) studies \textcolor{black}{who} use a pooled panel approach with nearly 2 million observations of stock returns from roughly 30,000 U.S.-listed companies,   with a corresponding feature set of over 100 company characteristics and macro-predictors.  This is also the backbone of \cite{CongEtAl2021}'s reinforcement learning approach.  A similar pooled panel with many cross-sectional and time series characteristics is also found for exchange rates in \cite{filippou2022out} and for cryptocurrencies in \cite{FilippouRapachThimsen2021}.   Given that MACE does not predict each stock separately,  and rather focuses on forecasting a single synthetic index with easily available time series,  it can be described,  at least in relative terms,  as a very low-maintenance strategy.   Moreover,  by construction,  it cannot rely on cross-sectional anomalies that  have already dissipated,  or focus on illiquid, once non-adequately priced stocks.  Obviously,  MACE is not exempt either from risking an eventual depletion of its sources of predictability.  Nonetheless,  those concerns are alleviated by the multi-solution nature of the algorithm and the opaque prediction function.  Indeed,  MACE's market timing comes from a (mostly) black box prediction function that cannot be easily deduced by other market participants,  and most importantly,  if a certain linear combination has been overharvested,  MACE can dig out others.  

\vskip 0.2cm

{\sc \noindent \textbf{Outline}.} This paper goes as follows.  Section \ref{sec:HNN} introduces MACE, motivates its structure,  and discusses practical aspects.  Section \ref{sec:app2} conducts the daily trading empirical analysis and section \ref{sec:app1} conducts a monthly frequency exercise.   Section \ref{sec:con} concludes.

\section{Multivariate Alternating Conditional Expectations}\label{sec:HNN}

The sophistication of MMLPs, particularly the use of nonlinear tree ensembles-based predictions,  necessitates the design of a vastly different framework for optimization than what prevailed for MPPs. 

\subsection{The Algorithm}\label{sec:algo}

MACE is, for the most part,  a conceptually trivial extension of ACE.  Its successful empirical development, however,  requires a fair amount of subtle machine learning craftsmanship.  \cite{BreimanFriedman1985}'s ACE  applied to a generic $h$-step ahead forecasting problem of a single target $Y_{t+h}$ reads as
		\begin{align}\label{ACE}
			g(\underbrace{Y_{t+h}}_{1 \times 1}) = f(\underbrace{\mathbf{X}_t}_{1 \times K}) \; + \; \varepsilon_{t+h}
		\end{align}
		where $g$ and $f$ are unknown functions,  $\varepsilon_{t+h}$ is the prediction error,  $\mathbf{X}_t$ is the matrix of $K$ available predictors at time $t$ (which may include lags or various indicators).   Thus, the sole deviation from the textbook predictive regression setup is the introduction of $g$.  ACE's goal is to find the optimal transformation of $g$, in the sense that it is maximally predictable by the output of $f$.   \cite{BreimanFriedman1985} show that $\hat{g}$ and $\hat{f}$ can be obtained from running an iterative algorithm that alternates between obtaining the conditional expectation of $g(Y_{t+h})$ given $\mathbf{X}_t$ for a fixed $g$ and the conditional expectation of $f(\mathbf{X}_t)$ given $Y_{t+h}$ for a fixed $f$.  Following the original incarnation,   $g$ and $f$ typically consist of backfitted polynomial functions,  which \textcolor{black}{used to be a popular} nonparametric ML approach -- before being outshadowed by the advent of tree ensembles in the 1990s and the resurrection of neural networks  in the mid-2000s.   Nonetheless,  the polynomial approach still \textcolor{black}{remains} the predictive function in recent ACE applications \citep{gopinathan2019stock,rao2022nonstationary}.    
		
This paper extends ACE in three ways so that it can uncover a modern brand of maximally predictable portfolios: \textbf{(i)} $Y_{t+h}$ is replaced by $\mathbf{Y}_{t+h} \in {\rm I\!R}^N$ and $g : {\rm I\!R} \rightarrow {\rm I\!R}$ by $g : {\rm I\!R}^N \rightarrow {\rm I\!R}$,  \textbf{(ii)} we impose a series of constraints on $g$ so that its output is a portfolio (up to a scaling constant),   and \textbf{(iii)} $f$ is a high-performing  off-the-shelf modern ML tool.  All three are vital to the current application,  to a varying degree of obviousness.  \textbf{(i)} puts the M in MACE by making it  a multivariate problem and therefore allowing for $g$'s input to be,  for instance,  a panel of stock returns.  From this,  \eqref{ACE} becomes the general MACE problem
								\begin{align}\label{MACE}
	\underbrace{g(\underbrace{\mathbf{Y}_{t+h}}_{1 \times N})}_{1 \times 1} = f(\underbrace{\mathbf{X}_t}_{1 \times K}) \; + \;  \varepsilon_{t+h}
\end{align}
which lies within the broad class of nonparametric canonical correlation problems \citep{michaeli2016nonparametric}.  This is also contained within the class of models for which \cite{makur2015efficient} develop theoretical guarantees for generic ACE-type algorithms.  Now,  \textbf{(ii)} restricts $g$'s original nonparametric ambitions to that of learning \textcolor{black}{a} linear combination of \textcolor{black}{$\mathbf{Y}_{t+h}$'s} so $\hat{g}(\mathbf{Y}_{t+h})$ is a portfolio return series --- as opposed to being literally anything,  which could nevertheless be of interest in other financial applications.  The minimization problem that ensues is
				\begin{align}\label{MACEpp}
\min_{\boldsymbol{w}, \phantom{.} f} \sum_{t=1}^T (\boldsymbol{w}'\boldsymbol{r}_{t+h} - f (\mathbf{X}_t))^2 +\lambda ||\boldsymbol{w}||^2 \quad \enskip \text{such that} \enskip \boldsymbol{w} \geq 0  \enskip  \text{and} \enskip \mathrm{Var}[\boldsymbol{w}'\boldsymbol{r}_{t+h}]=1
	\end{align}
		where $\mathbf{Y}_{t+h}$ is hereafter also assumed to be a panel of stock returns $\boldsymbol{r}_{t+h} $.  \textcolor{black}{The $\mathrm{Var}[\boldsymbol{w}'\boldsymbol{r}_{t+h}]=1$ constraint (also part of ACE) avoids degenerate solutions and pushes the algorithm towards maximizing the scaleless $R^2$.  The sum-to-one constraint ($\boldsymbol{w}'\iota=1$) can be applied on the resulting $\boldsymbol{w}$ after estimation so the allocation matches a portfolio investing one unit. } The addition of $\lambda ||\boldsymbol{w}||^2 $ provides $l_2$ regularization with intensity $\lambda$ (an hyperparameter)  that will guard against overfitting \textit{and} non-realistic allocations \citep{carrasco2011optimal}.  The non-negativity constraint may or may not be activated.  For instance,  it will be turned off in our daily application.  Its activation  implies additional shrinkage beyond that of the $l_2$ norm by inducing some sparsity (some weights will be constrained to 0).   Lastly,  \textbf{(iii)} is what will provide MACE with forecasting power.  $f$ is chosen to be a Random Forest (RF) for various reasons, some more subtle than others (see section \ref{sec:whynot}).  Surely,  what we want first and foremost, is $f$ to be a solid off-the-shelf predictive model handling nonlinearities and high-dimensional data while keeping overfitting in check without extensive hyperparameter tuning \citep{MSoRF}.  Clearly,  RF checks the first two boxes, along with Boosted Trees and (Deep) Neural Networks \citep{ESL}.  The last requirement is met by RF,   but not nearly as much by the other two well-known families of ML algorithms.  As will become apparent in section \ref{sec:whynot},  due to the iterative nature of the MACE (and,  in general,  the idea of having a function on each side of the equation),  RF's easily obtainable out-of-bag predictions, that are resilient to overfitting, will be a key ingredient in our routine.

\begin{algorithm}[tb]
\begin{algorithmic}[1]
\STATE Initiate  $\hat{z}_{0,t+h}$ as the scaled equally weighted portfolio,  learning rate is $\eta$,   
\FOR{$s=1$ to $s_{\text{max}}$}
\vspace{0.25em}
\STATE	\textbf{\textit{The Random Forest Step}}
\vspace{-0.7em}
$${f}_s^* = \argmin_{\enskip f \in \mathcal{F}_\text{RF}} \sum_{t=1}^T (\hat{z}_{s-1,t+h} - f (\mathbf{X}_t))^2 $$ 
where we keep $ f_s^*(\mathbf{X}_t)$, a time series of block out-of-bag predictions. 
\vspace{0.2em}
\STATE Update the RHS: $\hat{f}_s(\mathbf{X}_t)=\eta \times f_s^*(\mathbf{X}_t)+(1-\eta) \times \hat{f}_{s-1}(\mathbf{X}_t)$
\vspace{0.65em}
\STATE \textbf{\textit{The Ridge Regression Step}} 
\vspace{-0.7em}
$${\boldsymbol{w}}_s^* = \argmin_{\boldsymbol{w}} \sum_{t=1}^T ( \hat{f}_s(\mathbf{X}_t) - \boldsymbol{w}'\boldsymbol{r}_{t+h})^2 + \lambda ||\boldsymbol{w}||^2 \quad  \text{st} \enskip \boldsymbol{w} \geq 0  $$ 
where we keep in-sample predictions ${z}_{s,t+h}^*={{\boldsymbol{w}}_s^*} '\boldsymbol{r}_{t+h}$. 
\vspace{0.2em}
\STATE Update the LHS: $\hat{z}_{s,t+h}= \texttt{scale}(\eta \times {z}_{s,t+h}^*+(1-\eta) \times \hat{z}_{s-1t+h})$
\vspace{0.25em}
\ENDFOR $\enskip$ if $\boldsymbol{w}_s \approx \boldsymbol{w}_{s-1}$,  or  if some early stopping criterion has been met.
\end{algorithmic}
\caption{\color{dukeblue} \textit{\textbf{MACE}}}
\label{MACEalgo}
\end{algorithm}

{\vspace{0.15cm}}
{\noindent \sc \textbf{Initialization.}} Algorithm \ref{MACEalgo} is divided into two key steps, which are,  intuitively,  the updating of the right hand and left-hand side parameters,  respectively.  We initialize $\hat{z}_{0,t+h}$ as a plausible portfolio.  When $\boldsymbol{w} \geq 0$ is activated,  such a portfolio is the equally-weighted one (as used in section \ref{sec:mres}).  When it is not (as in section \ref{sec:dres}),  one can use the solution to the classic (and static) global minimum variance portfolio problem, which is an equally intuitive initialization point,  especially given the forthcoming discussion in section \ref{sec:mvrela}. Regarding $\hat{f}_{0}(\mathbf{X}_t)$,  it is set to 0 and $\eta=1$ for line 4 in iteration $s=1$.   This simply means MACE is initiated at the equally-weighted portfolio (or else) and its corresponding RF conditional mean.

Given the inherent non-convexity of the objective and the plethora of possible solutions,  initialization can matter.  This is especially true in extremely low SNR environments and when regressors are generated endogenously -- like in our daily returns prediction application.   Our approach to the multiplicity-of-solutions problem is in the spirit of deep learning rather than classical econometrics.   Indeed,  a fair amount of ink has been spent on devising efficient algorithms to uncover the global optimum within a very restricted class of MPP problems,  in part because the attained $R^2$ both in-sample and out-of-sample was regarded as a metric of market inefficiency.   We deviate from the statistical philosophy of going at lengths to obtain ``true parameters'' and rather look for ``useful parameters",  that is, any solution that can generate value for wealth management strategists.  Of course,   the two objectives are surely not mutually exclusive,  but they entail a different focus.   Thus,  in our applications,  we do not especially care for $f$ being unique nor the truest solution to anything,  but rather aim at building a portfolio and a predictive function that \textit{generalizes} well--- that is, it maximizes  $R^2_{\text{train}}$ \textit{and} $R^2_{\text{test}}$.  In that spirit, devising mechanisms to maximize $R^2_{\text{train}}$ remain essential,  but they are coupled with equally relevant algorithmic elements to insure such feats can be reproduced out-of-sample.   Limiting overfitting in both $f$ and $\boldsymbol{w}$ \textit{is} a necessary condition to successfully trade such a portfolio.  Thus,  in the coming paragraphs, we explain the nuts and bolts of MACE,  and how it spreads the ML gospel of \textcolor{black}{the} bias and variance trade-off to the MMLP problem. 

{\vspace{0.2cm}}

{\noindent \sc \textbf{The Random Forest Step and Block Out-of-Bag Subsampling.}} In many ways,  MACE is a traditional EM algorithm,  where we optimize certain parameters while keeping others fixed,  reverse roles in the following step,  and alternate until some stopping criterion is met.   Accordingly,  the first step predicts a fixed portfolio using RF.   Then,  predictions are updated as a convex combination of $f_s^*(\mathbf{X}_t)$ (current predictions) and the previous iteration's predictions $\hat{f}_{s-1}(\mathbf{X}_t)$,  where the speed of adjustment is determined by the learning rate $\eta$.  

When constructing  $f_s^*(\mathbf{X}_t)$,  it is imperative that one does not use RF's \textit{fitted values}, which are inevitably prone to immense overfitting.  Indeed,  RF's fit always delivers $R^2_{\text{train}}$ close to 1 (for any standard tuning parameters combinations) even though the true $R^2$ is nowhere near that (see \cite{MSoRF} for an explanation and a barrage of examples with classic datasets).  This does not prevent RF from delivering stellar \textcolor{black}{$R^2_ {\text{test}}$'s} -- the traditional object of interest --  and it is why the  $R^2_{\text{train}}>R^2_{\text{test}}$ differential  has mostly stayed under the radar of the ML community.  Whenever RF's in-sample predictions are required, one shall use the so-called out-of-bag (OOB) predictions,  which are,  by construction, immune to overfitting in a \textit{cross-sectional} context -- in the sense that their predictive accuracy will be exactly aligned with what one should expect out-of-sample \citep{breiman2001,ESL}.  In other words,  such predictions include the conditional mean,  whatever its quality may be,  and little to none true error term \citep{MSoRF}.  This is particularly crucial in MACE given that such predictions are to be fitted by another ML algorithm in a subsequent step.  In a manner,  this approximates the ideal ``cross-fitting'' solution where,  in our context,  the Ridge Regression step would be conducted on  one half of the sample,  and the RF \textcolor{black}{step} on the remaining half.  The latter (rather demanding) scheme is the backbone of so-called honest causal forest  \citep{athey2019grf}  in heterogeneous treatment effect estimation.  In fact,  it can be shown that OOB sampling and variants provide a convenient approximation when sample sizes are limited or other practical aspects render plain splitting nonoperational \citep{chen2022debiased}. 

The only thing standing in the way of such properties to be applied to our problem is the time series nature of our data.  Indeed,  time series dependence in the left-hand side \textcolor{black}{(LHS)} or right-hand \textcolor{black}{ side} \textcolor{black}{(RHS) }variables, which creates major complications for classical bootstrap inference,  generates similar hurdles for the validity of out-of-bag predictions.  While that in $\boldsymbol{r}_{t+h}$ is negligible for $h=1$,  it is certainly not so when considering $\boldsymbol{r}_{t+12}$,  the average return between $t$ and $t+12$ (in effect,  a sliding moving average).  This is even more prevalent in the case of $\mathbf{X}_t$ where predictors, while being stationary, may be quite persistent. This persistence will break the non-overfitting properties of OOB predictions -- with an immediate consequence that  $ f_s^*(\mathbf{X}_t)$ includes overfitted elements  of $\hat{z}_{s-1,t+h}$,  \textcolor{black}{and thus} failing  to approximate the LHS and RHS being trained on \textcolor{black}{``truly''} separate data sets.  All this is obviously related to how time series dependence biases downward bootstrapped standard errors used in small sample frequentist inference \citep{kreiss2012bootstrap},  and the solution to the aforementioned problem -- block bootstrapping or subsampling -- is backed out from the wide literature on the subject \citep{hyndman2018forecasting}.   Thus,  to obtain a $ f_s^*(\mathbf{X}_t)$ which is plausibly exempt from overfitting,  we will use  \textit{block} out-of-bag in-sample predictions.  Such techniques have been used to reliably extract more ``structural'' quantities like various macroeconomic latent states in  \cite{MRF} and \cite{HNN}.

{\vspace{0.15cm}}
{\noindent \sc \textbf{The Ridge Step.}}  The Ridge step takes RF predictions as given and optimizes $\boldsymbol{w}$ so that $\boldsymbol{w}'\boldsymbol{r}_{t+h}$ matches as closely as possible the predictions, in essence, collaborating with $f$ so as to maximize association.  The Ridge Regression apparatus comes with trivially implementable, yet healthy and necessary sources of regularization \citep{carrasco2011optimal}.   First,  there is $\lambda$ penalizing extreme allocations and shrinking $\boldsymbol{w}'\boldsymbol{r}_{t+h}$ to the equally-weighted portfolio -- as opposed to 0 in a typical Ridge Regression.  This is due to the unconditional variance of the portfolio  \textcolor{black}{being} fixed to 1 (line \texttt{6}) for identification purposes \textit{during estimation}.\footnote{Indeed,  it is easy to see in \eqref{MACE} why this is necessary : replacing $f$ by $\zeta \times f$ and $g$ by $\zeta \times g$ (where $\zeta$ is an arbitrary scalar) gives rise to the same likelihood.  Naturally,  this cannot occur when $g$ is the identity function, as in typical regression problems, but is inevitable within ACE and its descendants. }  Thus,  everything being shrunk to  \textcolor{black}{the same} value is what remains of the original ridge ``prior'' that every coefficient is shrunk to (\textcolor{black}{the same} value of) 0.  Given that the resulting portfolio will eventually be rescaled to satisfy  the capital budget constraint ($\boldsymbol{w}'\boldsymbol{\iota}=1$),  $\sfrac{1}{N}$ is the value towards which the shrinkage is effectively pointing at.  

Another source of regularization in the Ridge step is obviously the long-only constraint $\boldsymbol{w} \geq 0$.  It embeds the prior knowledge that we ``unconditionally'' expect the market to follow an upward trajectory,  and that MACE should preferably focus on portfolios of which it will most often hold a long position.  Additionally,  for our monthly rebalancing application,  limiting the occurrences of overall short positions is desirable from a risk management perspective \citep{JagannathanMa2003}.   This frequently imposed constraint in mean-variance optimization problems also plays here the additional role of limiting the Ridge's step expressivity by chopping out a wide space of potential $\boldsymbol{w}$'s.  As in anything,  good regularization balances bias and variance wisely by imposing constraints that will contort our likelihood the least.   Accordingly,  the implicit prior motivating $\boldsymbol{w} \geq 0$ for one-month ahead forecasts may not always be as well motivated for much shorter horizons -- \textcolor{black}{and indeed, we will relax that restriction in the daily application}.

{\vspace{0.15cm}}
{\noindent \sc \textbf{Learning Rate.}} Among the few more subtle technical extensions to ACE,  we use out-of-bag block-subsampled predictions and introduce a learning rate $\eta$ -- whose combined action is mostly to curb overfitting and facilitate optimization.  Their importance in practice is paramount since we are dealing with a high-dimensional Ridge Regression on one side and a RF on the other,  with both having the ability of overfitting the \textit{training} data, even if the other side of the aisle remains static.   Directly inspired from Boosting and Neural Networks,  the use of a learning rate curbs this problem and avoids zigzagging optimization paths.   There is an obvious trade-off between $\eta$ and $s_ {max}$,  with a lower  $\eta$ necessitating a larger $s_ {max}$.   In our experience,  anything above 0.2 can quickly lead to unstable computations,  learning rates above 0.1 will often lead to overfitted solutions (when the SNR is very low),  and the optimization of larger portfolios may get suck with too small of a learning rate (like anything below 0.01).   What lies within the 0.01-0.1 range usually provides interchangeable results and the symptoms of an impotent learning rate can easily be diagnosed from looking at the path of the in-sample loss.

\subsection{Relationship to Mean-Variance Portfolio Optimization}\label{sec:mvrela}

\textcolor{black}{
Suppose we \textcolor{black}{created} a synthetic security, \textcolor{black}{i.e.,  a portfolio,} $z_{t+h}(\boldsymbol{w})$, characterized by a relative weight vector $\boldsymbol{w}$ combining single security returns. \textcolor{black}{Now, we may want to trade this portfolio. Our trading position is determined by $\omega_{t+h}$, which is the absolute position over the portfolio, a scalar, which is either $\omega_{t+h} > 0$ if we hold a long position, respectively $\omega_{t+h} < 0$ if we short our synthetic security}.  Relative weights are fixed (unless re-estimated),  but \textcolor{black}{for an investor with mean-variance preferences}, absolute weights are changing in every period based on the  forecast for $z_{t+h}(\boldsymbol{w})$,  denoted $\hat{z}_{t+h}(\boldsymbol{w})$,  and its volatility,  $\hat{\sigma}^2_{t+h}(\boldsymbol{w})$.  Setting $h=1$ for simplicity,  the problem looks like
 		\begin{align}\label{eq:r2proof}
		\underset{\omega_{t+1}, \phantom{.}\boldsymbol{w}}{\text{max}} \quad E_t \left[ \omega_{t+1} {z}_{t+1}(\boldsymbol{w}) \; - \; 0.5\,\gamma\, \omega^2_{t+1}\, {\sigma}^2_{t+1}(\boldsymbol{w})  \right] 
		\end{align}
where $\gamma$ is the risk aversion parameter.  Plugging in the solution for  $\omega_{t+1}$ conditional on $\boldsymbol{w}$ and information available up to $t$ (i.e.,  $ \omega_{t+1} = \frac{1}{\gamma} \; \frac{\hat{z}_{t+1}(\boldsymbol{w})}{\hat{\sigma}^2_{t+1}(\boldsymbol{w})}$) gives 
 		\begin{align}\label{eq:r2proof}
		\underset{\boldsymbol{w}}{\text{max}} \quad \frac{1}{ \gamma} \times E_t \left[ \frac{\hat{z}_{t+1}(\boldsymbol{w}){z}_{t+1}(\boldsymbol{w})}{\hat{\sigma}^2_{t+1}(\boldsymbol{w})} \; - \; 0.5\,\frac{\hat{z}^2_{t+1}(\boldsymbol{w}){\sigma}^2_{t+1}(\boldsymbol{w})}{\hat{\sigma}^4_{t+1}(\boldsymbol{w})}  \right] 
		\end{align}
		which,  by linearity of expectation,  noting that $E_t \left[ {z}_{t+1}(\boldsymbol{w})  \right] = \hat{z}_{t+1}(\boldsymbol{w})$ as well as $E_t \left[ {\sigma}^2_{t+1}(\boldsymbol{w})  \right] = \hat{\sigma}^2_{t+1}(\boldsymbol{w})$,  and regrouping,  is
		 		\begin{align}\label{eq:r2proof2}
		\underset{\boldsymbol{w}}{\text{max}} \quad \frac{1}{ 2 \gamma} \times  \frac{\hat{z}^2_{t+1}(\boldsymbol{w})}{\hat{\sigma}^2_{t+1}(\boldsymbol{w})} \quad \Leftrightarrow \quad \underset{\boldsymbol{w}}{\text{max}} \quad  R^2_{t+1} (\boldsymbol{w}) 
		\end{align}
		thereby justifying on economic grounds the optimization of $\boldsymbol{w}$ (and $f$) so to maximize the resulting security's $R^2$.\footnote{\textcolor{black}{To avoid any confusion,  note that $R^2$ denotes the statistical/econometric measure for the goodness-of-fit of a regression model and not -- as often in finance papers -- squared (total) returns of some asset.}} The equivalence to maximizing a traditional $R^2$ (average of $R^2_{t+1}$ over many time periods) as MACE is using is exact under homoscedasticity.  Otherwise,  \eqref{eq:r2proof2} suggests a weighted least-squares $R^2$.  In principle, it is the unpredictable variance (not total variance) which should enter the denominator,  ultimately changing the formula to $\frac{ R^2_{t+1} (\boldsymbol{w}) }{1- R^2_{t+1} (\boldsymbol{w}) }$, which for small $R^2_{t+1}$,  like what prevails at the daily frequency,  alter in a very minor fashion the overall utility.  Note that whatever the variance entering the denominator is, the arg max remains the same since both are monotonically increasing functions of $R^2_{t+1}$.}
		
MACE constructs such a portfolio to be actively traded.  Nonetheless,  as we will see later,  the raw MMLP portfolio  often has nice properties when combined with much more passive trading (e.g.,  using a prevailing mean instead of RF).   In population,  when $f$ is a non-trivial function and the ``true'' $R^2$ is larger than zero,  MACE solves
		\begin{align}\label{death}
\min_{\boldsymbol{w}, \enskip f} \mathrm{Var}[z_{t+h}(\boldsymbol{w}) \perp f(\mathbf{X}_t) ] + \lambda ||\boldsymbol{w}||^2 \quad \enskip  \text{such that} \enskip \boldsymbol{w} \geq 0  \enskip  \text{and} \enskip \mathrm{Var}[z_{t+h}(\boldsymbol{w})]=1
	\end{align}
	where $z_{t+h}$ is the portfolio return $h$-steps ahead.  Thus, minimizing the error term in \eqref{MACEpp} is equivalent to minimizing the residual variance of the portfolio, that is,   the share of variance unexplained by the conditioning information. The apparition of the $\mathrm{Var}[z_{t+h}(\boldsymbol{w}) \perp \enskip f(\mathbf{X}_t)]$ term highlights an opportunity:  handling more \textit{unconditional} variance in the buy-and-hold return if some of it is  neutralized by trading based on informative signals.   This formulation highlights an imminent tension for $\boldsymbol{w}$ in the MACE problem because of its dual mandate:  keeping  $ \mathrm{Var}[z_{t+h}(\boldsymbol{w}) ]$ to a pre-specified level (up to a scaling constant) and minimizing $\mathrm{Var}[z_{t+h}(\boldsymbol{w}) \perp \enskip f(\mathbf{X}_t)] $.  Those may push $\boldsymbol{w}$  in the same direction, or not -- depending on what lies in $\mathbf{X}_t$ and the shape of $f$.  The benefits of all forms of regularization on risk-reward ratios also become obvious from \eqref{death}.  An overconfident MACE allocation will inflate $\mathrm{Var}[z_{t+h}(\boldsymbol{w})|\enskip f(\mathbf{X}_t)] $ in-sample.  If it fails to replicate predictive gains out-of-sample,  it is likely left with a portfolio which orthogonal variance may not be lower than that of simpler benchmarks.  Thus,  as is well known in the return predictability literature,  generalizable patterns deliver large utility gains,  but are not easily found  \citep{WelchGoyal2008}.  Note that,  in the absence of any predictability (i.e.,  in the true DGP,  the conditional mean is the unconditional mean ($f (\mathbf{X}_t)=\mu \enskip \forall  t)$),   \eqref{death} collapses to 
		$$\min_{\boldsymbol{w}} ||\boldsymbol{w}||^2 \quad \enskip \text{such that} \enskip \boldsymbol{w} \geq 0  \enskip  \text{and} \enskip \mathrm{Var}[z_{t+h}(\boldsymbol{w})]   =1$$ 
	which, in this limit case,   delivers the equal weights allocation.  As is well \textcolor{black}{documented} in the literature \citep{DeMiguelEtAl2009},   equal weights is a competitive benchmark,  and therefore,  a natural direction \textcolor{black}{which to shrink towards}.

	\subsection{MACE vs. Predicting Single Stock Returns Separately}\label{sec:ssrsrela}


When it comes to time series predictions of stock returns,  a popular ML approach is to conduct a pooled (nonlinear nonparametric) regression for a panel of stocks and their corresponding characteristics.  \cite{gu2020empirical} is the prime example, and they translate their predictions into returns via a long-short portfolio strategy.  Alternatively,  one can model each stock return separately with  its own time series regression, but this has important limitations.  In contrast to the above,  MACE forecasts a single series,  the portfolio's return.  From the linearity of the portfolio,  we have that ${\rm I\kern-.3em E}[z_{t+h}(\boldsymbol{w})|\mathbf{X}_t] = \boldsymbol{w}'{\rm I\kern-.3em E}[\boldsymbol{r}_{t+h}|\mathbf{X}_t]$ where ${\rm I\kern-.3em E}[\boldsymbol{r}_{t+h}|\mathbf{X}_t]$ is a vector of conditional expectations for each stock.    Thus,   one can legitimately wonder why not simplifying the algorithm considerably by (i) getting expectations from pooled or individual predictive regressions and then (ii) running the global minimum variance problem on residuals from such regressions.  Precisely,  solving    
		\begin{align*} 
\min_{\boldsymbol{w}} \sum_{t=1}^T \left(\boldsymbol{w}'(\boldsymbol{r}_{t+h} -{\rm I\kern-.3em E}[\boldsymbol{r}_{t+h}|\mathbf{X}_t])\right)^2 + \lambda ||\boldsymbol{w}||^2 \quad \enskip  \text{such that} \enskip \boldsymbol{w} \geq 0  \enskip  \text{and} \enskip \boldsymbol{w}'\boldsymbol{\iota}=1
	\end{align*}
after obtaining ${\rm I\kern-.3em E}[\boldsymbol{r}_{t+h}|\mathbf{X}_t]$ externally.  There are quite a few reasons not to consider such a route,  some conceptual and others, practical.  \textcolor{black}{All of them are worth mentioning here} because they highlight some of MACE's advantage that may so far \textcolor{black}{have gone} unnoticed.

${\rm I\kern-.3em E}[{r}_{i,t+h}|\mathbf{X}_t]$ is arguably much harder to learn than typical $z_{t+h}$ candidates from aggregate data, simply by the virtue of the latter being a portfolio.  Single stock returns contain a lot of variation that cannot be captured by macroeconomic predictors,  and with a small fraction of it being explainable by micro-level firm characteristics,  often for low-capitalization stocks.  What remains is a large amount of noise weakening the potential $f$ through an unappealing SNR and increased estimation error.  This crucially matters because (i) the extremely low SNR for separate stock returns is a serious impediment to any algorithm attempting to learn ${\rm I\kern-.3em E}[{r}_{i,t+h}|\mathbf{X}_t]$ and (ii) the chosen \textit{individual} model will often be one that puts a higher weight on minimizing estimation variance rather than entertaining ambitions to tackle bias$^2$.  Thus,  it can easily turn out that the selected/cross-validated $f$ is the null function (or close to it) whereas the true DGP does, in fact,  have an $f$ yielding a positive  $R^2$.    In other words,  choosing and optimizing ML (or any) models to predict ${r}_{i,t+h}$ separately might be at odds with the final objective of getting a fine estimate of ${\rm I\kern-.3em E}[z_{t+h}(\boldsymbol{w})]$.  And because of that,  a positive  $R^2$ remains unattainable without exceedingly large samples.  

One way out is the pooled (or global) regression approach with firm-level characteristics,  where $f$'s potency is revived through much more data and information on cross-sectional variation.  Another route is to predict directly what one will end up trading,  that is,  the portfolio return.  By the joint optimization of $\boldsymbol{w}$ and $f$,  $\boldsymbol{w}$ provides $f$ a more easily-forecastable target.  Hence,  the previously undetected $R^2>0$ becomes an attainable target because $\boldsymbol{w}$ collaborates in making $f$ win at the bias-variance trade-off.   This is convenient since getting a  ${\rm I\kern-.3em E}[\boldsymbol{r}_{t+h}|\mathbf{X}_t]$ vector worthy of use is not necessarily an easy task,   requiring large amounts of micro-level data not always easily accessible in real time,  and a fair amount of computing resources.  In comparison,  MACE finds profitable predictability in a convenient low-maintenance setting.

Given the attention they get,  predicting indexes such as the S\&P 500 rarely deliver sizable $R^2$s at short horizons.  But there are numerous ways into which stocks can be assembled,  and some of those blends may be more promising than others from a predictive viewpoint.  Linking it back to the econometric literature on the benefits and costs of aggregation (forecasting aggregates vs. aggregating components' forecasts),  MACE can be seen as finding the optimal aggregation that keeps variance low (by aggregating) and yet keeps bias$^2$ similarly  low by creating an aggregate with limited aggregation bias from neglecting heterogeneity \citep{lutkepohl2011forecasting}.

\subsection{Why Random Forest?}\label{sec:whynot}

A natural question to ask is: why Random Forest? In principle,  RF could be replaced by any ML algorithm.  In practice,  not quite so, and for many reasons.   First,  RF is the only algorithm which provides \textit{internally} out-of-bag predictions.   Obviously,  nothing prevents a very patient researcher to bootstrap-aggregate Boosting and Neural Networks {at every iteration} $s$ and incur a substantial computational burden.  This is especially true of applications with many observations and regressors.  Putting things in  perspective,  to obtain rightful $ f_s^*(\mathbf{X}_t)$'s from Boosting and NN,  it would take approximately 500 times (a reasonable number of bootstraps) longer than RF,   assuming that the three algorithms have a roughly similar computational time (which is quite generous to Boosting and NN in this application).  Alternatively,  one can ditch any call on to OOB predictions,  and extremely carefully tune hyperparameters.  Given the impracticability of such an approach (for anything more complicated than Ridge, Lasso, and derivatives) and known results about the virtues of cross-fitting and analogous methods \citep{chernozhukov2018double},  it appears that  justifying the costs of going for such an alternative route would require glowing expected benefits.  

There aren't.  Boosting, which, is often seen as marginally superior to RF in tabular data tasks,  often does so by providing mostly small improvements in high SNR environments -- a far cry from our financial application.  In fact,  with RF being less capricious tuning-wise,  it has been reported in many low SNR applications to be equally if not more competitive than Boosting  (see \cite{gu2020empirical} and  \cite{krauss2017deep} for returns,  and \cite{GCLSS2018} and \cite{MDTM} for macroeconomic forecasting).  Additionally,  extensive tuning is often required for Boosting to have an edge on RF,   which is highly impractical within an iterative procedure. 

Deep Neural Networks,  which incredible merits in non-tabular data tasks are indisputable \citep{deeplearning},  are known to still take the backseat to tree-based methods when it comes to tabular data \citep{JanuschowskiEtAl2022, grinsztajn2022tree}.  Moreover,  it has been the subject of considerable discussion that basic feature-engineering (like creating lags) combined with tree-based methods may outperform NNs with architectures tailored for time series data \citep{elsayed2021we}.  Of course,  none of this rejects the possibility that letting $f$ be constructed from some sophisticated deep recurrent network of any breed (like those in \cite{babiak2020deep}) could further improve results.  Rather, what it suggests is that this paper's results will not be severely handicapped by leaving the aforementioned extensions for future research.

A last deep learning-based alternative is to consider a neural network with two hemispheres as in \cite{HNN} -- one linear for the LHS and one for the RHS -- with a loss function being the squared distance between their respective outputs,  reminiscent of developments in \cite{andrew2013deep} and \cite{michaeli2016nonparametric}.  This ditches the need for alternating anything and can be optimized directly through gradient descent.  There are quite a few complications,  however.  The first, quite subtle in nature,  is that modern neural networks,  very large and deep,  vastly overfit the data in-sample, yet produce stellar out-of-sample performance for many tasks.  This phenomenon has many names -- double descent and benign overfitting among them -- and now has a theoretical literature of its own \citep{belkin2019reconciling,hastie2019surprises,bartlett2020benign}.  The problem this poses for building a MMLP is that,  if the most promising neural network attains a $R^2_{\text{train}} \approx 1$ fitting what is almost pure noise,  there is \textcolor{black}{neither} room nor need \textit{in-sample} to have the LHS collaborate in increasing the fit.  Given that our application has a SNR which is a far cry from 1 and \textcolor{black}{those of other} typical successful deep learning applications, it is a severe complication.  Nevertheless,  using (often unstable) small networks,  carefully crafting their design,  and considering an extensive hyperparameters search -- all this to avoid the slightest bit of overfitting in $ f_s^*(\mathbf{X}_t)$ --  could maybe make the derivation of MMLPs from such an approach less ill-fated.  In the concluding remarks,  we provide additional thoughts and suggestions on how this could perhaps be done in future research.

\subsection{Setting Hyperparameters}\label{sec:hyp}

Given that MACE incorporates two ML algorithms,  it inevitably has quite a few hyperparameters (HP),  which,  for convenience,  are summarized in Table \ref{hptable}.  Fortunately,  RF is well known to be very robust to tuning parameters choices (with default values often very hard to beat,  see \cite{MSoRF} and references therein),  and ridge is sparsely hyperparametrized.  We opt for setting tuning parameters to fixed values,  \textcolor{black}{with their} calibration  \textcolor{black}{being} motivated \textcolor{black}{by} domain knowledge and observations of the (block) out-of-bag error metric ($RMSE_{\text{OOB}}$).  

Given the nature of the MMLP problem --  the prediction of a non-fixed target -- considering a validation set as is often seen in ML forecasting studies in economics and finance \citep{gu2020empirical,GCLSS2018} is an avenue with strong headwinds and likely limited benefits.  This is due to the multiplicity of solutions where identical HPs,  when re-estimating the model with more data (e.g., reincorporating the validation data),  can lead to different solutions.  This is not unique to MACE in the ML realm,  as this is a commonly known feature of modern neural networks.  Also, there is an imminent tension between maximizing the reliability of the validation set and minimizing the likelihood of moving to a different optimum than what we optimized the hyperparameters for.  The former calls for a longer validation set and the latter for a shorter one.  In the light of all that,  it appears more reasonable to rely on common sense whenever possible,  and on the blocked $RMSE_{\text{OOB}}$,   which is a proper CV metric for time series \citep{bergmeir2018note},  whenever data-driven guidance is needed.

  \begin{table}[t] 
		\vspace*{0.25em}
	\centering
	\rowcolors{2}{white}{gray!15}
\small
\caption{Summary of Tuning Parameters and Their Values in Applications}\label{hptable}
	\begin{threeparttable}
	\vspace*{-0.75em}
 \begin{tabular}{l| c c c c c c } 
\toprule \toprule
\addlinespace[1pt]
& &  Monthly Data  &  &  Daily Data ($N \in \{20,50\}$) &  & Daily Data ($N = 100$)  \\
\midrule
\addlinespace[5pt] 

$\eta$ & & 0.1  &  & 0.01  &  & 0.05   \\ \addlinespace[2pt]
$s_{\text{max}}$  & & 100  &  & 250  &  & 500  \\ \addlinespace[2pt]
\texttt{stopping.rule}  & & $s=s_{\text{max}}$  &  & early stopping  &  & early stopping  \\ \addlinespace[2pt]
$\lambda$ & & $ R^2_{s,\text{train}}(\lambda)=0.05$  &  & $ R^2_{s,\text{train}}(\lambda)=0.01$ &  & $ R^2_{s,\text{train}}(\lambda)=0.01$ \\ \addlinespace[2pt]
\midrule
\texttt{mtry} & & $\sfrac{1}{3}$ &  &  $\sfrac{1}{10}$  &  & $\sfrac{1}{10}$  \\ \addlinespace[2pt]
$\texttt{minimal.node.size}$  & & 20  &  & 200  &  & 200  \\ \addlinespace[2pt]
\texttt{block.size} & & 24 months  &  & 2 months  &  & 2 months  \\ \addlinespace[2pt]
\texttt{subsampling.rate} & & 80\%  &  & 80\%  &  &  80\%  \\ \addlinespace[2pt]
\texttt{number.of.trees} & & 500  &  & 1500  &  &  1500  \\ \addlinespace[2pt]

\bottomrule \bottomrule
	\end{tabular}
	\begin{tablenotes}
	\scriptsize
		\item[] \hspace*{-0.5cm}{Notes}: Hyperparameters in the upper part relate to the algorithm as a whole, while those in the bottom part pertain exclusively to RF.
	\end{tablenotes}
\end{threeparttable}
\end{table}

We first concentrate on those HPs pertaining to MACE's iterative optimization itself.  The learning rate $\eta$ is set at 0.1 for monthly data,  which always delivers a stable solution and never gets stuck.  For daily data,  more care is needed given that $\boldsymbol{X}_t$ is not fixed.  The smallest learning rate always appears desirable -- as often observed for Boosting \citep{ESL} -- but we have noticed that too small of a $\eta$ coupled with large portfolios may lead to the algorithm not optimizing at all in-sample (analogous to exaggeratedly tiny learning rates for deep learning).  Thus,  it is set to 0.01 for $N \in \{20,50\}$,  but 0.01 being not large enough for $N=100$ in-sample loss to start decreasing,  we increase it to 0.05.

Closely related are the choice of $s_{\text{max}}$, the maximal number of iterations, and the stopping method.  For monthly frequency,  we find that $s_{\text{max}}=100$ is well enough for MACE to converge and that performance never seems to deteriorate substantially with $s$,  even after some plateau is achieved.  Thus,  the stopping criterion is simply $s^* = s_{\text{max}}$. Things are not so easy with the daily application where regressors are created endogenously.   First,  we set $s_{\text{max}}=250$ since $\eta$ is considerably smaller.  Since optimization is much more demanding in this environment,  it is not impossible for the OOB error to start increasing beyond a certain $s$ -- i.e.,  suggesting the LHS is starting to overfit.  Hence,  we set $s^* = \argmin_s RMSE_{\text{OOB}}(s)$, which can be seen as some form of internal early stopping,  a key player in the regularization arsenal of modern deep neural networks \citep{deeplearning}.  Early stopping is typically implemented using a validation set,  but here,  for aforementioned reasons,  it is  more preferable to use RF's internal error metric. 

The next four hyperparameters in Table \ref{hptable} are those of RF.   For monthly data,  $\texttt{mtry}$ is set to the default value of $\sfrac{1}{3}$,  one that is typically hard to beat,  except in extremely low SNR environments \citep{ESL,olson2018making}.  \textcolor{black}{At the daily frequency},  we noticed that $\texttt{mtry}=\sfrac{1}{3}$ could never deliver $RMSE_{\text{OOB}}(s)<1 $ for any $s$.  Given that the daily application has a much lower SNR and a sparser $\boldsymbol{X}_t$ (hence,  less potential for diversification in RF \citep{MSoRF}), it is  not entirely surprising that  $\texttt{mtry}=\sfrac{1}{3}$  might be too large and lead to early overfitting.  Thus,  we set $\texttt{mtry}=\sfrac{1}{10}$ which kills two birds with one stone by decreasing computing time sharply.  

In a similar spirit,  $\texttt{minimal.node.size}$ is set to a very high value of 200 in the daily application (nonetheless $\approx \sfrac{1}{25}$ of the training sample size),  which greatly helps in easing the daily application's computational burden all the while helping improving performance as measured by the OOB.  Default values for $\texttt{minimal.node.size}$ usually go up to 10.  However,  when faced with a low SNR,  deep trees are either redundant or harmful,  because additional splits allowed by $\texttt{minimal.node.size}=10$  vs. $\texttt{minimal.node.size}=200$ are typically fitting the noise and cancel out through bagging in the out-of-sample projection \citep{MSoRF}.  Limiting the expressivity of RF is not without precedent for predicting returns,  as \cite{gu2020empirical} report using trees of very limited depth.   For the monthly frequency application, we set it to $\texttt{minimal.node.size}=20$,  a moderately high value that eases computations without any apparent effect on $RMSE_{\text{OOB}}(s)$ (vs. default values).

The next two tuning parameters of RF are \texttt{subsampling.rate} and \texttt{block.size}.  We set $\texttt{subsampling.rate}=80\%$ for all applications,  which is standard.   \texttt{block.size},  on the other hand,  needs to be chosen slightly more carefully to balance two goals.  As already hinted at above,  too low of a block size will lead to $ f_s^*(\mathbf{X}_t)$ including fitted noise,  to be subsequently fed into the Ridge Regression.  An overkill block size will seriously handicap bagging by limiting the number of different block combinations used to construct the trees in the ensemble,  ultimately weakening RF (on the variance side) by decreasing the diversity of trees.  Thus, we set  $\texttt{block.size}$ to be a window size within which any form of meaningful dependence between the first and the last observations will have faded away,  for both $\boldsymbol{X}_t$ and $\boldsymbol{r}_{t+h}$.  For the monthly application, we set it to two years,  which very well cover the dependence in $\boldsymbol{r}_{t+h}$ and the stationarized  \cite{WelchGoyal2008} predictors in $\boldsymbol{X}_t$.  
In the daily application,  since $\boldsymbol{X}_t$ and $\boldsymbol{r}_{t+h}$ are daily returns with minimal time series dependence,  two business months appears more than sufficient to maintain the interchangeability of the blocks, and leaving plenty of room for bagging to fulfill its task. 

A last hyperparameter for RF is the number of trees.  It is \href{https://philippegouletcoulombe.com/blog/the-number-of-trees-in-random-forest-is-not-a-tuning-parameter}{not a tuning parameter} \textit{per se} because there is no statistical trade-off for its choice: the larger the better, with the only constraint being computational burden \citep{ESL}.  Given that RF predictions usually stabilize (by the law of large numbers for an average) well before 500 trees,  500 is usual the default setting for \texttt{number.of.trees} in many RF implementations and is what is used for the monthly application.  Yet, there is a subtle twist, that leads us to increase \texttt{number.of.trees} up to 1500 for the daily application.  We are generating regressors endogenously -- in essence using lags of a continuously updated target --rather than taking $\boldsymbol{X}_t$ to be fixed observed predictors.  In doing so, we may incur attenuation bias in RF attributable to the generated regressor problem.   More precisely,  ``measurement error'' can impair RF's ability to detect nonlinear time-series dependence because,  $ f_s^*(\mathbf{X}_t)$ being out-of-bag predictions,  there are an average of $(1-\texttt{subsampling.rate}) \times \texttt{number.of.trees}$ trees,  which falls to 100 with  $\texttt{number.of.trees}=500$.  Bumping \texttt{number.of.trees} to 1500 makes it an average of 300 single tree predictions \textcolor{black}{for $ f_s^*(\mathbf{X}_t)$},  which is enough to curb measurement error problems without exploding computational costs.

Finally,  we must set a value for $\lambda$ in the Ridge Regression step.  While it could be tempting to cross-validate $\lambda$ internally at each step,  the optimally chosen $\lambda$ at each  $s$ may be well off from that of the final $s$, and lead optimization in a poor direction.  For instance,  in early steps,  cross-validation can easily choose a $\lambda$ that shrinks the portfolio \textcolor{black}{excessively},  because, at that early stage $s$,  there is, indeed, very little or no predictability being detected.   Moreover,  on top of withstanding the additional computational demand,  changing  $\lambda$ may lead to certain steps not improving the loss, \textcolor{black}{thereby} impairing the EM-style algorithm's ability to minimize the overall loss.  Therefore,  for the monthly application,  we set $\lambda$ such that it attains an in-sample $R^2$ of 0.05, which is a high  yet not unreachable mark.   In the daily application,  facing the inevitability that $R^2_{s,\text{test}}$ is unlikely to stand above 1\%,  $\lambda$ is chosen at each step so to target $R^2_{s,\text{train}}=0.01$.

 \section{Application I: Daily Stock Returns Prediction}\label{sec:app2}                    

We begin our exploration of MACE by constructing maximally ML-predictable portfolios at  the daily frequency.  The high frequency brings both opportunities and difficulties.  Among the former is the availability of more data points,  and a lessened need to rely on data going way back to the late 1950s,  as is typically the case in monthly exercises.   In terms of the latter,  an even more hostile signal-to-noise ratio comes to mind,  as well as the scarcity of freely available predictors at such frequency.

Here, $\boldsymbol{r}_{t+1}$ comprises individual stock returns for firms listed on the NASDAQ.  We keep \textcolor{black}{$N \in\{20,50,100\}$} of them with highest market capitalization on January 3$^\text{rd}$  2017, which is the date of the beginning of our test sample.   Hence,  there is no look ahead bias for the \textit{test} sample and the stocks are as liquid as it gets.   For computational reasons,  we do not re-estimate models  and consider a fixed train-test split of the data,  as done in,  e.g.,  \cite{ChenPelgerZhu2021} and \cite{fallahgoul2020asset}.  The training set is thus 2000/03/02--2016/30/12 and the test set 2017/03/01--2022/07/12.  Thus, the test set includes a fair  level of variety in terms of ``financial regimes''.  In chronological order, we have: a relatively quiet period,  a crisis period with extremely high volatility,  an unprecedentedly bullish bull market,   and a long-lasting bear market.   

As mentioned above,  gathering many exogenous predictors available in real-time for $\mathbf{X}_t$ is not easy and often not cost-free.  In this application,  we will see whether MACE can generate predictability with nothing more than time-series properties of the portfolio.  Creating portfolios or stock combinations that have exploitable persistence properties (for mean reversion- or momentum-based trading  strategies) has been studied,  for instance, in \cite{Aspremont2011}.  Nonetheless,  the near universe of following studies, by the limitations of relying on variants of canonical correlation analysis,  are bounded to consider only linear autoregressive properties.  Needless to say,  if there is any remaining form of mean reversion that has not been drilled out yet,  it will need to be complex in $f$ or $g$ -- or both.  Thus,  $f$ being a RF able to estimate any form of nonlinear nonparametric dependence may help in finding mean reversion patterns that remained undetected to the naked eye or simpler algorithms.  

Beyond its coverage of heterogeneous financial conditions,  the test set is interesting in its own right simply by the virtue of being \textit{recent}.  \cite{krauss2017deep} finds substantial gains from a simple daily  long-short strategy using signals from predicting single stocks separately via tree-based techniques (among others) with lag returns as predictors (as we use).  However,  and now a recurring theme,  the improvements are circumspect to the pre-2010 era,  which, as the authors note,  is likely due  to the widespread dissemination in the 2010s of the very techniques they use.  Thus,  an interesting question is whether MACE (and other less obvious uses of ML) will find exploitable nonlinear mean reversion in a period where simpler methods could not.    In a similar spirit,  localized episodes of predictability,  reasonably frequent before 2000,  are found to be a much rarer event afterwards \citep{FarmerEtAl2022}.  A related question is whether MACE,  through ML-based nonlinearities in $f$,  can capture and exploit those in real time rather than only observing them ex-post.

It is natural to wonder whether MACE-based predictability will outlive its test set,  that is,  after the profitable pattern is openly communicated to other market participants.  MACE's design partly protects against early depletion by (i) forecasting a synthetic security (rather than highly scrutinized stocks or portfolios) and (ii),  doing so with a mostly opaque model.  Additionally,  as discussed in section \ref{sec:HNN},  MACE can generate plenty of solutions,  with most of them achieving the predictability goal in different ways.  Hence,  in principle, there should be no shortage of possibilities for enlarging the set of nonlinearly mean-reverting synthetic securities,  especially keeping in mind that our application deliberately focuses on creating them from a narrow set of well-known stocks.

{\vspace{0.15cm}}
{\noindent \sc \textbf{Nonlinear Mean Reversion Machine.}}  In what follows, we describe the remaining building blocks necessary to go from plain MACE to a version specialized for daily data.  Throwing many lags of many stocks at the daily frequency directly in $\mathbf{X}_t$ will be at the nexus of computational and statistical inefficiency.   A more manageable and promising route is the parsimonious problem  
\begin{align}
	\min_{\boldsymbol{w}, \enskip f} \sum_{t=1}^T (\boldsymbol{w}'\boldsymbol{r}_{t+1} - f (\left[ \boldsymbol{w}'\boldsymbol{r}_{t-1},  \dots , \boldsymbol{w}'\boldsymbol{r}_{t-21} \right]))^2 +\lambda ||\boldsymbol{w}||^2 \quad  \enskip \text{such that}   \enskip  \mathrm{Var}[\boldsymbol{w}'\boldsymbol{r}_{t+1} ] =1
\end{align}
where features -- lags of the portfolio returns -- are created endogenously given the portfolio weights.\footnote{Note that for a linear $f$,  this optimization problem could be solved by nonlinear least squares or some sort of generalized eigenvalue problem as studied in \cite{Aspremont2011} and others.  } There are two substantial modifications  to Algorithm \ref{MACEalgo}.  The first is that we drop the $\boldsymbol{w} \geq 0$ constraint. \textcolor{black}{Keeping such a constraint in place  -- as in the monthly application of section \ref{sec:mres} -- would push} MACE to find portfolios for which it will most often go long with,  and avoid relying extensively on short-selling to turn a profit.   Such a restriction \textcolor{black}{comes} with the prior that,  at the lowest frequencies,  the market is always expected to go upward and that a successful strategy should not immensely deviate from this evident fact about the unconditional mean.   At the daily frequency,  however,  the importance of the low-frequency component justifying a long position for the long-run is much tinier-- and we want the configuration of the daily MACE to absorb this knowledge.  More importantly,  $\boldsymbol{w} \geq 0$ appears unnecessary given that the portfolio will be bought or sold plausibly every day, and single-stock short positions will be short-lived by construction.  The relaxation of this constraint now allows MACE to simultaneously hold short and long positions over single assets,  even though we may go long or short with the overall portfolio.   

The second modification is, obviously,  the inclusion of an additional step before the RF step that creates lags of the portfolio as it is constituted at iteration $s$.   Given the SNR faced in this application,  cleverly designed regularization is key.   For that sake,  we apply the MARX (Moving Average Rotation of $X$) transformation of such lags \citep{MDTM} and stack those in $\mathbf{X}_t$.  As argued in \cite{MDTM},  the transformation implies more approximate implicit regularization at high frequencies than raw lags themselves.   Precisely,  in a linear model with an $l_2$ or $l_1$ norm on coefficients,  this switches the model from shrinking  each coefficient towards zero to being shrunk to one another successively.   For more complex ML methods where the implicit regularization (almost always entailing the prior that each feature should contribute,  but marginally) cannot easily be altered by changing the penalty (like RF or Neural Networks),  MARX is a trivial additional step that can help bolster predictability by embedding a more appropriate prior in the model.  Its implementation is simply a one-sided moving average of the lagged portfolio return (of  increasing length,  up to a month) instead of raw lags --- which, in the current application, is analogous to a basket of momentum indicators. \footnote{Note that in a linear model with no regularization, by the virtue of MARX being a rotation of $\mathbf{X}_t$,  it  does not alter the span of predictors and yield identical fitted values.  However,  this is not true of regularized and/or nonparametric methods, which is our prediction tool in this paper.}

{We benchmark MACE for each $N \in \{20,50,100\}$ against a set of relevant and informative competitors: equal weights\textcolor{black}{, i.e. $w_t = \frac{1}{N}$,} and the minimum variance portfolio.  The latter correspond to the initialization values of MACE.  We also include the S\&P 500.  Those are all predicted with a prevailing mean and RF.  Going forward we denote the prevailing mean models as EW (PM), MinVar (PM), S\&P 500 (PM) and the RF models EW (RF), MinVar (RF), and S\&P 500 (RF) respectively.   RF models are also procured with MARX-transformed features.}  Finally,  we complement those with MACE (PM),  which is a portfolio constructed with Algorithm \ref{MACEalgo} but where RF predictions are substituted out-of-sample by MACE's portfolio prevailing mean. This serves the purpose of evaluating MACE's raw portfolio return (since, in effect,  it comes from a well-defined mean-variance problem as per section \ref{sec:mvrela}'s discussion) and quantifying how much of MACE's success comes form leveraging predictability.

{\vspace{0.15cm}}
{\noindent \sc \textbf{Trading.}} Relative weights are fixed,  but absolute weights (i.e.,  the overall position on the synthetic asset or portfolio) are changing in every period.  To transform predictions into trading positions for economic evaluation metrics,  we solve the prototypical mean-variance problem for a single return $y_{t+1}$
 		\begin{align}\label{eq:unimv}
		\underset{\omega_{t+1}}{\text{arg max}} \quad \omega_{t+1} \hat{y}_{t+1} \; - \; 0.5\,\gamma\, \omega^2_{t+1}\, \hat{\sigma}^2_{t+1} \quad \Rightarrow \quad \omega_{t+1} = \frac{1}{\gamma} \; \frac{\hat{y}_{t+1}}{\hat{\sigma}^2_{t+1}} \; ,
		\end{align}
		as is laid out in \cite{FilippouRapachThimsen2021} and many others.  The risk aversion parameter $\gamma$  is set to 5 and $\hat{\omega}_{t+1}$ is constrained to lie between -1 for 2 for reasonable allocations.

{\vspace{0.15cm}}
{\noindent \sc \textbf{Evaluation Metrics.}} 
\textcolor{black}{The evaluation metrics are the out-of-sample $R^2$,  average annualized return ($r^A$),  and the annualized Sharpe Ratio ($SR$) -- where returns are collected from the trading exercise as described in Equation \eqref{eq:unimv}.  Following \cite{campbell2008predicting} and the ensuing literature, the out-of-sample $R^2$ of model $m$ for forecasting portfolio return $y_{t+1}$ is defined as $1-\sfrac{MSE_m^{\text{OOS}}}{MSE_{\text{PM}}^{\text{OOS}}}$ where PM stands for the prevailing mean and $MSE_{m}= { \frac{1}{\#\text{OOS}}\sum_{t \in \text{OOS}} (y_{t+1}-\hat{y}_{t+1|t})^2}$.  PM is specified to be the historical mean of the training sample. 
Given the inherent unpredictability of financial markets,  this not-so naive benchmark is in fact one that is notoriously difficult to beat. } 

To bring further enlightenment,  we also report  $R^2$ for key subsamples in Table \ref{tab:summstats_daily}.  Namely,  we report $R^2_{\text{CovidW1}}$,  which is the $R^2_{\text{OOS}}$ for the onset of the first wave of Covid-19,  defined as February,  March,  and April 2020.  Those 3 months were characterized by a level of volatility unseen since the financial crisis,  and consists of the only recession in our test set.  It is well documented that predictability is more likely to be found during bad economic times (see \cite{LiZakamulin2020} and the many references therein).  Thus,  we wish to investigate whether (i) MACE  follows that rule and (ii) if it can find predictability outside of the recessionary episode.  Accordingly,  $R^2_{\neg \text{CovidW1}}$ is the $R^2_{\text{OOS}}$ excluding those three months.  	In a similar spirit to $R^2_{\text{CovidW1}}$,   we  include $R^2_{\text{2022}}$ (the $R^2_{\text{OOS}}$ from the first business day of 2022 until the end of our sample in December 2022) and $r^A_{\text{2022}}$, which  is the corresponding annualized return for the same era.  This allows to shed light on (i) whether there was any meaningful predictability to be found in the long-lasting bear market of 2022 and (ii) whether active daily trading with MACE or other RF-based strategies can avoid the sharp losses of the US stock market in 2022.     

We complement this extended set of metrics with \cite{keating2002universal}'s Omega Ratio ($\Omega$),  an increasingly popular measure of the risk-reward ratio that leverages all the moments of the distribution of returns (whereas $SR$ only exploits the first two).  This measure is particularly useful in situations where the distribution of returns is skewed,  as only negative deviations from \textcolor{black}{a certain} threshold -- \textcolor{black}{e.g. the mean return of a benchmark, or an investor's desired average return} -- contribute to the risk component.  First,  we do find \textit{positive} skewness in MACE-based returns.  Second,  and in a more striking fashion,  RF-based returns, when predictability is non-negligible,  are found to be much more leptokurtic (i.e.,  Laplacian-looking) than those of other strategies, even when excluding CovidW1.  This is not unheard of for ML-based strategies \citep{ChenEtAl2021}.  Thus,  to compare apples with apples without the inherent assumption of normality in $SR$ and to avoid penalizing similarly both large positive and negative returns,  we include $\Omega$ in Table \ref{tab:summstats_daily} as a supplementary risk-reward ratio.\footnote{The expected benchmark model return in $\Omega$ used as a cutoff is the mean return of the S\&P 500 in the training set \citep{balder2017risk}.  Doubling it does not alter rankings. }

\subsection{Results}\label{sec:dres}

We report relevant summary statistics for our daily exercise in Table \ref{tab:summstats_daily}.  Log cumulative return plots for the small ($N=20$) and large ($N=100$) portfolios are shown in Figure \ref{cumret_daily} and $R^2$ Comparison of MACE to Random Alternatives plots are available in Figure \ref{r2_comp_daily}.  To alleviate notation,  MACE ($N=100$, PM) will be written as   MACE$_{100}$ (PM) and other accordingly.  Additionally,  MACE using RF for prediction is simply denoted MACE.

   \begin{table}[t!]
	\footnotesize
	\centering
	\begin{threeparttable}
	\caption{\normalsize {Summary Statistics for Daily Stock Returns Prediction} \label{tab:summstats_daily}
		\vspace{-0.3cm}}
		\setlength{\tabcolsep}{0.61em}
				  \setlength\extrarowheight{2.5pt}
 \begin{tabular}{l| r r r r r r r r r | r r r r r r r r  } 
\toprule \toprule
\addlinespace[2pt]
& &  $R^2_{\text{OOS}}$ &  & $R^2_{\text{CovidW1}}$ &  &  $R^2_{\neg \text{CovidW1}}$ &  &  $R^2_{\text{2022}}$ & & & $r^A$\phantom{ } & &$SR$  &  &  $r^A_{\text{2022}}$  &  &  $\Omega$\phantom{  }  \\
\midrule
\addlinespace[5pt] 
\rowcolor{gray!15} 
 ${\mathbf{N=20}}$ & &  & &  & &   & & &  &&    &&  && &  &\cellcolor{gray!15}  \\ \addlinespace[2pt]
MACE & & \textbf{3.42}   & & \textbf{7.86}  & &  \textbf{0.56}  & &  -0.55  &  && \textbf{23.10}    && 0.99 & & 5.03 & & \textbf{1.18}  \\ \addlinespace[2pt]  
MACE (PM) & & 0.01   & & -0.05  & &  \phantom{}0.04  & &  -0.12  &  && 18.71    && \textbf{1.04} & & 0.75 & & 1.13  \\ \addlinespace[2pt]  
EW (RF) & & -4.71   & & -9.39  & &  -1.21  & &  \textbf{\color{ForestGreen} 0.33}  &  && 9.15    && 0.34 & & \textbf{\color{ForestGreen} 36.43} & & 1.02  \\ \addlinespace[2pt]  
EW (PM) & & -0.04   & & -0.06  & &  -0.03  & &  -0.10  &  && 14.28    && 0.78 & & -10.23 & & 1.08  \\ \addlinespace[2pt]  
MinVar (RF) & & -1.57   & & -2.69  & &  -0.80  & &  -0.61  &  && 7.55    && 0.42 & & 2.94 & & 1.01  \\ \addlinespace[2pt]  
MinVar (PM) & & -0.01   & & -0.04  & &  0.02  & &  -0.13  &  && 18.13    && 1.00 & & -3.32 & & 1.12 \\ \addlinespace[2pt]  
 \midrule 
 \rowcolor{gray!15} 
 ${\mathbf{N=50}}$ & &  & &  & &   & & &  &&    &&  && &  &\cellcolor{gray!15}  \\ \addlinespace[2pt]
MACE & & \textbf{0.89}   & & \textbf{3.88}  & &  -1.07  & &  -0.14  &  && \textbf{20.42}    && 0.91 & & \textbf{18.41} & & \textbf{1.14}  \\ \addlinespace[2pt]  
MACE (PM) & & -0.04   & & -0.04  & &  -0.04  & &  -0.03  &  && 14.52    && 0.78 & & 4.66 & & 1.08  \\ \addlinespace[2pt]  
EW (RF) & & -1.01   & & -0.95  & &  -1.06  & &  -0.87  &  && 9.82    && 0.37 & & 4.55 & & 1.03  \\ \addlinespace[2pt]  
EW (PM) & & -0.05   & & -0.05  & &  -0.05  & &  -0.06  &  && 13.23    && 0.73 & & -8.30 & & 1.07  \\ \addlinespace[2pt]  
MinVar (RF) & & 0.75   & & 1.94  & &  \textbf{-0.01}  & &  -0.25  &  && 17.32    && \textbf{0.96} & & 3.48 & & 1.11  \\ \addlinespace[2pt]  
MinVar (PM) & & -0.04   & & -0.04  & &  -0.03  & &  \textbf{-0.02}  &  && 14.39    && 0.83 & & 7.19 & & 1.08  \\ \addlinespace[2pt] 
 \midrule 
  \rowcolor{gray!15}
 ${\mathbf{N=100}}$ & &  & &  & &   & & &  &&    &&  && &  &\cellcolor{gray!15}  \\ \addlinespace[2pt]
MACE & & \textbf{\color{ForestGreen}4.05}  & & \textbf{\color{ForestGreen} 12.20}  & &  \textbf{\color{ForestGreen} 0.86}  & &  \textbf{0.23}  &  && \textbf{\color{ForestGreen} 41.36}    && \textbf{\color{ForestGreen} 1.59} & & \textbf{23.93} & & \textbf{\color{ForestGreen} 1.33}  \\ \addlinespace[2pt]  
MACE (PM) & & -0.02   & & 0.01  & &  -0.03  & &  -0.22  &  && 15.20    && 0.91 & & -16.81 & & 1.09  \\ \addlinespace[2pt]  
EW (RF) & & 0.00   & & 1.17  & &  -0.99  & &  -0.94  &  && 9.88    && 0.38 & & -10.48 & & 1.03  \\ \addlinespace[2pt]  
EW (PM) & & -0.05   & & -0.04  & &  -0.06  & &  -0.03  &  && 12.62    && 0.69 & & -10.26 & & 1.06  \\ \addlinespace[2pt]  
MinVar (RF) & & 0.96   & & 2.06  & &  0.25  & &  -0.16  &  && 18.87    && 0.99 & & -7.01 & & 1.12  \\ \addlinespace[2pt]  
MinVar (PM) & & -0.02   & & -0.05  & &  0.00  & &  0.06  &  && 10.18    && 0.58 & & -6.39 & & 1.04  \\ \addlinespace[2pt]  
 \midrule 
S\&P 500 (RF) & & 2.88   & & 7.41  & &  -0.14  & &  0.09  &  && 13.29    && 0.64 & & 2.80 & & 1.09  \\ \addlinespace[2pt]  
S\&P 500 (PM) & & -0.01   & & -0.06  & &  0.03  & &  -0.32  &  && 11.65    && 0.69 & & -17.98 & & 1.06  \\ \addlinespace[2pt]  
\bottomrule \bottomrule
	\end{tabular}
	\begin{tablenotes}[para,flushleft]
	\scriptsize 
		\textit{Notes}: The first column-wise panel consists of out-of-sample $R^2$'s for different test (sub-)samples.  The second are economic metrics, where $r^A$ := Annualized Returns, $SR$ := Sharpe Ratio,  $r^A_{2022}$ := Annualized Returns for 2022,  $\Omega$ := Omega Ratio.  All statistics but $SR$ and $\Omega$ are in percentage points.   Returns and risk-reward ratios are based on trading each portfolio using a simple mean-variance scheme with risk aversion parameter $\gamma=5$.  PM means the prediction is based on the respective prevailing mean with a lookback period of ten years,  while RF means using that of a Random Forest.  The 4 row-wise panels are for portfolios of $N\in\{20,50,100\}$ stocks and the S\&P 500 index.  Numbers in \textbf{{bold}} are the best statistic within portfolios of the same size.  Numbers in \textbf{\color{ForestGreen} green} are the best statistic of the whole column (that is,  across all portfolio sizes and including  S\&P 500).
	\end{tablenotes}
\end{threeparttable}
\end{table}

{\vspace{0.15cm}}
{\noindent \sc \textbf{Statistical Results.}} For all three MACE specifications,  we find strong evidence of predictability through time-series dependence at the daily frequency.  Out-of-sample $R^2$'s are abnormally high and distance most of the competitors for all subsamples but 2022.  In the latter case,  only MACE$_{100}$ achieves a positive $R^2$ that narrowly beats that of S\&P 500 (RF).   The bulk of predictability is indeed found during the first wave of the Covid-19 pandemic,  with local $R^2$ for the three MACEs ranging from 3.88\%  to a stunning 12.20\%.   While MACE hits the two highest marks for the era,  unusually high  $R^2$'s (taking \cite{FarmerEtAl2022}'s local predictability results as a reasonable yardstick) are not its exclusivity.   S\&P 500 (RF) also delivers a large $R^2$ during the era (7.4\%),  and EW$_{100}$ (RF) and MinVar$_{100}$ (RF) are getting 1.17\% and  2.06\%, respectively.  

What is more exclusive, however, is predictability outside of the turbulent spring of 2020.  MACE$_{20}$ and MACE$_{100}$ achieve it both with 0.56\% and 0.86\%, which are sizable at the daily frequency, especially in \textit{good} times \citep{LiZakamulin2020,FarmerEtAl2022}.    All other RF-based models fail to deliver $R^2_{\neg \text{CovidW1}}>0$,  all situated in the vicinity of -1\% ,   except for MinVar$_{100}$  (RF) at 0.25\%,  the closest competitor to MACE on this metric.  Obviously,  negative $R^2$'s at such a frequency and with so little conditioning information are what one would expect. \footnote{In a recent evaluation from an out-of-sample period overlapping with ours,  \cite{haase2022predictability} get nearly all negative $R^2$ for 20 models using vast conditioning information for prediction at the weekly frequency. } Nonetheless,  two MACEs out of three outperform this predicament.  And it holds up in 2022.  We see that MACE$_{100}$ has a marginal outperformance during the bear market at 0.23\%.  For other models,  negative $R^2$  are again the norm rather than the exception,  with notable deviations by  S\&P 500 (RF) at 0.09\% and EW$_{20}$  (RF) at 0.33\%.  However,  in the latter case,  it delivers the worst $R^2$ of any model for the other three subsamples.  In fact,  MACE$_{100}$ is the only model with four positive $R^2$ out of four.

It is interesting to get a sense where MACE's out-of-sample $R^2$'s stand with respect to random alternatives,  like single stocks (in effect corner solutions of the MACE) and random portfolios.  Especially,  in the latter case,  it can be seen as an implicit out-of-sample statistical test for the procedure itself.  It aims at answering: if we were to draw random stock combinations and predicting them with RF rather than running MACE,  how many of those would fare better out-of-sample? The location and shape of the distribution will also be informative about how much room MACE had to find a MLPP.   We draw 150 such random portfolios with $\boldsymbol{w} \geq 0$ imposed,  and 150 without.  

Figure \ref{r2_comp_daily}  reports such results for $R^2_{\neg \text{CovidW1}}>0$ and $R^2_{ \text{CovidW1}}>0$ (forming two complementary samples).  First,  we observe how the prospects of predictability change from \text{CovidW1} to $\neg \text{CovidW1}$,  with the random portfolios distributions showing negative means and clear negative skewness without CovidW1 data for both $N=20$ and $N=100$.   $R^2_{ \text{CovidW1}}>0$'s for random portfolios see a sharp decrease in negative skewness for both portfolio size.  In fact,  for $N=100$, skewness visibly becomes mildly positive.  Additionally,  we can see a shift in location, with $N=20$ random portfolios' mean being approximately 0 and that of $N=100$ being mildly positive.  A similar pattern is observed for single stocks, but is inevitably rougher given $N<300$ and single stocks having higher variance than linear combinations of them.

\addtocounter{figure}{-1}
\begin{figure}[t!]
\captionsetup{skip=5mm}
\setlength{\lineskip}{2.2ex}
  \begin{subfigure}[b]{0.5\textwidth}
  \captionsetup{skip=0.1mm}
\hspace{-0.25cm}  \includegraphics[width=\textwidth, trim = 0mm 10mm 0mm 10mm, clip]{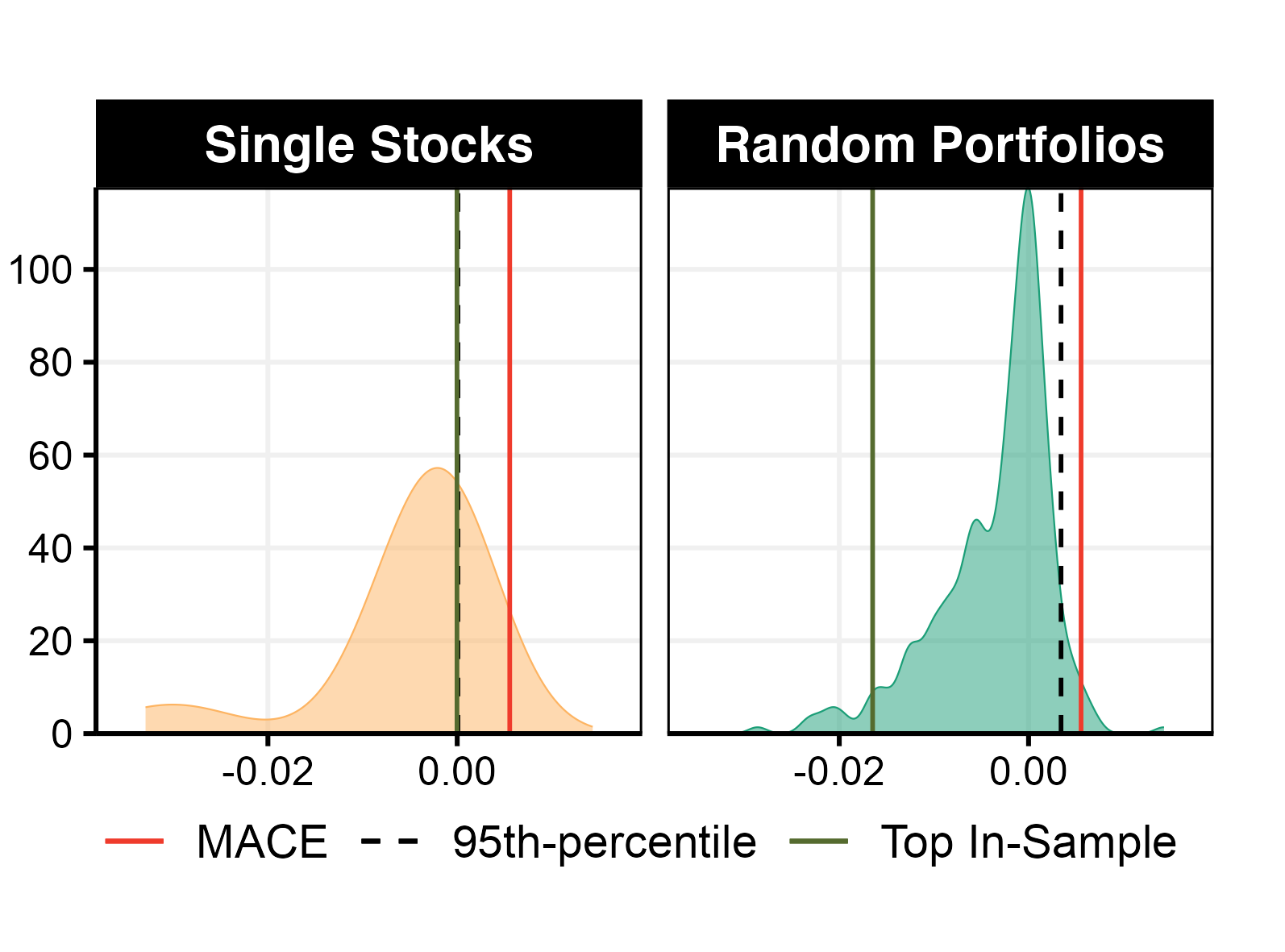}
\caption{\footnotesize  $N=20$,  $R^2_{\neg \text{CovidW1}}$}\label{r2_comp_daily_n20_nocovid}
      \end{subfigure}
  \begin{subfigure}[b]{0.5\textwidth}
  \captionsetup{skip=0.1mm}
        \includegraphics[width=\textwidth, trim = 0mm 10mm 0mm 10mm, clip]{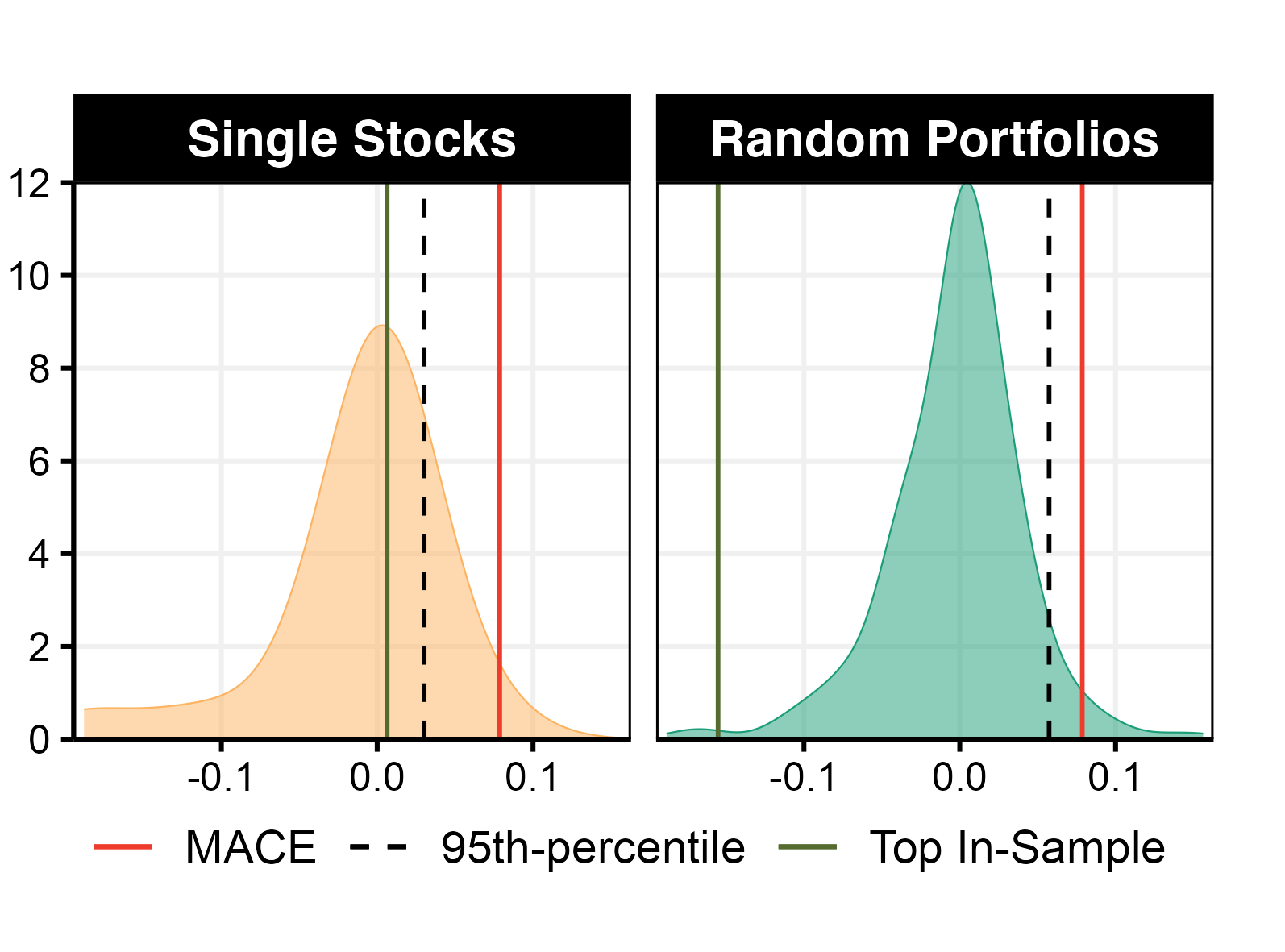}
\caption{\footnotesize  $N=20$,  $R^2_{\text{CovidW1}}$ }\label{r2_comp_daily_n20_covid} 
      \end{subfigure}
        \begin{subfigure}[b]{0.5\textwidth}
        \captionsetup{skip=0.1mm}
\hspace{-0.25cm}  \includegraphics[width=\textwidth, trim = 0mm 10mm 0mm 10mm, clip]{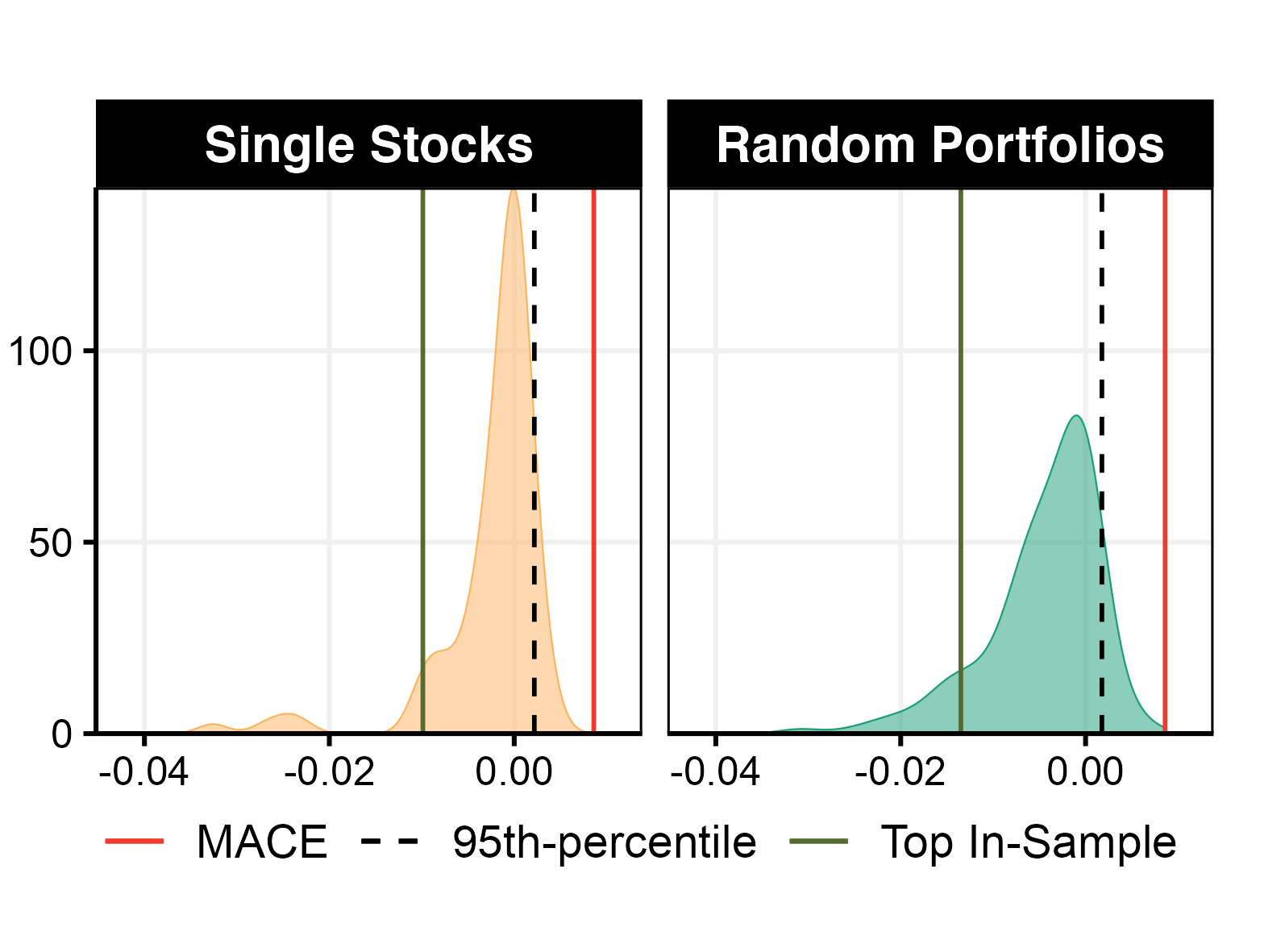}
\caption{\footnotesize  $N=100$,  $R^2_{\neg \text{CovidW1}}$ }\label{r2_comp_daily_n100_nocovid}
      \end{subfigure}
  \begin{subfigure}[b]{0.5\textwidth}
  \captionsetup{skip=0.1mm}
        \includegraphics[width=\textwidth, trim = 0mm 10mm 0mm 10mm, clip]{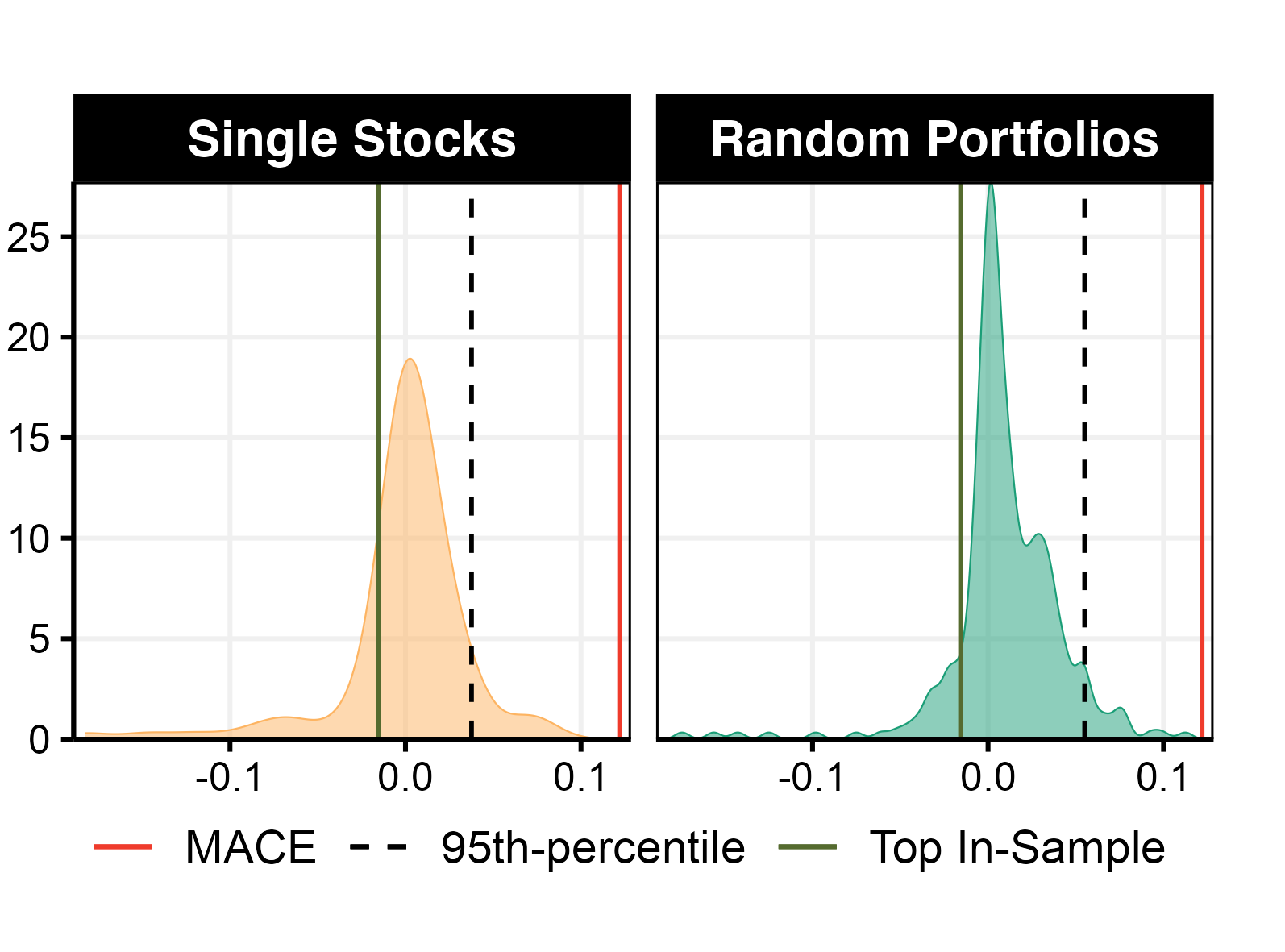}
\caption{\footnotesize  $N=100$,  $R^2_{\text{CovidW1}}$}\label{r2_comp_daily_n100_covid} 
      \end{subfigure}
      
      \begin{threeparttable}
 \begin{tablenotes}[para,flushleft]
 \setlength{\lineskip}{0.2ex}
	\scriptsize 
		\textit{Notes}: This plot shows distributions of OOS-$R^2$'s,  for different portfolio sizes and subsamples.  The "Single Stocks" panel reports the distribution of the $N$  $R^2$'s obtained from predicting each stock in the panel separately with RF.  The "Random Portfolios" panel shows the distribution of 300 $R^2$ obtained from predicting randomly drawn portfolios with RF.   \textit{Top In-Sample} denotes the OOS-$R^2$ of the single stock, random portfolio respectively, that achieved the highest $R^2$ during training. \textit{95$^\text{th}$-percentile} denotes marks the 95$^\text{th}$ percentile of the corresponding distribution shown in the graphs.
	\end{tablenotes}
	 \end{threeparttable}
	 \vspace*{-0.6cm}
  \caption{$R^2$ Comparison of MACE to Random Alternatives for the Daily Application }\label{r2_comp_daily}
\end{figure}

Those distributions, in themselves, highlight some things we already know,  like the immense difficulty of finding predictability outside of ``bad'' economic times: almost 95\% of random portfolios have a $R^2_{\neg \text{CovidW1}}<0$,  yet $R^2$'s greater than 0 constitute approximately 50\%  of the $R^2_{\text{CovidW1}}$. Obviously,  MACE's objective is not only to beat those odds by systematically landing on the ``good'' side of the distribution,  but also to strive for the rightmost part of it.  And it is what we see in all four cases of Figure \ref{r2_comp_daily}.  Indeed,  the red line is to the right of the dashed line,   meaning we can reject the null (at 5\% confidence level,  against a one-sided alternative) that MACE is randomly drawn portfolio.  Also,  the red line is always to the right of the best in-sample single stock or random combination.

{\vspace{0.15cm}}
{\noindent \sc \textbf{Economic Results.}} From Table \ref{tab:summstats_daily},  we see that MACE generates the highest return for all portfolio class sizes,  with MACE$_{100}$ leading the march with a massive 41.36\% annualized return over 2016-2022.  MACE$_{20}$ and MACE$_{50}$  deliver more ``reasonable'' returns of 23.1\% and 20.42\%,  which are nonetheless meaningfully higher than alternatives.   Closest contenders include MACE (PM) itself and sometimes MinVar (PM or RF,  depending on $N$).  Hence,  only MACE appears to consistently deliver the highest return.  Obviously,  that could all be at the expense of significantly more risk. 

Indeed,  we see that the variance of MACE's returns often appears to be higher than that of alternatives since its $SR$ is narrowly behind that of alternatives for $N=20$ (0.99 vs. 1.04) and $N=50$ (0.91 vs 0.96).  Note,  however,  that MACE is consistently among top contenders whereas, e.g. ,  MinVar (RF),  delivers the leading 0.96 for $N=50$ and the second-to-last 0.42 for $N=20$.  For $N=100$,  MACE gets a dominating annualized $SR$ of 1.59,  suggesting most of the 41.36\% return it gets is not due to unshackled variance.   The closest alternative among all $N$'s is $SR=1.04$ for MACE (PM,  $N=20$).

Figure \ref{cumret_daily} is telling about how that came about.   First,  there are the eye-grabbing flash gains during the onset of the Pandemic.  These are unpacked in their own section \ref{sec:covid}.  Given that major crisis only appear to occur on semi-decadal frequency,   it is natural to wonder whether MACE$_{100}$'s $SR$ would still be as startling without its miracle run in March 2020.  By peaking at Figure \ref{cumret_daily_n100},  it seems so:  translating downward the red line by 0.6 from 2020 onward still make it land comfortably above competitors in terms of final cumulative returns.  Moreover,  those are increasing in a nearly linear fashion starting from 2018.  The $SR$ excluding the highly profitable month is 1.32,  which confirms visual observations.  Looking at MACE$_{100}$  (PM)'s cumulative returns is also revealing for MACE$_{100}$'s overall performance.   The latter is magnified during the early stages of the Pandemic because the former (i.e.,  the raw portfolio itself) suffers minimal losses.  However,  we see that MACE$_{100}$  (PM) proves inferior to its RF-based counterpart by delivering approximately 0 returns in 2018, 2020 as well as 2021,  and losses in 2022.  

The $r^A_{2022}$ column also helps in understanding overall returns.  The variance among results for this metric is vast.  Some strategies suffered important losses,   yielding returns that are quite correlated with the overall bear market.  Others turned \textcolor{black}{into} a massive profit.  All three MACEs do fine,  with MACE$_{100}$  delivering almost 24\%, with apparently (Figure \ref{cumret_daily})  higher variance than previous years,  however.  MACE$_{20}$ is lowest among the three,  with 5.03\% and facing a major setback midyear by losing 10\% in a bit more than a week.  We will see in section \ref{sec:refine} that such setbacks can be smoothed out,  most notably,  by ``bagging strategies''.  While MACE$_{20}$ and MACE arguably get a headstart for 2022 with raw portfolios (PM) delivering marginally positive returns,  that of MACE$_{100}$  (PM) is a dramatic -16.81\%.  In all three cases,  it is the effect of active trading using RF predictions that avoids the failure of many strategies in 2022.  It is also true,  to a lesser extent,  for S\&P 500 (RF) which turns in 2.80\% while the PM version suffers major losses.  In fact, there are quite a few RF-based strategies that are profitable in 2022, however, unlike MACEs,  those are not accompanied with enviable returns in less turbulent years. 

\begin{figure}[t!]
  \begin{subfigure}[b]{0.5\textwidth}
  \captionsetup{skip=-0.2cm}
\hspace{-0.25cm}  \includegraphics[width=\textwidth, trim = 0mm 0mm 0mm 0mm, clip]{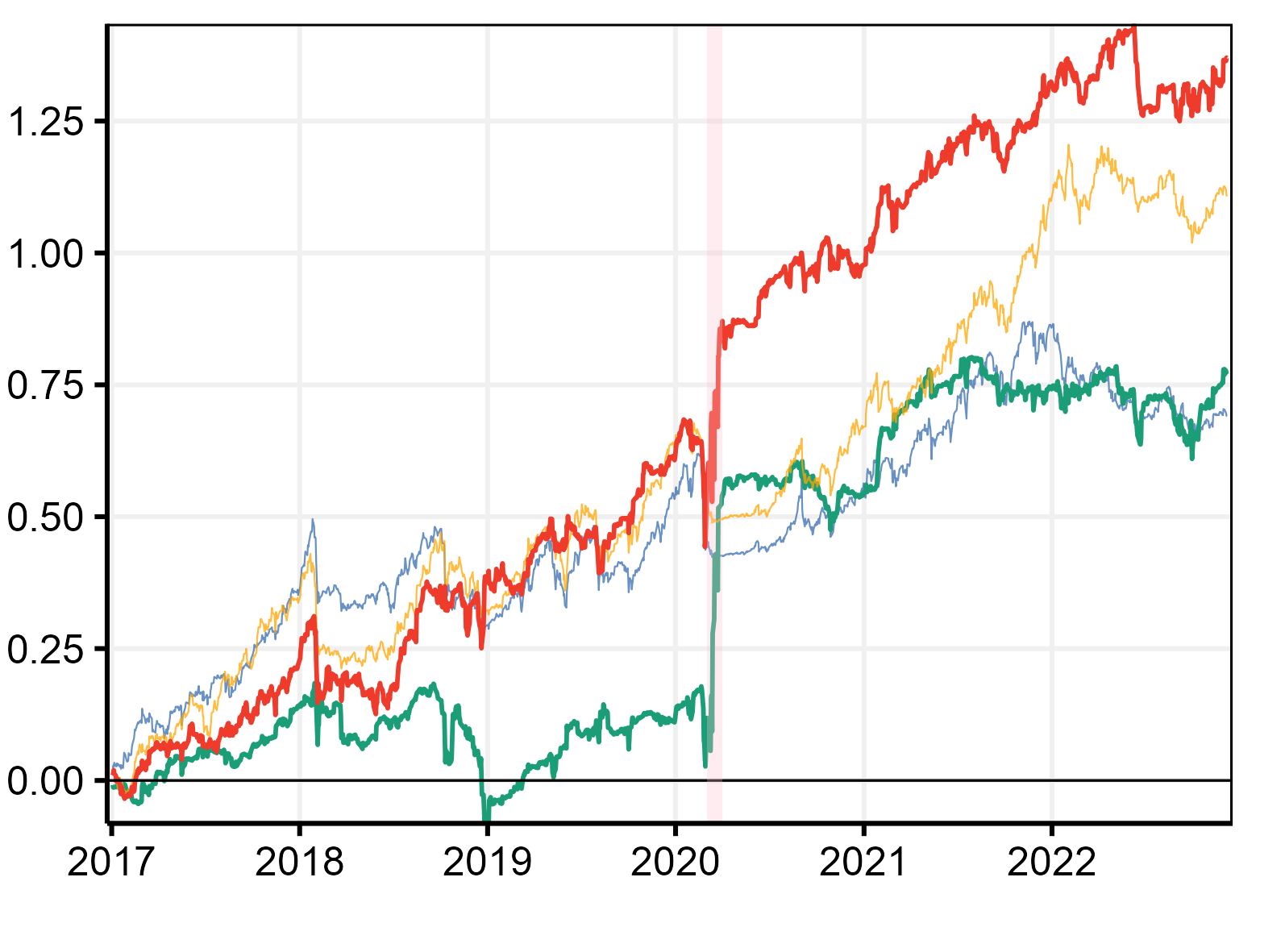}
\caption{\footnotesize  $N=20$ }\label{cumret_daily_n20}
      \end{subfigure}
  \begin{subfigure}[b]{0.5\textwidth}
  \captionsetup{skip=-0.2cm}
        \includegraphics[width=\textwidth, trim = 0mm 0mm 0mm 0mm, clip]{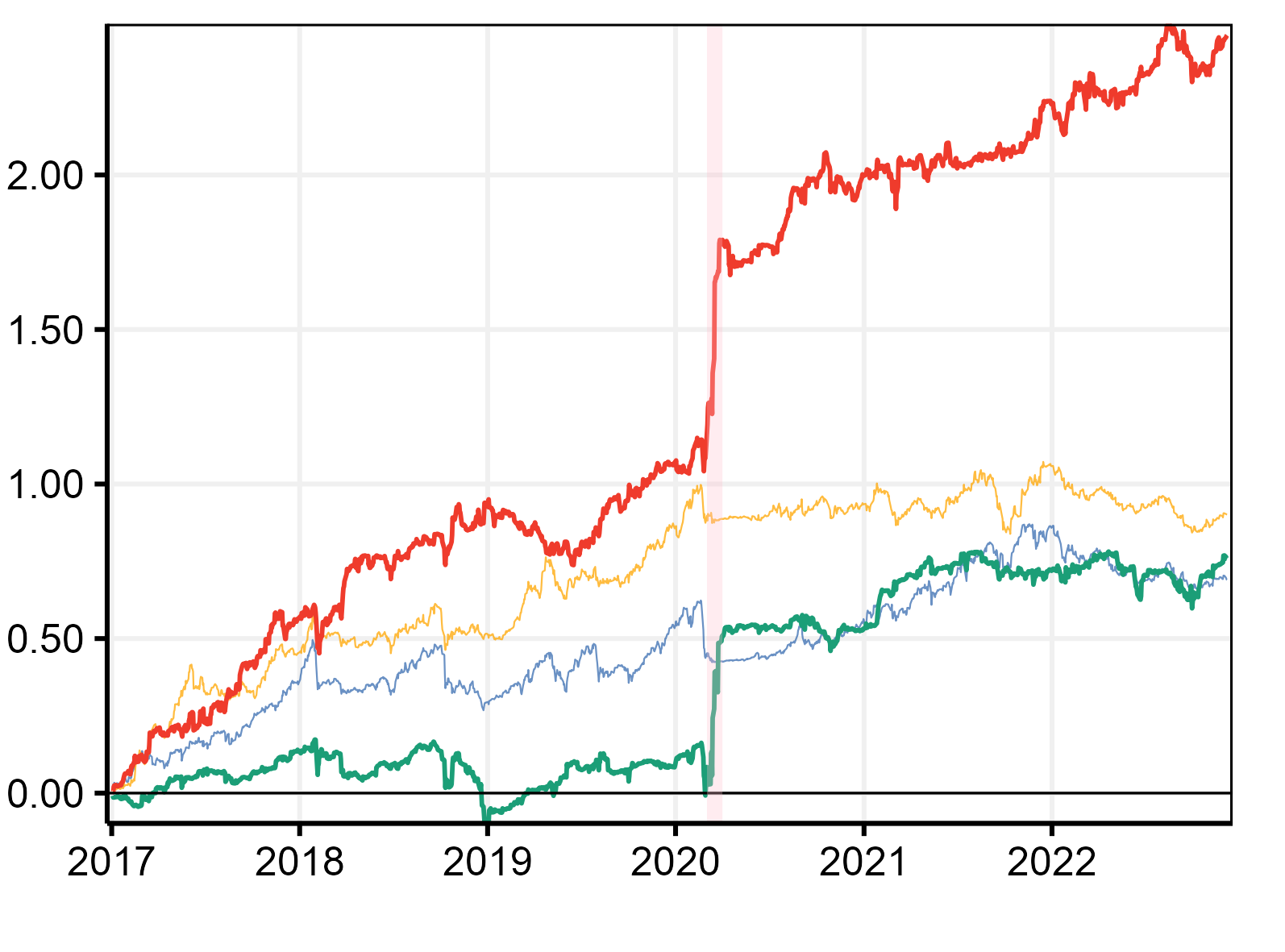}
\caption{\footnotesize  $N=100$ }\label{cumret_daily_n100} 
      \end{subfigure}\ \\
       
 \begin{subfigure}[t]{\textwidth}
	\centering
  \includegraphics[width=0.75\textwidth, trim = 0mm 0mm 0mm 0mm, clip]{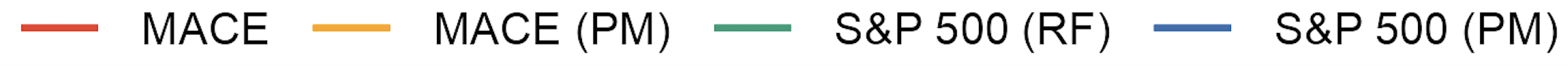}
	\end{subfigure}
       
  \caption{Cumulative Returns }
  \label{cumret_daily}
\end{figure}

The large positive jumps in returns that many RF-based methods experience may be seen unfavorably by  $SR$.   From our inspections,  many RF-based strategies generate returns that are always more leptukurtic  -- almost Laplacian-looking -- than strategies based on the prevailing mean.   For instance,  excluding CovidW1,  MACE$_{100}$ has a kurtosis of 5.96 vs  2.43 for MACE$_{100}$ (PM),  and S\&P 500 (RF) has 9.52 vs 6.87 for its prevailing mean version.  Again with the aforementioned exclusion zone,  we also notice that MACE$_{100}$ returns have positive skewness (0.2), which is highly preferable,  whereas all other strategies have negative skewness.  The other two RF-based MACE's do not have positive skewness, but it is always the least negative among its portfolio size group.  Including CovidW1 only magnifies (substantially) the extent of such findings.  Thus,  a performance measure that appreciates such subtleties about return distribution,  going beyond mean and variance,  may give a different assessment of portfolios predicted by RF.   This is indeed what we find.  MACE (non-PM) portfolios always deliver the highest $\Omega$ within $N$-wise groups.  Notable changes in ranking with respect to $SR$ are MACE$_{20}$ going from (narrowly) third to clearly first within its group,  MACE$_{50}$ moving from second to first with the $N=50$ group,  and S\&P 500 (RF) beating S\&P 500 (PM). 

Among other stylized facts, the variance of MACE portfolio return  \textit{before trading} is higher than all other portfolios.  For instance,   MACE$_{100}$ unconditional standard deviation is 1.54 out-of-sample excluding CovidW1 and 1.44 in-sample,  whereas the MinVar portfolio (effectively MACE$_{100}$ at $s=0$) has 0.56 and 0.69.  Thus,  in its quest for higher predictability and returns,  MACE creates a synthetic asset which unconditional volatility is higher and which is tamed through more accurate predictions.  We see that this higher volatility is no out-of-sample ``surprise'' as the training and test nearly coincides on this metric.  This is not true for MinVar.  Those observations highlight both opportunities and perils for the MACE. The opportunities have already been documented widely.  The peril is that of overfitting: letting MACE overfit will lead to higher unconditional variance out-of-sample than what can be inferred from the in-sample performance.  From a risk-reward perspective,  this could be a lose-lose situation.  Clearly,  the MACEs studied here are doing fine in that regard, but one should always bear in mind the dual costs of overfitting in the MMLP problem.

In Appendix \ref{sec:TC},  we report how and MACE$_{20}$ and MACE$_{100}$  performances are affected from introducing various levels of transaction costs (TC).  \textcolor{black}{We derive TC as some multiple of portfolio turnover, where we estimate the multiple ($\textfrak{c}$) as one-half the bid-ask spread \citep{DuTepperVerdelhan2018}.  In particular, we compute the distribution of $\textfrak{c}$ based on the bid-ask spreads of all our 100 stocks, observed on a daily basis between 2017-01-01 and 2022-12-31. We then take three candidates for $\textfrak{c} \in \{0.01\%, 0.015\%, 0.03\% \}$, which correspond to the 50$^\text{th}$, 75$^\text{th}$, and 90$^\text{th}$  percentiles respectively.  Even though we are trading at a daily frequency, MACE remains highly competitive, both in absolute, as well as risk-adjusted returns.  TC start to bite for $\textfrak{c} = 3$ basis points,  which is however derived from the 90$^\text{th}$ percentile of the bid-ask-spread distribution and thus a rather conservative estimate.}

\subsection{Understanding March 2020}\label{sec:covid}

\begin{wrapfigure}{r}{0.41\textwidth} 
\begin{center} 
\vspace*{-0.85cm}  
\hspace*{-0.3cm}\includegraphics[width=.41\textwidth]{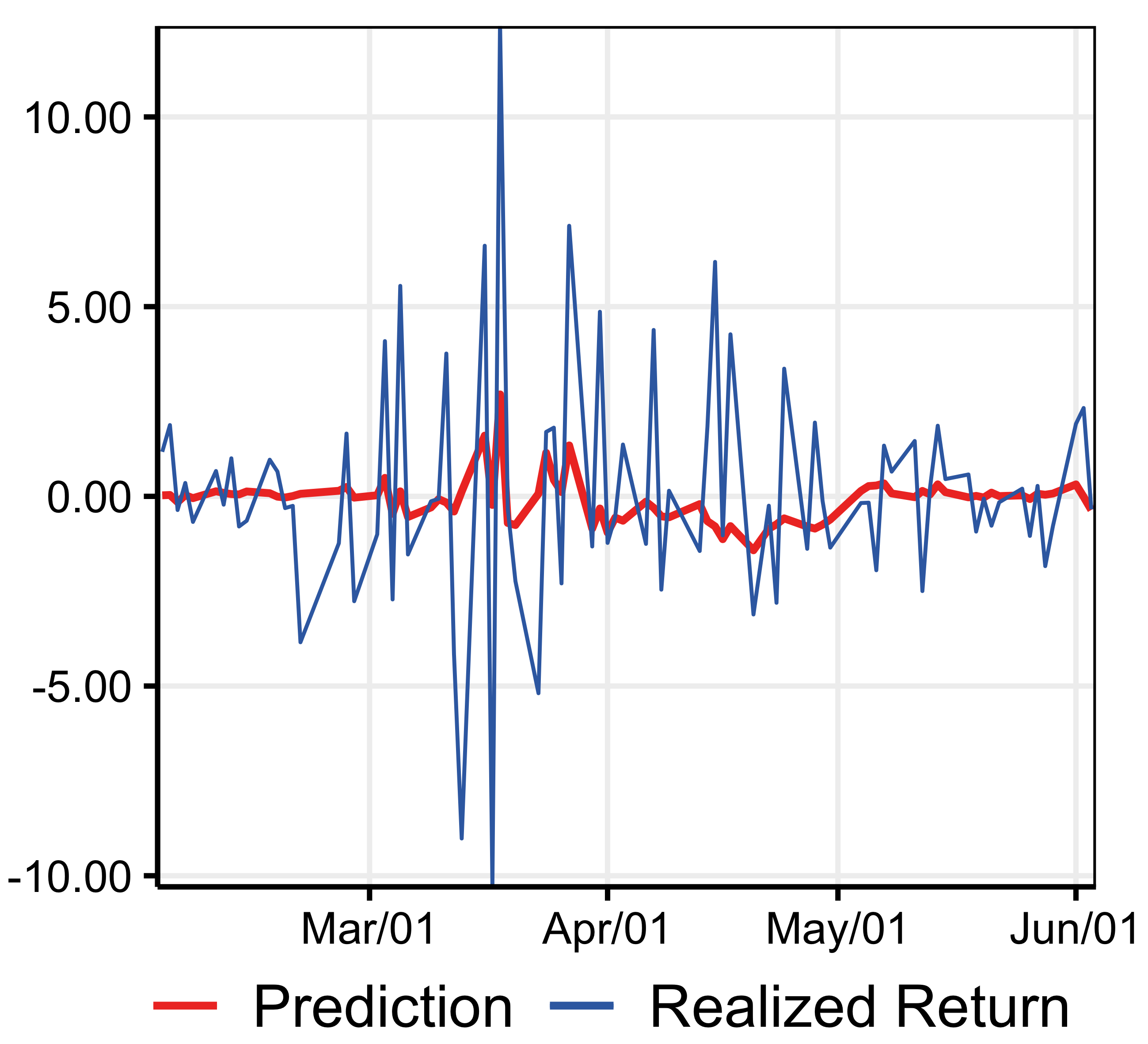}
\vspace*{-0.15cm}  
\caption{\footnotesize MACE$_{100}$ during CovidW1}
\label{covid_mace}
\vspace*{-0.58cm}  
\end{center}
\end{wrapfigure}

It is quite a sight in Figure \ref{cumret_daily} that MACEs  \textcolor{black}{generate} massive gains during the short-lived Pandemic recession.  These can obviously be linked back to  the abnormally high $R^2$'s observed for that era.\footnote{Increased predictability in the early Covid era has also been noted in \textit{in-sample} analyses such as  \cite{LalwaniMeshram2020} who found that the stock returns in certain industries are statistically significantly more predictable during 2020Q1 than a few months prior.} A similar result is found for S\&P 500 (RF).  However,  S\&P 500 (RF) is brought down by a dismal performance for the three years preceding 2020.  A similar pattern holds for EW (RF) (unreported),  where returns are dismal  for the entirety of the test set.  This highlights that RF-based models can most definitely fail to outperform basic strategies  and that the MACE ``treatment effect'' is paramount in giving them consistently the upper hand.

MACEs' and S\&P 500 (RF)'s winning streak (both for $N=20$ and $N=100$) spans approximately 21 business days starting from early March 2020.  Its intensity is greater for MACE$_{100}$ and S\&P 500 (RF) where wealth is approximately tripled in one month.  Indeed,  zooming in on these precise days,  MACE$_{100}$ \textbf{hits a local }$R^2$ \textbf{of 20\% and its sign prediction accuracy is 78 \%}.  Needless to say,   losing money only once every five days, thus compounding returns most days of the week in a time of high volatility,  is what generates the rocket increase in profits during March 2020.  

Predictability patterns are clearly visible in Figure \ref{covid_mace},  where the RF embedded in MACE predicts many of the bounce backs and the overall zigzagging nature of the market in times of crisis.  To the naked eye,  it looks like RF, in part, uses strong negative autocorrelation from one day to the next--thus,  fast-paced mean reversion.  Table \ref{covid_ar1} verifies such intuition with a first order approximation to nonlinear day-to-day dynamics.  We find that the AR(1) coefficient for both  MACE$_{100}$ and S\&P 500 are strongly negative and highly statistically significant ($t$-stats above 4) during the first four months of the Pandemic.  RF predictions provide significant economic value in this period because they precisely capture just  that---those are strongly negatively correlated with yesterday's return.   

With $\text{Corr}(\hat{r}_t^{\text{RF}}, r_{t-1})$ for the two targets in vicinity of -0.6,  it is natural to ask where RF may have learned that.  It is equally natural to conjecture it did so during the last major financial crisis.  The third panel of Table \ref{covid_ar1} verifies that.  We see, albeit in a marginally more subtle way,  that (i) returns are significantly negatively autocorrelated on a daily basis and (ii) RF predictions leverage the phenomenon to a non-trivial extent. 

While some statistically  significant mean-reversion remains in MACE$_{100}$ outside of the pandemic,  it is smaller by orders of magnitude.  For the S\&P 500,  it is completely gone,  one would expect.  Thus,  given the wide heterogeneity,  a successful model must detect \textit{ex-ante}  whether we are in a state of strong daily mean reversion or one where there is little to none at all.    It is an aspect in which the nonlinear nature of RF becomes key.  This state-dependence is obviously only one of the many time series nonlinearities that tree ensembles can capture \citep{MRF}.  Finally,  it is worth remembering that,  even within those two hypothetical states,  the linear approximation in Table \ref{covid_ar1}  only captures  a fraction of RF nonlinear predictive dynamics.  This is particularly true for ${\neg \text{CovidW1}}$ where RF's prediction correlates very little with the first lag.  Even during CovidW1,  it is worth noting that the first lag explains less than 40\% of the variance of RF predictions.

\begin{table}[t!]
\centering
		\vspace*{0.45em}
  \begin{threeparttable}
\centering
\footnotesize
\caption{First Order Approximation to Nonlinear Dynamics in Returns \label{covid_ar1} \vspace{-0.3cm}}
\setlength{\tabcolsep}{1.9em}
\begin{tabular}{ l *{7}{c} }
\toprule
\rowcolor{white}
\hspace*{0.075cm} &
\multicolumn{2}{c}{ ${ \text{CovidW1}}$} &
\multicolumn{2}{c}{${\neg \text{CovidW1}}$}&
\multicolumn{2}{c}{2008}  \\
\cmidrule(lr){2-3} \cmidrule(lr){4-5} \cmidrule(lr){6-7} 
&
\makebox[3em]{MACE$_{100}$} &
\makebox[3em]{S\&P 500 (RF)} &
\makebox[3em]{MACE$_{100}$} &
\makebox[3em]{S\&P 500 (RF)} &
\makebox[3em]{MACE$_{100}$} &
\makebox[3em]{S\&P 500 (RF)} \\
\midrule
\hspace*{-0.5cm} Coefficient    & -0.451&-0.402&-0.074&-0.036  & -0.104 & -0.156 \tabularnewline \addlinespace[2pt]
\hspace*{-0.5cm} Standard Error    &\phantom{ }0.097&\phantom{ }0.100&\phantom{ }0.027&\phantom{ }0.027 & \phantom{ }0.063 & \phantom{ }0.062 \tabularnewline \addlinespace[2pt]
\hspace*{-0.5cm} $\text{Corr}(\hat{r}_t^{\text{RF}}, r_{t-1})$   &-0.616 &-0.577&\phantom{ }0.102&\phantom{ }0.014 & -0.243 & -0.173 \tabularnewline
\bottomrule
\end{tabular}
\begin{tablenotes}[para,flushleft]
\scriptsize \textit{Notes}: This table reports the AR(1) coefficient and its standard error for two different return series on two non-overlapping subsamples of the test set spanning from 2016 to 2022 as well as 2008 from the training sample.  $\text{Corr}(\hat{r}_t^{\text{RF}}, r_{t-1})$ is the correlation between Random Forest's prediction of the portfolio's return and the realized return on the previous business day.
  \end{tablenotes}
  \end{threeparttable}
  		\vspace*{-0.75em}
\end{table}

\subsection{Bagging Strategies and Other Algorithmic Refinements}\label{sec:refine}

In a real-life implementation of a daily trading strategy,  one may be more than willing to exchange transaction costs for computational costs.  This is particularly true for MACE-based strategies, where all the computational burden is incurred while finding $\boldsymbol{w}$, which remains fixed ever after,  limiting daily computing costs to that of making one prediction with a Random Forest. 

 In this subsection,  we activate two algorithmic refinements  that help in improving the $N=20$ results in terms of annualized returns and Sharpe Ratios.  First, we introduce a modification to  the Ridge Regression step that brings back part of the unconditionally expected return constraint that is usually part of classical mean-variance portfolio optimization.  In our setup,  it will have a second nature as unconditional return \textit{regularization}.   Second,  we introduce ``bagging of strategies'' as a reasonably intuitive way to (i) tame the variance of returns and (ii) decrease the dependence of the solution on initialization values.  This latter refinement,  basically ensembling strategies,  multiplies the computational cost by the size of the ensemble ($B$).

{\vspace{0.15cm}}
{\noindent \sc \textbf{The Return of the Minimum Return Constraint.}} Given that MACE has a conditioning set and builds a portfolio purposely for the sake of actively trading it,  imposing an unconditional expected return constraint in the Ridge step does not nearly appear as natural as it would in a traditional mean-variance optimization setup.  Notwithstanding,  the close relationship between MACE and MinVar made explicit in section \ref{sec:mvrela} suggests that bringing back part of the constraint, in one form or another, could be beneficial.  One such scenario occurs if MACE's predictive power is overstated in-sample,  and leads it to overly rely on predictability to make an otherwise highly unprofitable portfolio profitable.  Mere overfitting can lead the out-of-sample solution of MACE to be closer to MinVar (where the conditional mean \textcolor{black}{is replaced by} an unconditional mean).  For those reasons,  unconditional return regularization may prove a helpful foolproof.  

The implementation is simple: in MACE$_{\mu \geq \underline{\mu}}$,  we turn off the intercept in the Ridge Regression step and add a positive value $\xi=1$ to $\hat{f}_s(\mathbf{X}_t)$. This has the effect of tilting the solution towards an allocation which has a higher unconditionally expected return in-sample.  Intuitively,  it pushes the ridge coefficients not only to reward each individual stock association with predictions $\hat{f}_s(\mathbf{X}_t)$, but also \textcolor{black}{to reward} historically higher growth stocks.  Conversely,  taking short positions on, e.g.,  Apple and Amazon,  is discouraged unless completely hedged with other assets.

{\vspace{0.15cm}}
{\noindent \sc \textbf{Bagging Strategies.}}  Interestingly,  and, in fact, as a matter of necessity,  we start by showing that the ensemble of strategies can be reduced to a single aggregate strategy,  hereby avoiding the bag of strategies to multiply transaction costs by $B$.  First,  it is worth remembering that,  unlike when predicting a fixed target, it is not possible here to merely average predictions.  Those are attached to inevitably heterogeneous targets, and there is  \textcolor{black}{no} guarantee that the average prediction will be appropriate \textcolor{black}{for} the average portfolio.  An extreme case is that, in an ensemble of two models with,  for estimation $b$ to be the mirror image of estimation $b'$ so that $\boldsymbol{w}_b+ \boldsymbol{w}_{b'}=\boldsymbol{0}$,  then the predictor and the predictand is 0 for all observations (even though each estimation separately had a positive $R^2$).  Thus,  the ensembling logic must be pushed further than merely averaging models, and rather look for bagging \textit{strategies}.

The proposed bagging scheme is the following: we run MACE $B$\textcolor{black}{-times} with different initializations, collect the $B$ predictions, translate this into $B$ positions through \eqref{eq:unimv},  and then,  finally,  average the returns of a total of $B$ strategies.  However,  stated as such, this would imply carrying at worst $N \times B$ trades a day instead of $N$.  Fortunately,  the bag of strategies can be rewritten so that it collapses to a single strategy.  Precisely, we have that 
\begin{align*}
r_t^{\text{bag}}  &= \frac{1}{B} \sum_{b=1}^B \omega_{t,b} \sum_{j=1}^N w_{j,b} r_{j,t}  =  \sum_{j=1}^N r_{j,t} \frac{1}{B} \sum_{b=1}^B \omega_{t,b} w_{j,b} =  \sum_{j=1}^N w_{j,t}^{\text{bag}} r_{j,t} 
\end{align*}
where $w_{j,t}^{\text{bag}}= \frac{1}{B} \sum_{b=1}^B \omega_{t,b} w_{j,b} $.  In words, the bag of strategies is equivalent to a single strategy where the daily weight on stock $j$ is $w_{j,t}^{\text{bag}}$,  implying at most $N$ transactions per day.   

We introduce two sources of randomization to make $\boldsymbol{w}_b$'s differ.  First,  the minimum variance solution used for initialization is randomized by estimating the covariance matrix on a random subsample of 70\% of the training data.  Second,  we use  decreasingly stochastic optimization steps via random observation weights which variance decrease with iterations ($\kappa_{t,s } \sim \text{exp}({s})$) in the Ridge part (Step 5 in Algorithm \ref{MACEalgo}).   This mild source of randomness is completely shut down when it becomes negligible ($\kappa_t =1 \enskip \forall t \text{  if  } s> \frac{s_{max}}{3}$).  The inspiration behind this randomization agent are some implementations of Boosting where trees are fitted on subsamples of the training data,  or simply stochastic gradient descent in neural networks.  Beyond creating a diversified ensemble,  it may help in avoiding early trivial overfitting solutions and in getting unstuck from local minima.  The choice of the exponential distribution (vs. subsampling) allows to keep all observations in at all times and is motivated from the Bayesian Bootstrap (see, e.g.,  the treatment in \cite{taddy2015bayesian} or \cite{MRF}) .

\vspace{1cm}
\hspace*{-1.25cm}
\begin{minipage}{1.05\linewidth}
		
\begin{minipage}{0.5\linewidth}
	\centering
	\rowcolors{2}{white}{gray!10}
\small
\hspace*{0.25cm}
	\begin{threeparttable}
	\captionof{table}{Refinements for MACE$_{20}$} \label{tab:summstats_daily_refine}
 \begin{tabular}{l| c c c c c c } 
\toprule \toprule
\addlinespace[1pt]
& &  $r^A$  &  & $SR$ &  &  $\Omega$  \\
\midrule
\addlinespace[5pt] 

MACE & & 23.10  &  & 0.99  &  & 1.18  \\ \addlinespace[2pt]
MACE$_{\text{bag}}$ & & 20.60  &  & 1.07  &  & 1.20   \\ \addlinespace[2pt]
MACE$_{\text{loose bag}}$ & & 23.03  &  & \textbf{1.36}  &  & \textbf{1.25}   \\ \addlinespace[2pt]
MACE$_{\mu \geq \underline{\mu}}$ & & \textbf{29.76}  &  & 1.17  &  & 1.19   \\ \addlinespace[2pt]

\bottomrule \bottomrule
	\end{tabular}
\begin{tablenotes}[para,flushleft]
	\scriptsize 
		\textit{Notes}: Economic metrics are $r^A$ := Annualized Returns, $SR$ := Sharpe Ratio,  $\Omega$ := Omega Ratio.  All statistics but $SR$ and $\Omega$ are in percentage points.   Returns and risk-reward ratios are based on trading each portfolio using a simple mean-variance scheme with risk aversion parameter $\gamma=5$.  Numbers in \textbf{{bold}} are the best statistic of the column.
	\end{tablenotes}
\end{threeparttable}
 \end{minipage}\hfill
 \begin{minipage}{0.5\linewidth}
 	\centering
 	\captionof{figure}{Cumulative Returns}\label{tab:cumret_daily_refine}
	\hspace*{-0.5cm}	\includegraphics[width=0.95\textwidth, trim = 0mm 0mm 0mm 2mm, clip]{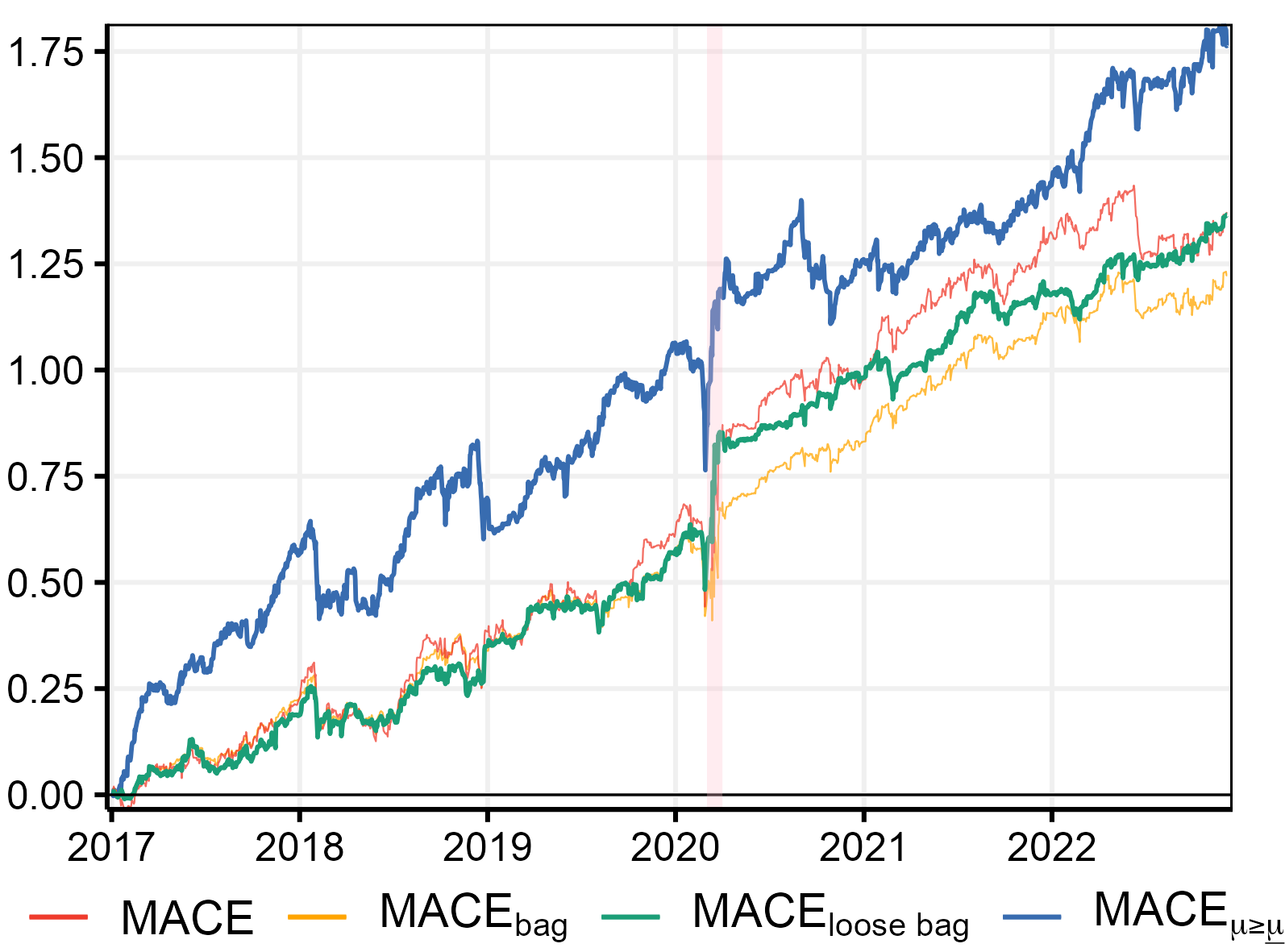}
		
 \end{minipage}          
\end{minipage}\ \\
\vspace*{0.25cm}

We use $B=50$.  Note that,  by construction,  the annualized return of MACE$_{\text{bag}}$ will be the average of the $B$ annualized returns (by the linearity of means).  However,  its volatility may be lower than the sum of each run's volatility,  resulting in improved Sharpe Ratios. \textcolor{black}{Another refinement is}   MACE$_{\text{loose bag}}$ \textcolor{black}{,which} is inspired from RF itself.  The rule for $\lambda$ is changed from $ R^2_{s,\text{train}}(\lambda)=0.01$ to $ R^2_{s,\text{train}}(\lambda)=0.02$ so to decrease the ``bias'' of base learners at the cost of increased variance,  and finally letting the ensembling step take care of bringing down the overall variance.

Results are reported in Table \ref{tab:summstats_daily_refine}.  All  \textcolor{black}{proposed extensions} refine the original MACE$_{20}$ results.  MACE$_{\mu \geq \underline{\mu}}$ increases dramatically expected returns,  which comes at the cost of reasonably increased volatility.  It outperforms the simpler MACE specification for both risk-reward ratios,  although the improvement is incremental for $\Omega$.  The bagging portfolios have a markedly different behavior: they have a marginally decreased $r_A$ but greatly decreased variance.  As a result,  MACE$_{\text{loose bag}}$ and MACE$_{\text{bag}}$ both ameliorate over the reference specification, with  MACE$_{\text{loose bag}}$ being the superior refinement both in  terms of $SR$ and $\Omega$.  The reasons behind such remarkable improvements are apparent in Table \ref{tab:summstats_daily_refine}: \textcolor{black}{both MACE$_{\text{bag}}$ and MACE$_{\text{loose bag}}$ follow an almost linear -- in log-terms -- trajectory without suffering from any outstanding drawdowns.  In particular,  the latter fares as well 2022 as in any other year,  and mid-March 2020 is only a momentaneous interruption of its otherwise steady exponential growth path.}

In Appendix \ref{sec:TC},  we report that MACE$_{\text{loose bag}}$ and MACE$_{20}$ outperformance is mostly unabated when accounting \textcolor{black}{for moderate to elevated levels of transaction costs. Only for very conservative estimates of TC does MACE's performance take more substantial hits.}

\subsection{Explainability by Factor-based Strategies}\label{sec:factor_d}

We investigate whether MACE can create a portfolio whose return profile cannot be replicated by a certain factor strategy. To do so,  we use MACE's returns as a target in the usual  factor regressions and check if a statistically  significant positive intercept ($\alpha$) ensues.  The regression uses the \cite{FF2015} five-factor model augmented by \cite{Carhart1997} momentum factor.   We source the constituents of the five-factor model -- market factor (MKT), size factor (SMB), value factor (HML), profitability (RMW), and investment factor (CMA) -- from Kenneth R. French's \href{https://mba.tuck.dartmouth.edu/pages/faculty/ken.french/data_library.html}{\color{blue} website}.  In contrast, we construct the momentum factor (MOM) based on the stocks we are trading on, i.e. ,  in the case of this daily application,  we construct MOM based on $N=20, 100$ stocks.

The full-page Table \ref{tab:FF5_d_Trading} in Appendix \ref{sec:tables} shows regression results for spanning regression of several MACE portfolios on the six-factor model.  Each of the three panels corresponds to the subsamples previously studied in this application.  The key observation is that whatever the portfolio or time period we are looking at,  trading according to signals coming from RF pushes the investor away from a factor-based strategy.  Indeed,  MACE (PM) returns are rather explainable by the 6-factor regression whereas MACE is not.  This is particularly evident from the \textit{No Covid} panel where MACE$_{20}$ (PM) gets an $R^2$ of 0.58 and MACE$_{20}$ reduces it to 0.16.  A similarly large wedge is observed for the MACE$_{100}$ case,   which magnitude,  in this case,  is unabated by including Covid observations.  We see that this loss in factor-based explainability for the RF-traded portfolio is mostly attributable to HML and RMW losing their predictive power on the resulting return.

Here are some additional observations.  Almost all MACE portfolios -- regardless of the sample period and whether they are traded or not according to RF -- covary positively with the market (MKT). Yet, with the exception of the Covid period, the market beta for MACE$_{20}$, MACE$_{100}$, and MACE$_{20,\text{loose bag}}$  barely exceeds 0.5,  which suggests that the MACE portfolio shuns the investor from being excessively exposed to overall movements in the market.  Comparing the various MACE vs MACE (PM), it is clear that this is attributable to both long and short positions being allowed in the construction of the portfolio rather than the action of Random Forest.  For $N=20$, MACE seems to exploit the strategy of the investment factor (CMA) by shorting firms that invest aggressively and taking a long position in firms that take a more conservative approach towards their investment decisions.  Also,  the negative loading on the size factor is not surprising given the fact that MACE$_{20}$ and MACE$_{100}$ are portfolios constructed from the 20, respectively 100, stocks with highest market capitalization at the beginning of 2016.  Lastly,  the momentum factor does not have any explanatory power for the MACE portfolios, which composition is actually a construct of the \textit{nonlinear mean reversion machine} as explained in Section \ref{sec:app2}, suggesting that accounting for nonlinearities is indispensable when trying to uncover exploitable autoregressive dynamics at the daily frequency.

Despite apparently exploiting some of the established investment strategies of the six-factor model, MACE is still able to find some unexplored alpha.   MACE$_{100}$ generates a daily unexplained excess return of about eight basis points, which would accumulate to 20\% annualized.  Making MACE work out a strategy with an elicit group of only $N=20$ stocks and requiring it to generate some statistically significant alpha is admittedly a daunting task.  Still, the \textit{loose bag} version finds a way to bundle these few stocks in a profitable manner that is not entirely explained by common factors.  We note that the point estimate for alpha is the same for both MACE and MACE$_\text{loose bag}$.  What makes the latter significant at the 5\% level and the former not is MACE$_\text{loose bag}$ attaining the same average return with a lowered variance, as reported in Section \ref{sec:refine}.

\section{Application II: Monthly Stock Returns Prediction}\label{sec:app1}                    

\textcolor{black}{Lastly, we test MACE on monthly data.} We shy away from any eccentricity and consider building portfolios  with individual stock returns from CRSP for firms listed on the NYSE,  AMEX, and NASDAQ \citep{gu2020empirical} using the 16 macroeconomic indicators of \cite{WelchGoyal2008} as the basis for $\mathbf{X}_t$.   We conduct a pseudo-out-of-sample expanding window experiment with a training set starting in 1957m3.  The test set originally starts in 2005m1 and ends in 2019m12.   We re-estimate MACE and the suite of competing models every 3 months and at each $t$,  $\boldsymbol{r}_{t+1}$ is comprised of all the stocks that have been continuously present in the dataset from 1957m3 until $t$.  Accordingly, the number of stocks included in 2005 is 192  and shrinks to 113  in 2019.\footnote{Naturally,  if the attrition were to become problematic, one could alleviate it by considering a rolling window instead.} We also report results when starting the test set in 1987 (as \cite{gu2020empirical} do) but regard those starting in 2005 as more indicative of performance since MACE not only needs to learn a complex $f$ (as in any ML-finance paper) but also $g$, and all that with time series of limited length.  Also,  predictability is known to be harder to find starting from the mid-2000s,  with many studies reporting important gains from ML-based stock returns forecasting, but those nearly all take place before the start of the new millennium  \citep{gu2020empirical,babiak2020deep}.

\begin{table}[t!]
	\footnotesize
	\centering
\caption{\normalsize {Summary Statistics for Monthly Stock Returns Prediction} \vspace*{-0.3cm}} \label{tab:summstats_monthly}

\begin{threeparttable}
		\setlength{\tabcolsep}{0.40em}
		  \setlength\extrarowheight{2.5pt}
 \begin{tabular}{l| rrrrrrrrrrr | rrrrrrrrrrr} 
\toprule \toprule
\addlinespace[2pt]
& & \multicolumn{9}{c}{01/2005 - 12/2019} & & & \multicolumn{9}{c}{01/1987 - 12/2004} \\
\cmidrule(lr){3-11} \cmidrule(lr){14-22} \addlinespace[2pt]
& &  $R^2_{\text{OOS}}$ &  & $r^A$ &  &$SR$ &  &$DD^{\text{MAX}}$ &  &$\Omega$ & & &
	  $R^2_{\text{OOS}}$ &  & $r^A$ &  &$SR$ &  &$DD^{\text{MAX}}$ &  & $\Omega$ & \\
\midrule
\addlinespace[5pt] 
\rowcolor{gray!15} 
 \multicolumn{22}{l}{\textbf{Main Results}} &\cellcolor{gray!15} \\ \addlinespace[2pt] 
MACE & & \textbf{4.13}   & & \textbf{18.70}  & &  \textbf{\color{ForestGreen} 1.05}  && \textbf{27.02}  & &  \textbf{\color{ForestGreen}1.89}  & && -7.99 & & 8.73 & & 0.39& & 71.80  & & 1.14  &  \\ \addlinespace[2pt]  
MACE (PM) & & -0.30   & & 13.29  & &  0.63  && 70.84  & &  1.36  & && -0.43 & & 8.89 & & 0.43 & & 84.57 & & 1.15 &  \\ \addlinespace[2pt] 

\midrule
\addlinespace[5pt] 
\rowcolor{gray!15} 
 \multicolumn{22}{l}{\textbf{Benchmarks}} &\cellcolor{gray!15} \\ \addlinespace[2pt] 
EW (RF) & & 1.88   & & 12.60  & &  0.71 && 42.73    & &  1.42  & && -8.24 & & 9.71 & & 0.47 & & 66.24 & & 1.21 &  \\ \addlinespace[2pt]  
EW (PM) & & -0.24   & & 10.74  & &  0.48 && 113.95    & &  1.23   & && \textbf{\color{ForestGreen}-0.39} & & \textbf{11.62} & & \textbf{0.58} & & \textbf{\color{ForestGreen}53.16} & & \textbf{1.30}  &  \\ \addlinespace[2pt] 
S\&P 500 (RF) & & -3.53   & & 11.98  & &  0.70&& 45.57    & &  1.40   & && -13.77 & & 5.26 & & 0.23& & 125.03  & & 1.01  &  \\ \addlinespace[2pt]  
S\&P 500 (PM) & & -0.52   & & 7.72  & &  0.42 && 101.70   & &  1.13  & && -0.48 & & 9.54 & & 0.54 & & 78.27 & & 1.23 &  \\ \addlinespace[2pt]

\midrule
\addlinespace[5pt] 
\rowcolor{gray!15} 
 \multicolumn{22}{l}{{\textbf{Refinements}}} &\cellcolor{gray!15} \\ \addlinespace[2pt] 
 MACE$_{\text{bag}}$ & & \textbf{\color{ForestGreen}4.84}   & & 16.90  & &  0.97  && \textbf{\color{ForestGreen}23.43}    & &  1.73 & && -7.61 & & 9.24 & & 0.43 & & 65.71  & & 1.17 &  \\ \addlinespace[2pt]
MACE$_{\mu \geq \underline{\mu}}$ & & 4.27   & & \textbf{\color{ForestGreen}19.42}  & &  1.01  && 38.32  & &  1.78  & && -4.04 & & \textbf{\color{ForestGreen}13.26} & & \textbf{\color{ForestGreen}0.61} & & 59.30  & & \textbf{\color{ForestGreen}1.34} &  \\ \addlinespace[2pt]

\bottomrule \bottomrule
	\end{tabular}
		\begin{tablenotes}[para,flushleft]
	\scriptsize 
		\textit{Notes}: The first column-wise panel consists of out-of-sample $R^2$'s for different test (sub-)samples.  The second are economic metrics, where $r^A$ := Annualized Returns, $SR$ := Sharpe Ratio,  and $DD^{MAX}$ = Maximum drawdown.  All statistics but $SR$ are in percentage points.   Returns and risk-reward ratios are based on trading each portfolio using a simple mean-variance scheme with risk aversion parameter $\gamma=3$.  PM means the prediction is based on the respective prevailing mean with a lookback period of twenty years,  while RF means using that of a Random Forest.  Numbers in \textbf{{bold}} are the best statistic within the first two panels (that is,  excluding MACE refinements).  Numbers in \textbf{\color{ForestGreen} green} are the best statistic of whole column.
	\end{tablenotes}
\end{threeparttable}
\end{table}

Regarding variable transformations,  we subtract the risk-free rate from each stock return in $\boldsymbol{r}_{t+1}$.  Considering the exact composition of $\mathbf{X}_t$, we first-difference  the clearly non-stationary series in  \cite{WelchGoyal2008} and include 12 lags of each.  The evaluation metrics are similar to those introduced in section \ref{sec:dres}.  We complement those with the maximum drawdown ($DD^{\text{MAX}}$) as in \cite{gu2020empirical}:
\vspace*{-0.25cm}
\[
	DD^{\text{MAX}} = \underset{0 \leq t_1 \leq t_2 \leq T}{\text{max}} \; (Y_{t_1} - Y_{t_2})
\]
where $Y_t$ is the log cumulative return from $t_0$ through $t$.  For computing $MSE_{\text{PM}}^{\text{OOS}}$ in the denominator of the out-of-sample $R^2$ we again compute PM as the historical mean of the training-set.

Given that available stocks are changing throughout the dataset, so does the composition of EW and MACE portfolios.  Moreover,  even with a fixed basket of available stocks,  MACE's portfolio weights can change slowly as new data points enter  the training set.  Accordingly,  the $R^2$ (and the other metrics as well) are not one for a fixed target, but rather for a fixed ``strategy''.  In other words,  when $y_{t'+1}$ and $y_{t''+1}$ enter $MSE_{\text{MACE}}$ and belong to two different estimation windows,  they are likely coming from two distinct time series.  Since those share the same unconditional variance by construction (1 or some other standardization necessary for identification in \eqref{MACEpp}),  they can be aggregated without window $t'$ errors driving results more than that of window $t''$ for mechanical reasons.  

When trading the MMLP, we again solve the prototypical mean-variance problem for a single return $y_{t+1}$ as stated in Equation \eqref{eq:unimv}. As in \cite{FilippouRapachThimsen2021},  the risk aversion parameter $\gamma$  is set to 3 and $\hat{\omega}_{t+1}$ is constrained to lie between -1 for 2 for reasonable allocations.

\subsection{Results}\label{sec:mres}

We report relevant summary statistics for our monthly exercise in Table \ref{tab:summstats_monthly}.  Similar to section \ref{sec:dres}, we show results for MACE and its corresponding modifications/refinements as well as EW and S\&P 500. We report the relevant statistics for two distinct out-of-sample periods as outlined above.  Log cumulative return plots are shown in Figure \ref{cumret_monthly} and $R^2$ comparison of MACE to Random Alternatives plots are available in Figure \ref{r2_comp_monthly}.

{\vspace{0.15cm}}
{\noindent \sc \textbf{Economic Results.}}
Table \ref{tab:summstats_monthly} shows that MACE is clearly outperforming any other model along each evaluation metric. The annualized average return of $r^A = 18.70\%$ beats the closest competitor (EW (RF)) by a stunning six percentage points. Yet, this achievement does not seem to come at the cost of extreme volatility with the Sharpe Ratio and the maximum drawdown both dominating the competitors by large margins. Even though the proposed refinements might even achieve superior performance in a specific metric -- MACE$_\text{bag}$ gives the investor a smaller max drawdown, whereas MACE$_{\mu \geq \underline{\mu}}$ can even boost the annualized return -- both modified MACEs cannot keep up with the Sharpe Ratio of $SR=1.05$. This suggests that MACE strikes an appealing balance between risk and reward.   In Appendix \ref{sec:TC}, we document that \textcolor{black}{MACE remains highly competitive} for various levels of transaction costs.

This dominance, however, fades for the earlier period between 1987 and 2004. As outlined above, the shorter $T$ dimension limits MACE's ability to learn a complex signal,  if there is any,  resulting in diminished out-of-sample performance. As discussed in section \ref{sec:refine}, in such a situation it may merit to tilt MACE away from pure predictability and towards a safeguard of higher unconditional return instead.  This might lead to sacrificing some in-sample predictability,  but prove beneficial out-of-sample.  We achieve such a tilting with MACE$_{\mu \geq \underline{\mu}}$, which generates both the highest annualized average returns and scores the highest Sharpe Ratio ($SR=0.61$).

\begin{figure}[t!]
\captionsetup{skip=2mm}
      \begin{subfigure}[b]{0.5\textwidth}
  \centering
  \captionsetup{skip=-0.5mm}
        \includegraphics[width=\textwidth, trim = 0mm 0mm 0mm 0mm, clip]{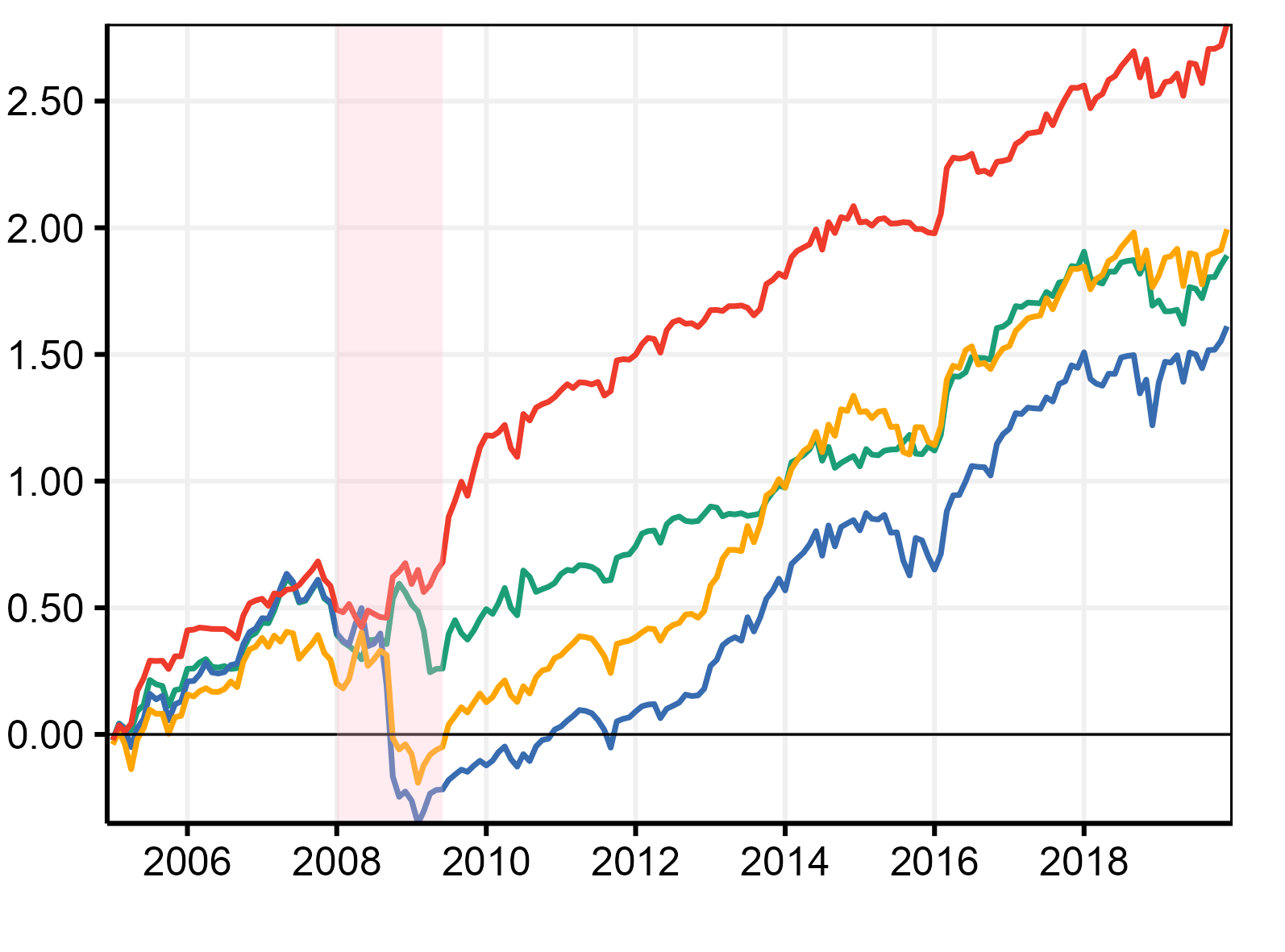}
\caption{01/2005-12/2019}
      \end{subfigure}
  \begin{subfigure}[b]{0.5\textwidth}
  \centering
  	\captionsetup{skip=-0.5mm}
	\includegraphics[width=\textwidth, trim = 0mm 0mm 0mm 0mm, clip]{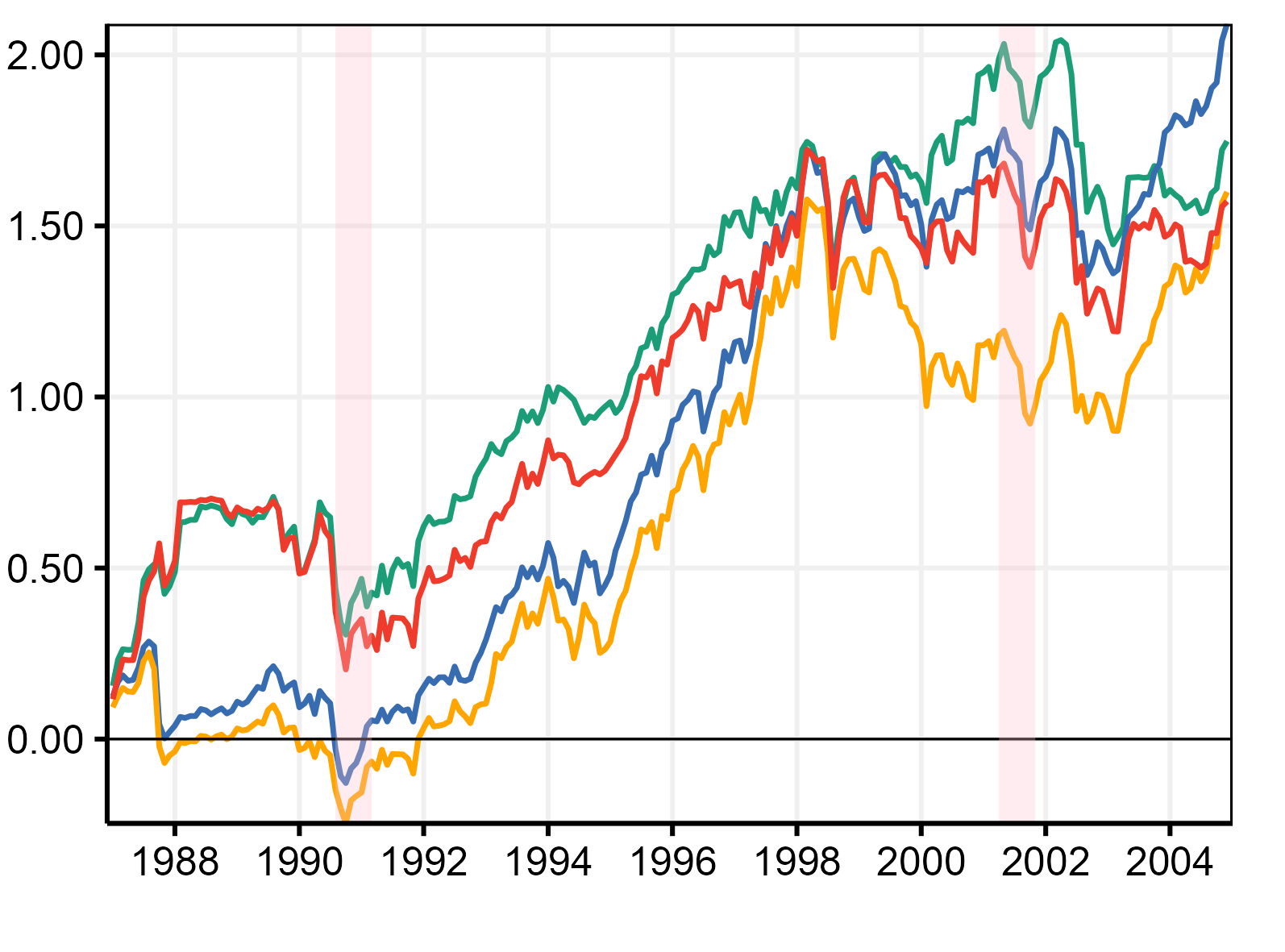}
\caption{01/1987-12/2004}
      \end{subfigure}  
      \begin{subfigure}[t]{\textwidth}
	\centering
 \vspace*{0.05cm}
  \includegraphics[width=0.7\textwidth, trim = 0mm 0mm 0mm 0mm, clip]{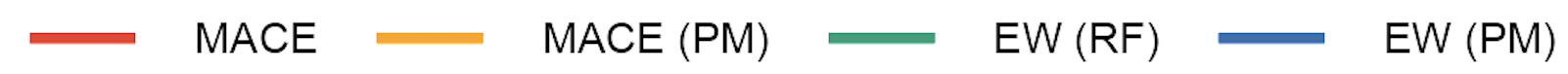}
	\end{subfigure}
  \caption{Cumulative Returns }
  \label{cumret_monthly}
\end{figure}

Figure \ref{cumret_monthly} gives us deeper insights into the underlying return dynamics that are cushioned by the single summary figures in Table \ref{tab:summstats_monthly}.  It is interesting to see how well MACE navigates the Great Recession (GR). While all competitors take either a deep hit or drift sideways (EW (RF)), MACE takes off already during the first half of GR and even accelerates its growth-rate at the outset.  This behavior is different from MACE's behavior during earlier recessionary periods in the early 1990s and 2000s where MACE takes visible hits. Yet, the results for the period 2005-2019 suggest that MACE has learned from earlier mistakes. 

With a $R^2$ of 1.88\% during the 2005-2019 subperiod,   EW (RF) also fares well,  most notably by avoiding heavy losses during GR.  However,  and as it often the case with more basic approaches,  predictability is heavily localized during tumultuous economic times,  leading  EW (RF) (and also similarly S\&P 500 (RF)) to overall underperform their prevailing mean counterparts both in terms of returns and volatility.  

{\vspace{0.15cm}}
{\noindent \sc \textbf{Statistical Results.}}
In Figure \ref{r2_comp_monthly} we see that the share of predictable Single Stocks is very similar across the two out-of-sample periods, whereas the distribution for Random Portfolios clearly shifts to the right during the 2005-2019 era.  Thus,  there \textit{is} exploitable predictability, and MACE's job is to find promising $\boldsymbol{w}$ ex-ante.  We see that it does so: it is superior to about 95\% of randomly drawn portfolios.  In the first subperiod,  where more than 95\% of such portfolios deliver negative $R^2$,  MACE suffers a setback along with the crowd.   With only  $\sim$5\% of portfolios found to be predictable ex-post, it is quite a daunting task to land in the promising region ex-ante.   As was also observed in daily results, MACE appears particularly apt at finding the MPP when there is a reasonable number of possibly successful candidates to work with.  In the opposite scenario where no or very few RFs attain any predictability,  MACE inevitably struggles.

\begin{figure}[t!]
\captionsetup{skip=2mm}
\setlength{\lineskip}{2.2ex}
  \begin{subfigure}[b]{0.5\textwidth}
  \captionsetup{skip=0.1mm}
\hspace{-0.25cm}  \includegraphics[width=\textwidth, trim = 0mm 10mm 0mm 10mm, clip]{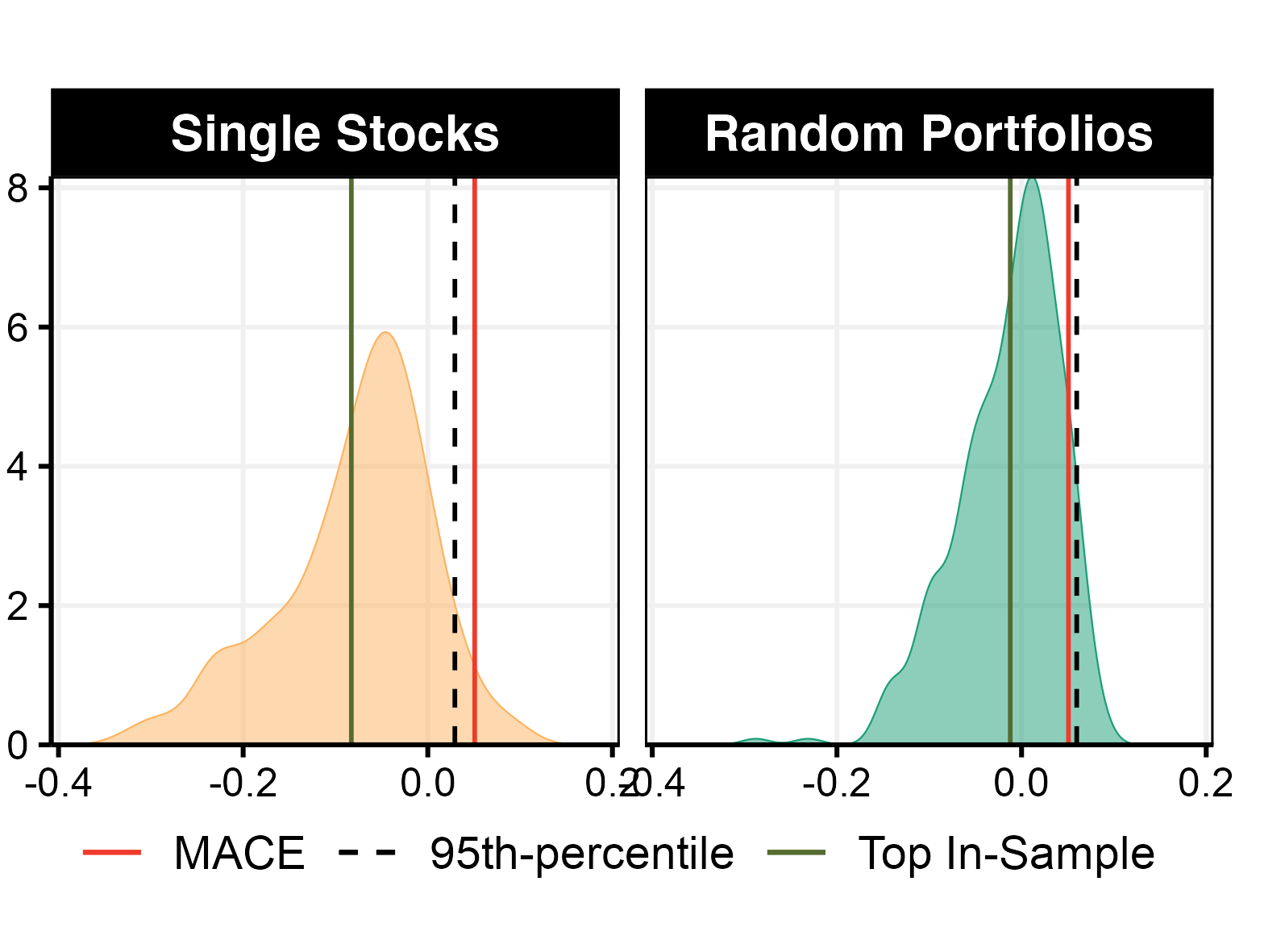}
\caption{01/2005-12/2019}\label{r2_comp_monthly_nAll_post2004}
      \end{subfigure}
  \begin{subfigure}[b]{0.5\textwidth}
  \captionsetup{skip=0.1mm}
        \includegraphics[width=\textwidth, trim = 0mm 10mm 0mm 10mm, clip]{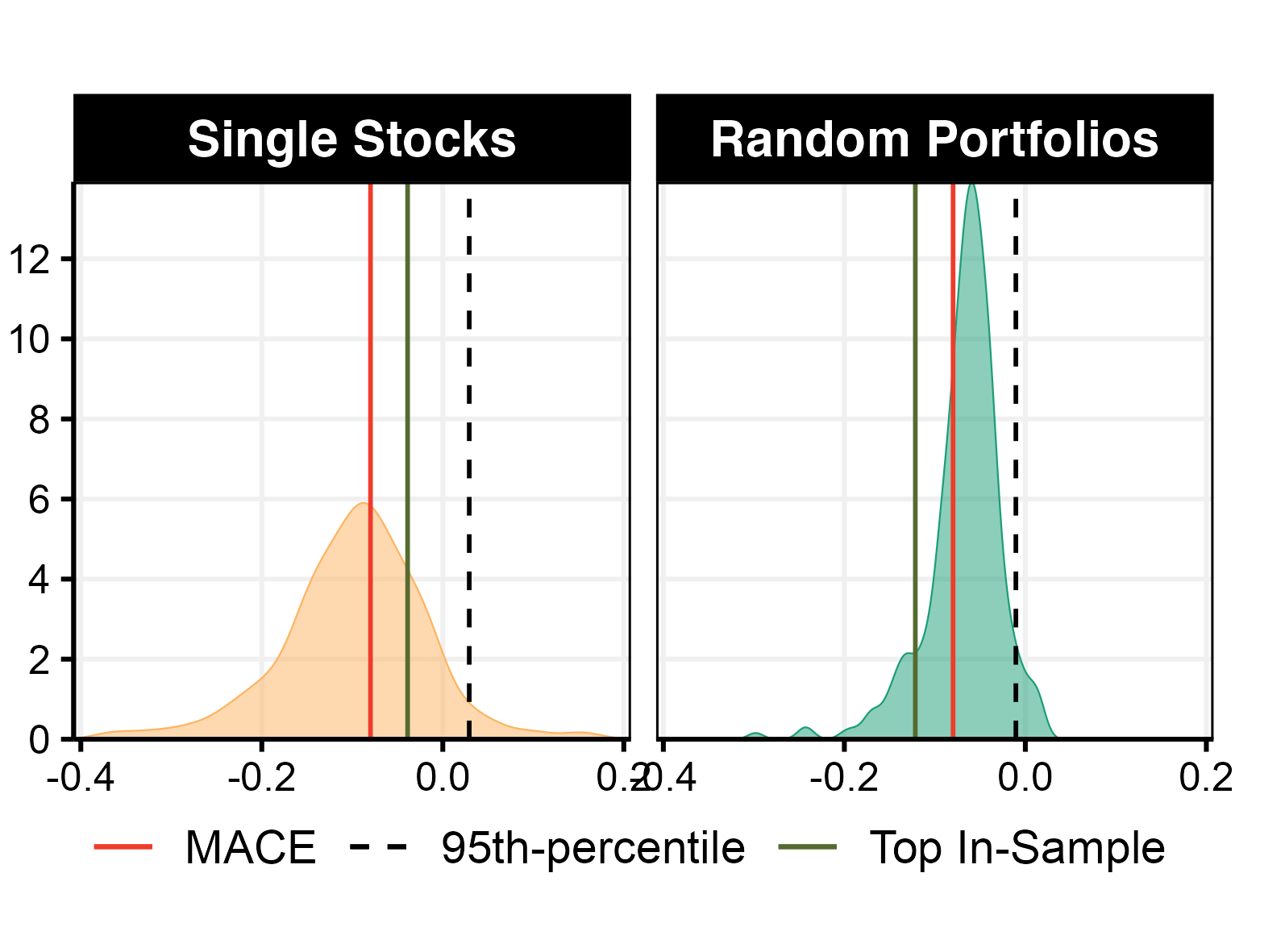}
\caption{01/1987-12/2004}\label{r2_comp_monthly_nAll_pre2005}
      \end{subfigure}
       \vspace*{-0.1cm}
        \begin{threeparttable}
 \begin{tablenotes}[para,flushleft]
 \setlength{\lineskip}{0.2ex}
	\scriptsize 
		\textit{Notes}: This plot shows distributions of OOS-$R^2$'s,  for different portfolio sizes and subsamples.  The "Single Stocks" panel reports the distribution of the $N$  $R^2$'s obtained from predicting each stock in the panel separately with RF.  The "Random Portfolios" panel shows the distribution of 300 $R^2$ obtained from predicting randomly drawn portfolios with RF.   \textit{Top In-Sample} denotes the OOS-$R^2$ of the single stock, random portfolio respectively, that achieved the highest $R^2$ during training. \textit{95$^\text{th}$-percentile} denotes marks the 95$^\text{th}$ percentile of the corresponding distribution shown in the graphs.
	\end{tablenotes}
	 \end{threeparttable}\ \\
	 \vspace*{-0.3cm}
	 \addtocounter{figure}{-1}
  \caption{$R^2$ Comparison of MACE to Random Alternatives for the Monthly Application}\label{r2_comp_monthly}
\end{figure}

Note that the large mass to the right of 0 in Figure \ref{r2_comp_monthly_nAll_post2004} is not necessarily indicative of market inefficiency  given that one still needs to find the relevant vectors ex-ante.   It, however, highlights a non-trivial number of possibilities for it to occur. Yet,  the absence of a mass on the right side of 0 (as in Figure \ref{r2_comp_monthly_nAll_pre2005}) is suggestive of efficiency,  conditional on choices for stocks,  information set,  and predictive function.

\subsection{Predictability through the (Nonlinear) Time-Varying Risk Premia?}

Figure \ref{cumret_monthly} makes clear that MACE's stellar performance results predominantly from not tanking during the Great Recession and climbing steeply at its outset.  This pattern is different from the other models. While EW (RF) did neither tank during the crisis, its engine sputtered post-crisis.  In contrast, the MACE (PM)'s and EW (PM)'s growth picked up relatively quickly in the aftermath of the Great Recession, however, only after having tanked deeply throughout the crisis period.  Only MACE seems to get the market timing right for both the crisis period and the subsequent recovery. 

To shed light on which economic indicators are driving this success, we use Shapley Values,  a well established and evermore popular tool to quantify the contribution of predictors in opaque models  \citep{Shapley1953,lundberg2017unified}.  We refer the reader to \cite{molnar2019interpretable} for a generic textbook treatment and \cite{anatomy} for a focus on its applicability to financial and macroeconomic forecasting.  Here, we dedicate our attention to the \textit{out-of-sample} period 01/2008 - 12/2009. Relevant details regarding the construction of our variable importance metric from expanding windows are relegated to Appendix \ref{sec:vi}.

\begin{figure}[h!]
\captionsetup{skip=2mm}
      \begin{subfigure}[b]{0.333\textwidth}
  \centering
  \captionsetup{skip=-0.5mm}
        \includegraphics[width=1\textwidth, trim = 0mm 0mm 0mm 0mm, clip]{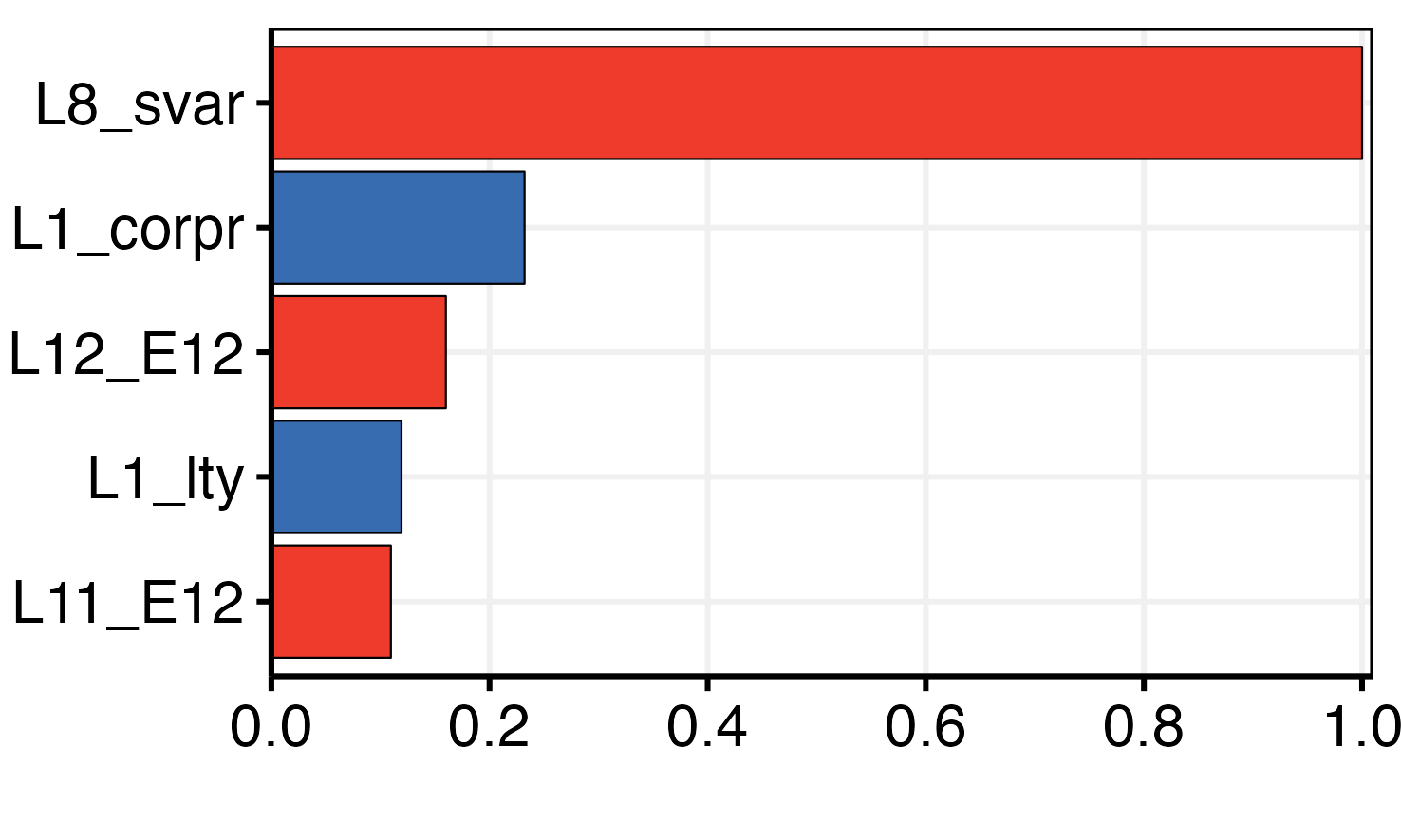}
\caption{MACE:  $VI^{oos}_i$}
      \end{subfigure}
  \begin{subfigure}[b]{0.33\textwidth}
  \centering
  	\captionsetup{skip=-0.5mm}
	\includegraphics[width=\textwidth, trim = 0mm 0mm 0mm 0mm, clip]{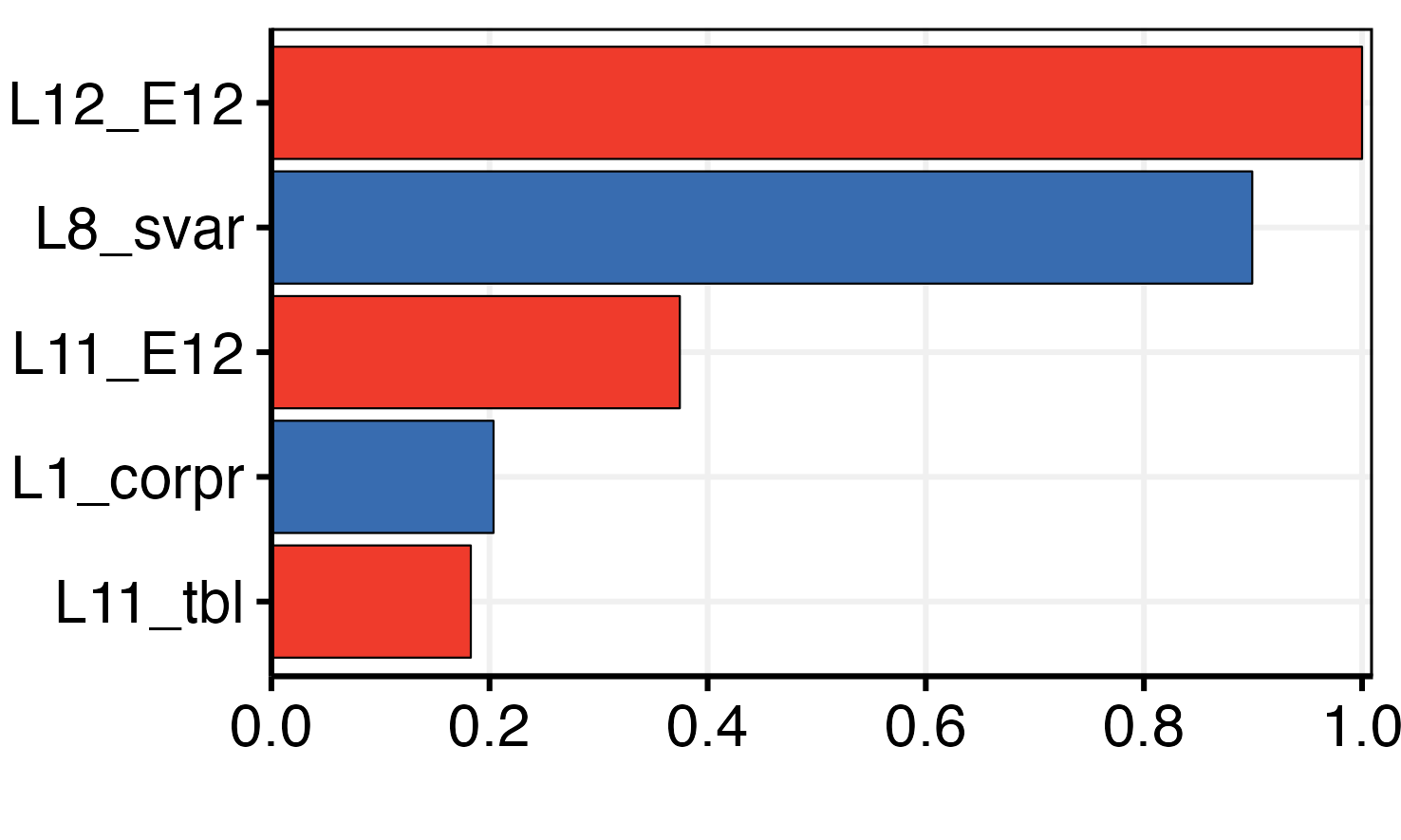}
\caption{EW (RF):  $VI^{oos}_i$}
      \end{subfigure}%
      \begin{subfigure}[b]{0.33\textwidth}
  \centering
  \captionsetup{skip=-0.5mm}
        \includegraphics[width=\textwidth, trim = 0mm 0mm 0mm 0mm, clip]{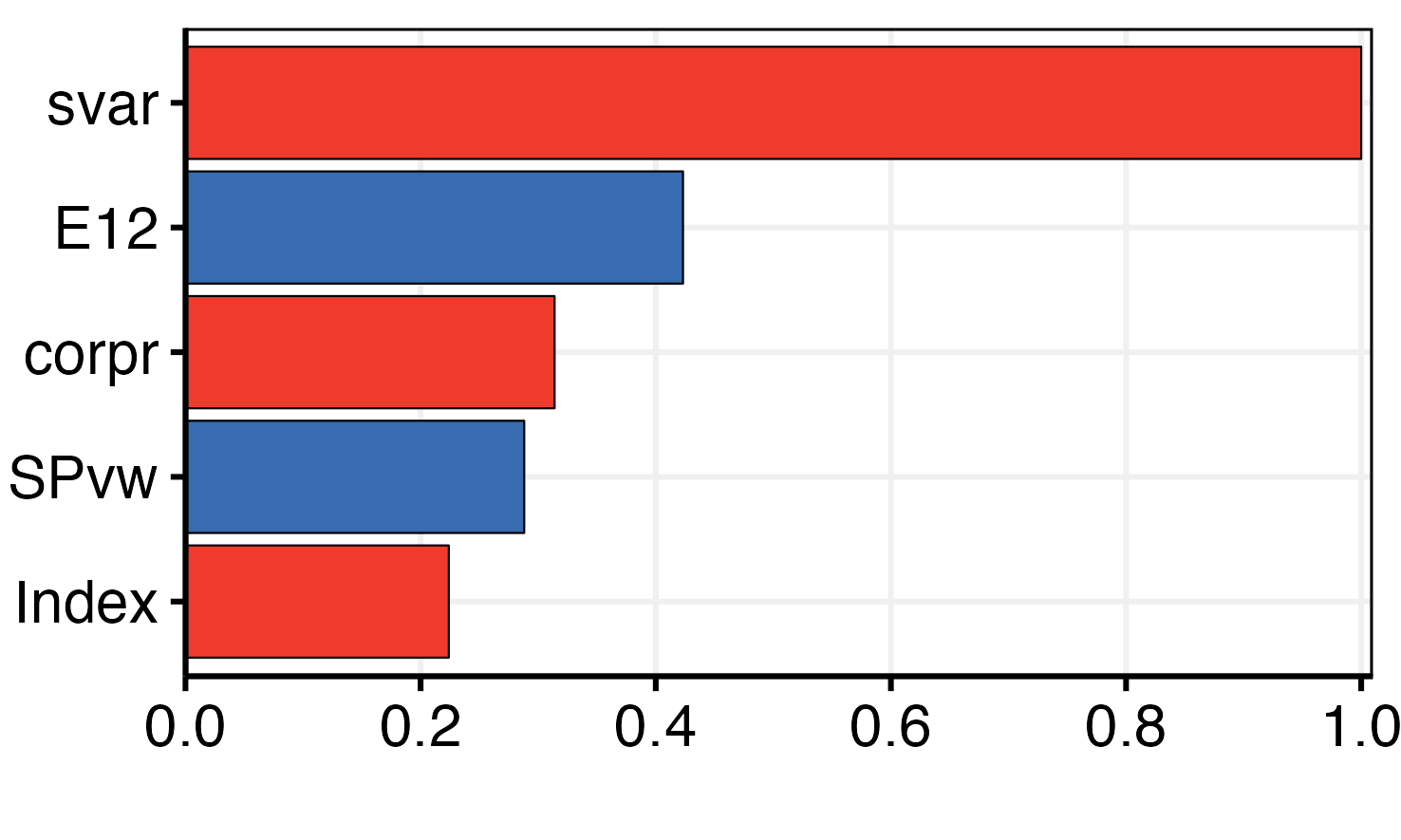}
\caption{MACE:  $VI^{oos}_g$}\label{fig:vig}
      \end{subfigure}%
 \vspace*{-0.1cm}
        \begin{threeparttable}
 \begin{tablenotes}[para,flushleft]
 \setlength{\lineskip}{0.2ex}
	\scriptsize 
		\textit{Notes}: The bars represent the $VI$ of predictor $i$ and grouped versions ($VI_g$, i.e. ,  summing the Shapley Values across all lags of variable $i$) as in Equation \ref{equ:VI_oos},  scaled by the corresponding maximum value.
	\end{tablenotes}
	 \end{threeparttable}\ \\
       \vspace*{0.02cm}
        \addtocounter{figure}{-1}
  \caption{Shapley Value Importance: 01/2008 - 12/2009}
  \label{fig:VI_monthly}
\end{figure}

Figure \ref{fig:VI_monthly} reports the five most important predictors for MACE and EW (RF) separately.   The last panel combines the importance of all the lags of a given indicator.   For MACE,  the picture is dominated by a single strong predictor: the eight-month lagged stock market volatility ($\mathtt{L8\_svar}$).\footnote{Of course,  the predictor itself had high variance during this period. In Appendix \ref{sec:Adj_VI}, we address this concern and show that $\mathtt{L8\_svar}$ stands the test of adjusting the Shapley Values for the indicator's volatility. } Grouping all lags together in Figure \ref{fig:vig} reinforces the case for the overall importance of $\mathtt{svar}$ itself.   This indicates that MACE may leverage a subtle form of time-varying risk premium,  as originally formalized in the ARCH-M model of \cite{EngleLilienRobins1987}, its extension with time-varying parameters (TVP ARCH-M, \citealt{ChouEngleKane1992}),  or the GARCH-in-mean of  \cite{french1987expected}.  These models allow for an asset's volatility to directly feed into the conditional mean of the asset's return,  allowing for time-varying risk premia.   "Subtle" refers here to the pattern being less evident than what one would expect from these classic models.  \textcolor{black}{The reason for this is threefold:} first,  there is a significant delay.  Second,  the volatility metric undergoes a highly non-linear transformation.  Third,  it is not MACE's previous volatility that enters the conditional mean, but that of the overall market as proxied by the S\&P 500. 

\begin{figure}[t]
\captionsetup{skip=2mm}
\hspace{0.4cm}
	\begin{minipage}{0.5\textwidth}
      \begin{subfigure}[t]{\textwidth}
  \centering
  \captionsetup{skip=-0.5mm}
    \vspace*{-0.1cm}
        \scalebox{1}[1]{\includegraphics[width=\textwidth, trim = 0mm 35mm 0mm 25mm, clip]{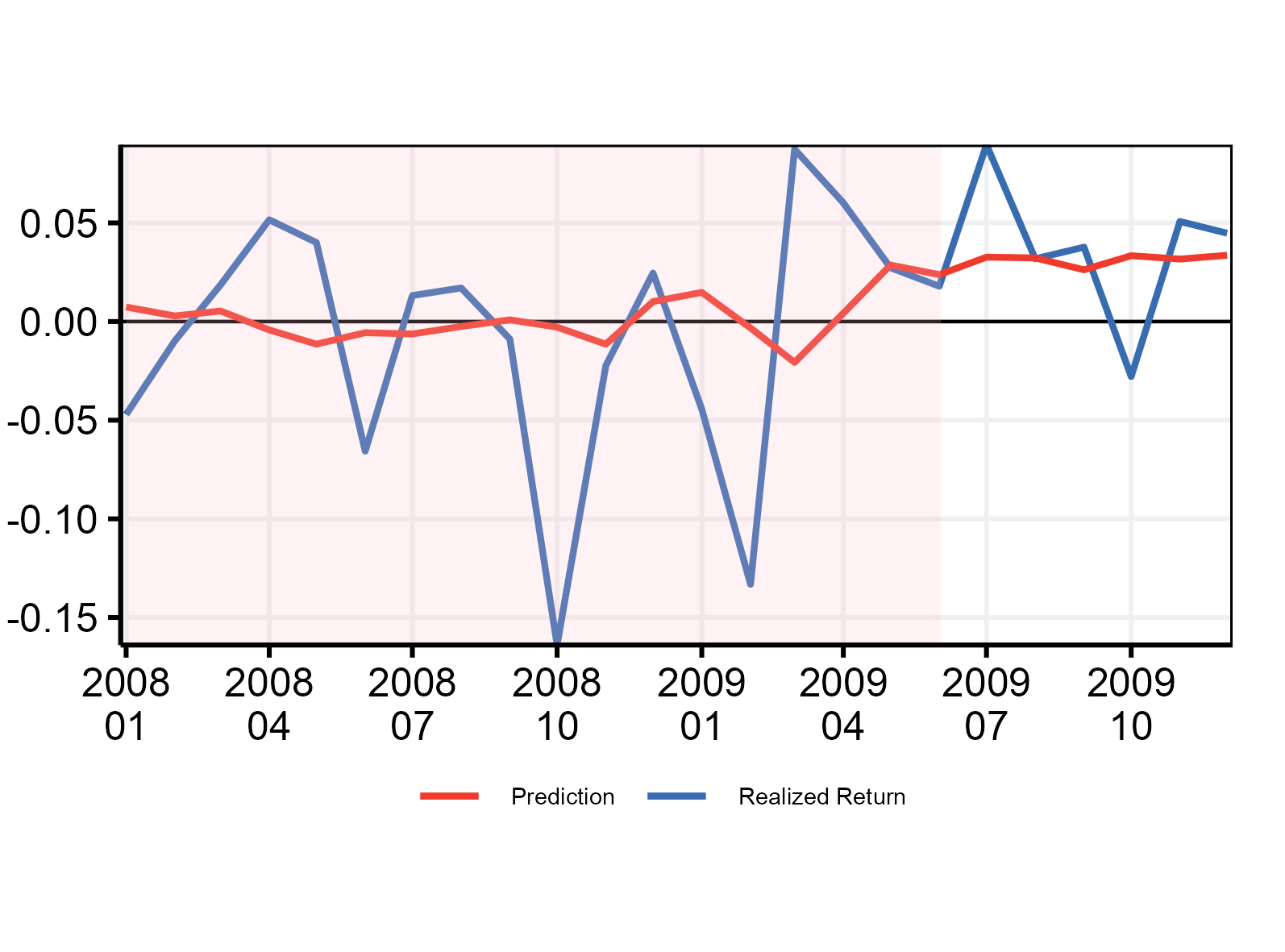}}
\caption{MACE} \label{fig:GR_monthly_MACE}
  \end{subfigure}
  \begin{subfigure}[b]{\textwidth}
  \vspace*{0.45cm}
  \centering
  	\captionsetup{skip=-0.5mm}
	\scalebox{1}[1]{\includegraphics[width=\textwidth, trim = 0mm 35mm 0mm 25mm, clip]{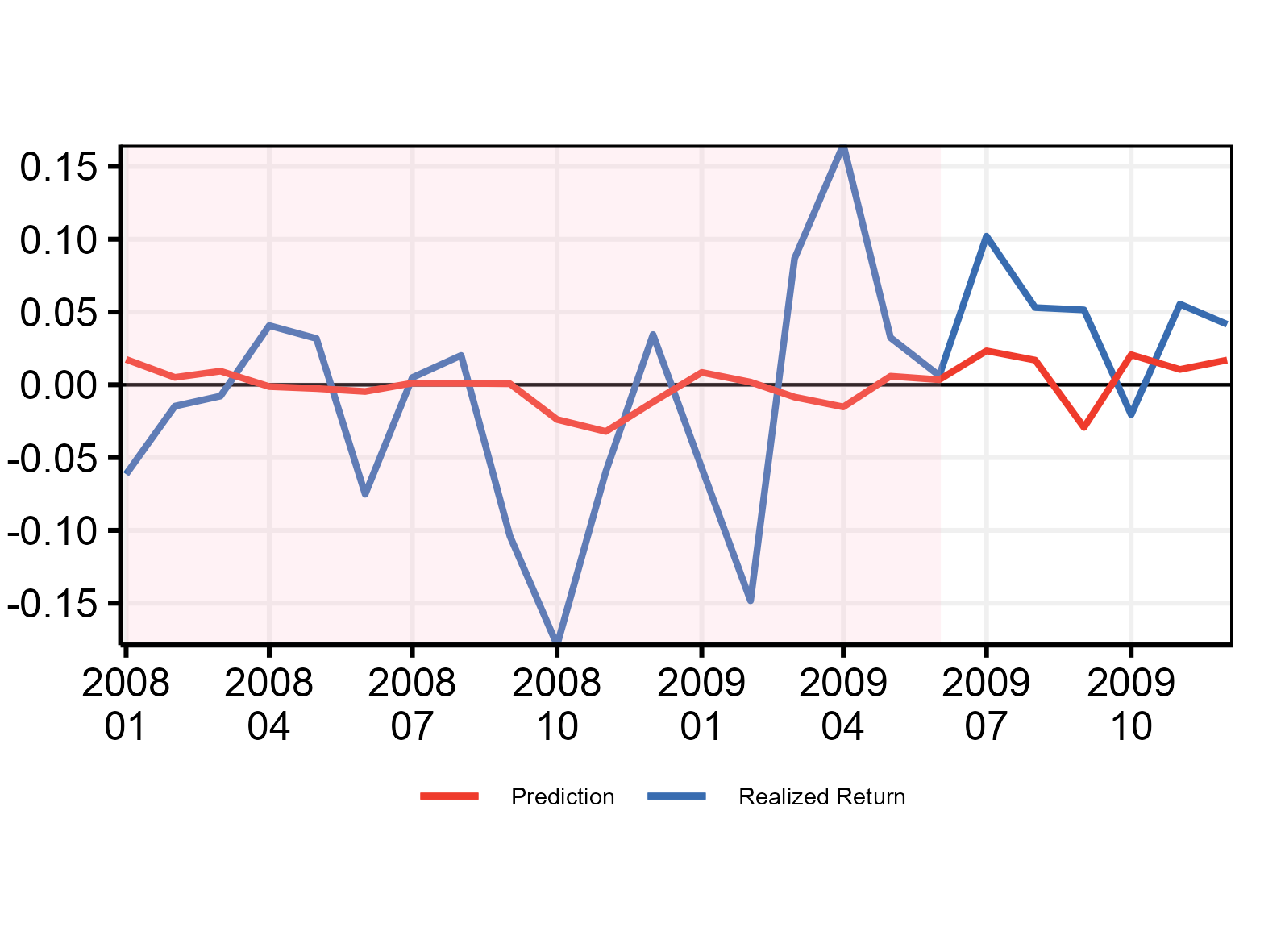}}
\caption{EW (RF)} \label{fig:GR_monthly_EW}
      \end{subfigure} 
       \begin{subfigure}[b]{\textwidth}
  \centering
  	\vspace*{0.3cm}
	\includegraphics[width=0.9\textwidth, trim = 0mm 0mm 0mm 0mm, clip]{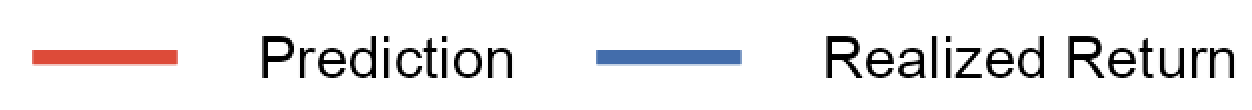}
      \end{subfigure} 
      
     \end{minipage}%
     \begin{minipage}{0.5\textwidth}
     \vspace*{-0.7cm}
     \captionsetup[subfigure]{oneside,margin={-0.3cm,0cm}}
        \begin{subfigure}[b]{\textwidth}
  \centering
  	\captionsetup{skip=5mm}
  	
	 \includegraphics[width=\textwidth, trim = 30mm 5mm 0mm 0mm, clip]{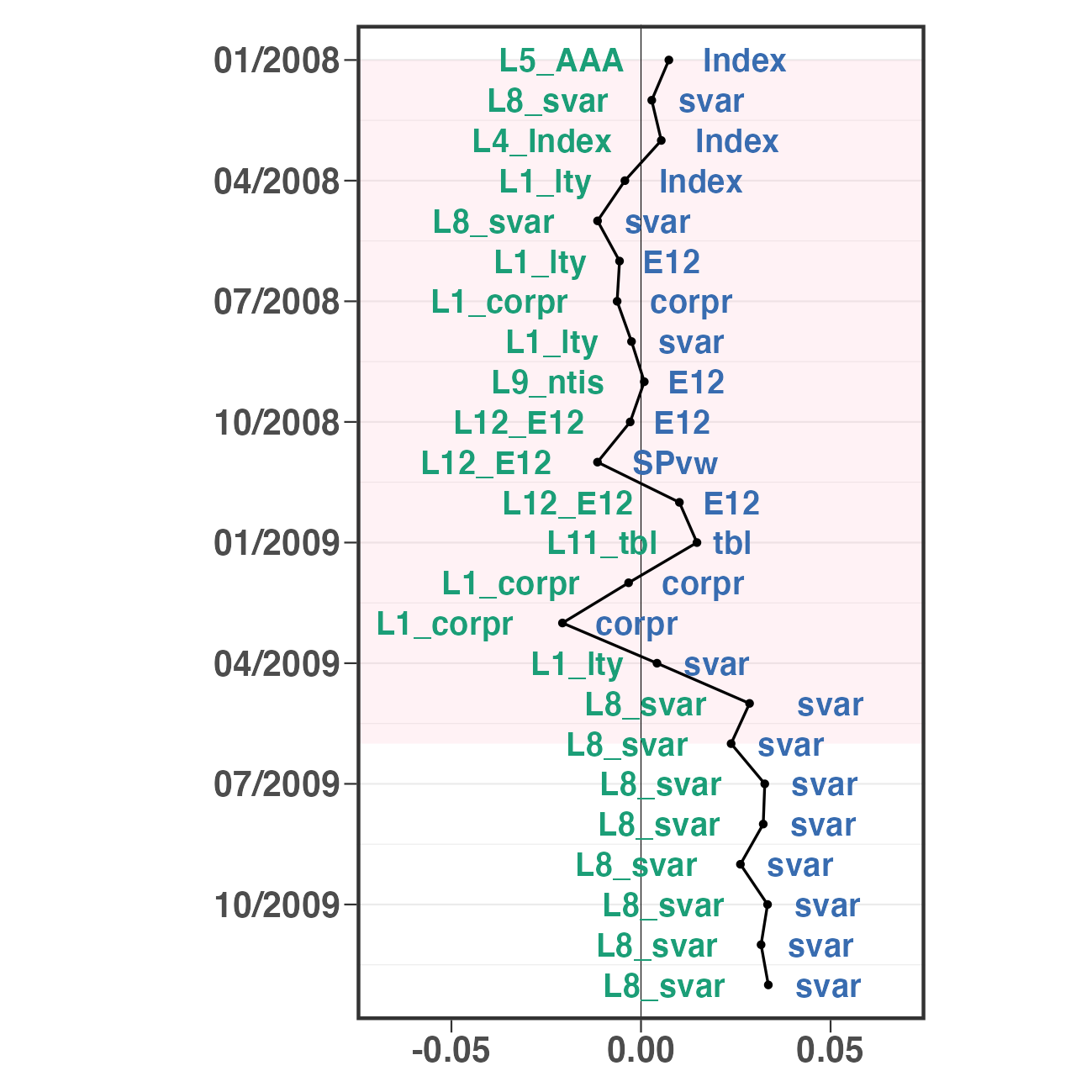}
	\caption{MACE: Most Important Predictors} \label{fig:GR_monthly_c}

      \end{subfigure}   
    \end{minipage} 
  \begin{minipage}{\textwidth}
 	\begin{threeparttable}
 \begin{tablenotes}[para,flushleft]
 \setlength{\lineskip}{0.2ex}
	\scriptsize
	\textit{Notes} Panels (a) and (b): we plot the realized return of the MACE and an equally weighted portfolio in blue. In red, we plot the corresponding predicted return of MACE and EW (RF).\ \\
		\textit{Notes} Panel (c): the solid line shows again MACE's prediction (see the solid blue line in panel (a)). To the left of the solid line, we show in green the most important single predictor in each month $t$ based on a Shapley Value decomposition. To the right of the solid line, we show in blue the most important grouped predictor in each month $t$.

	\end{tablenotes}
	 \end{threeparttable}\ \\
 \end{minipage}
      
        \addtocounter{figure}{-1}
  \caption{MACE and EW (RF) during the Great Recession}
  \label{fig:GR_monthly}
\end{figure}

\textcolor{black}{When it comes to EW (RF), a few differences stand out: 
the contributions are more evenly distributed, with $\mathtt{E12}$ coming in first,  followed by stock market volatility.}  The other features are mostly shared with  MACE's prediction.  Yet,  EW (RF) predictions partly go awry post 2008,  and MACE's peculiar use of $\mathtt{svar}$ is the most plausible explanation for it avoiding this predicament.  This is quite visible in Figure \ref{fig:GR_monthly_MACE} and \ref{fig:GR_monthly_EW}, where we show the predictions of MACE and EW (RF) compared to the corresponding realized return.  MACE's portfolio has a slightly less volatile return  starting from 2009 and the associated predictions lie confidently in positive territory.   EW's realized return has higher highs and lower lows and predictions are much more timid,  that is,  they are much closer to their unconditional mean.  

\textcolor{black}{In} Figure \ref{fig:GR_monthly_c}, we plot again the prediction for MACE and in each month,  we report the single most important feature to its left-hand side.  On the right,  it is the single most important group (of lags).\footnote{See Appendix \ref{sec:vi} for further details about the calculation.  For the grouped version, we sum the absolute Shapley Values at a given point in time  across all lags of a particular feature $i$.   This plotting scheme is inspired by \cite{anatomy}. }  It is striking that the string of positive predictions at the outset of the Great Recession are all attributable to $\mathtt{svar}$, and in particular,  its 8$^\text{th}$ lag. Yet,  the prior evidence on the relevance of $\mathtt{svar}$ is mixed.   In applications \textcolor{black}{with sample periods} ending prior to the Great Recession,  \cite{Guo2006} gets positive results while  \cite{WelchGoyal2008}  and \cite{RapachStraussZhou2010} get negative ones.   Using data through 2013,  \cite{LimaMeng2017} find $\mathtt{svar}$ gaining forecasting power for S\&P 500 excess returns post-1985.  MACE differs by being nonlinear and not looking at a pre-specified index.  However,  nonlinearity by itself appears to be insufficient as per EW (RF) not leveraging $\mathtt{svar}$ in any meaningful way.

\begin{figure}[t!]
\hspace*{-1cm}
\captionsetup{skip=2mm}
      \centering
      \begin{minipage}{0.5\textwidth}
      \captionsetup[subfigure]{oneside,margin={1cm,0cm}}
          \begin{subfigure}[t]{\textwidth}
  \centering
  \vskip -17pt
      \includegraphics[width=\textwidth, trim = 0mm 8mm 0mm 3mm, clip]{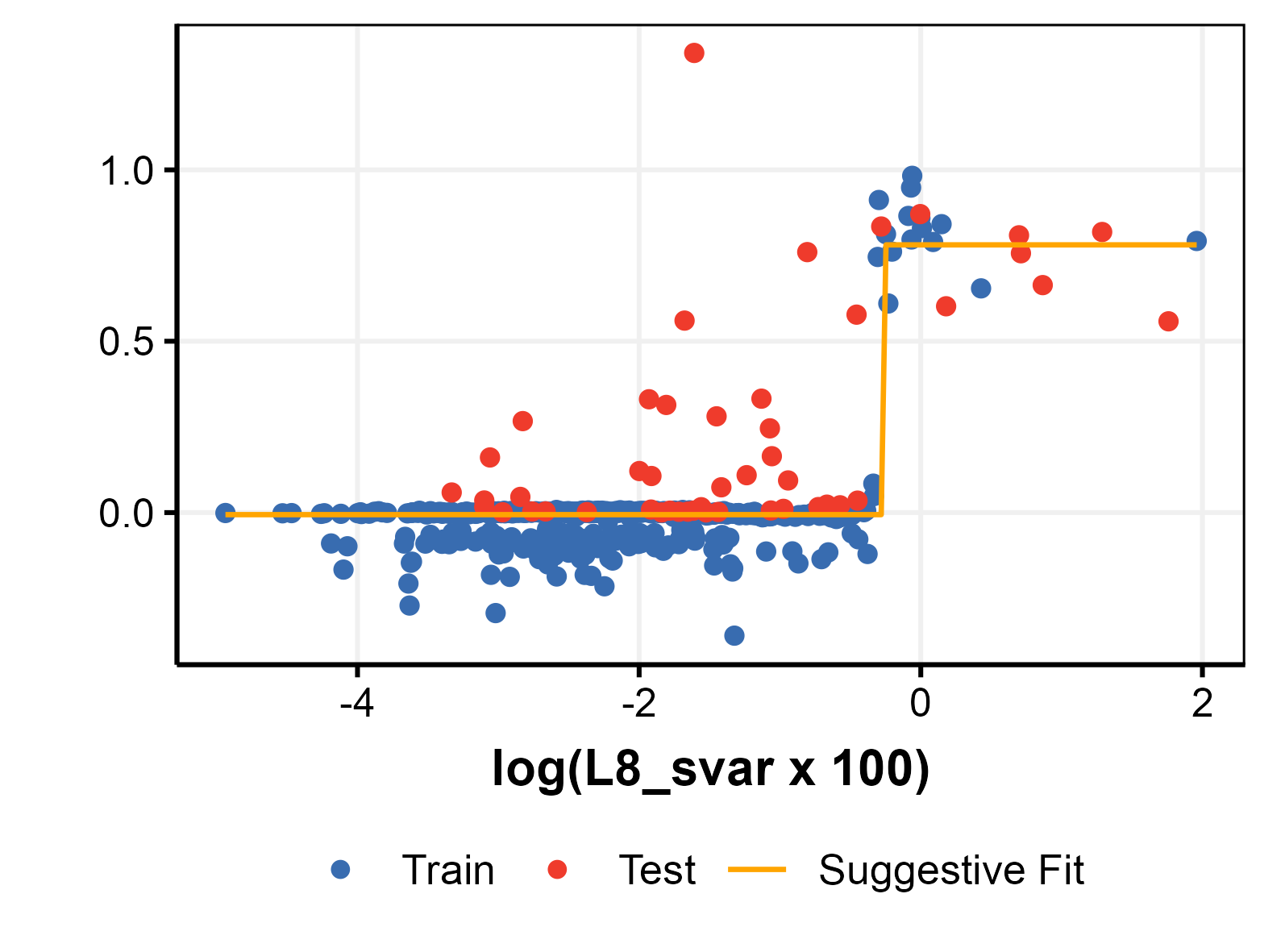}
      \vskip -1pt
      \caption{\footnotesize Shapley Values: $\mathtt{L8\_svar}$ }\label{fig:svar_Shap_TS_scatter_sc}
  \end{subfigure}
  \end{minipage}%
  \begin{minipage}{0.5\textwidth}  
  \captionsetup[subfigure]{oneside,margin={1cm,0cm}}
  \begin{subfigure}[t]{\textwidth}
  \centering
      \includegraphics[width=\textwidth, trim = 0mm 24mm 0mm 23mm, clip]{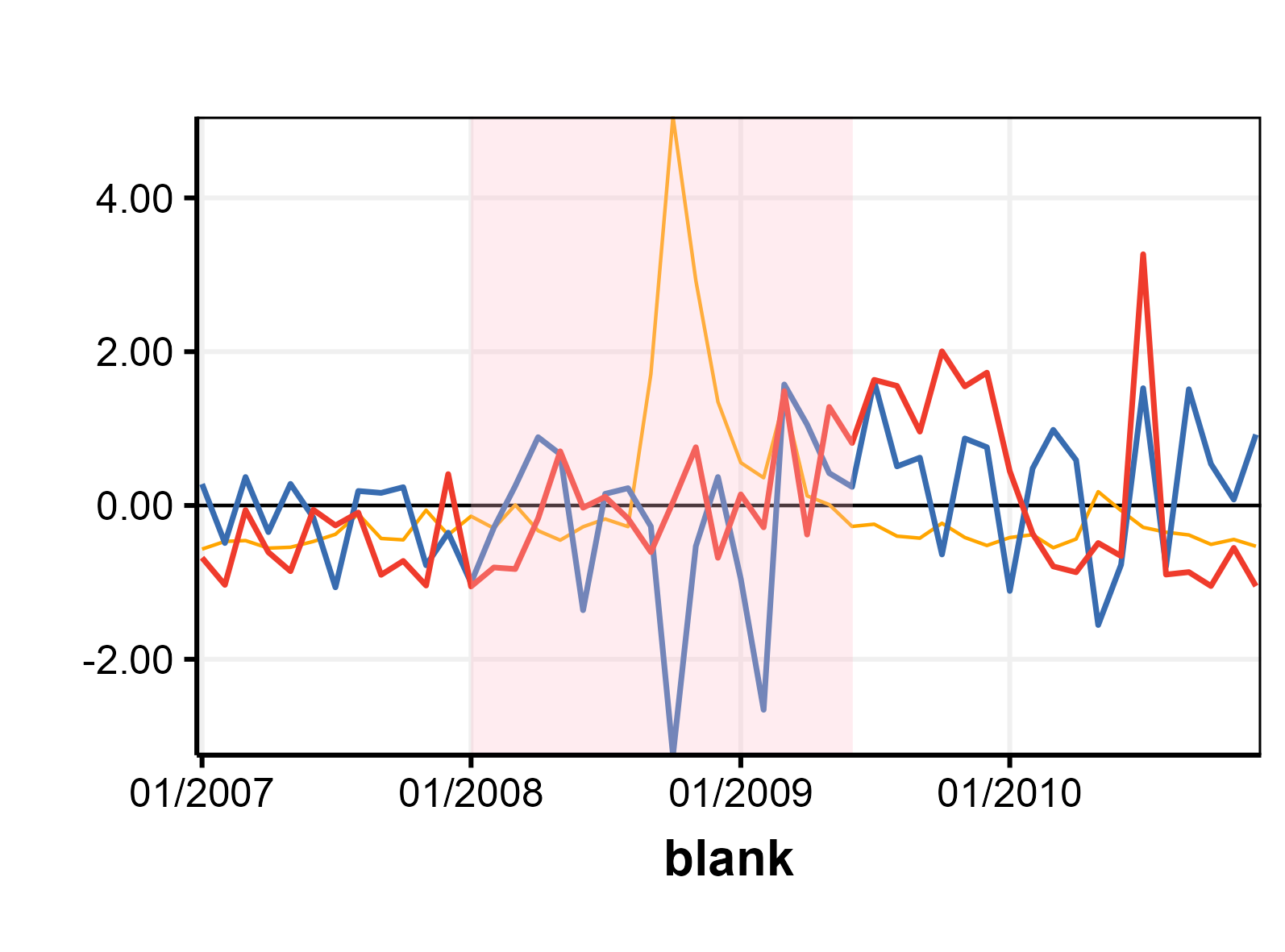}
     \hspace*{1cm} \includegraphics[width=0.85\textwidth, trim = 0mm 0mm 0mm 0mm, clip]{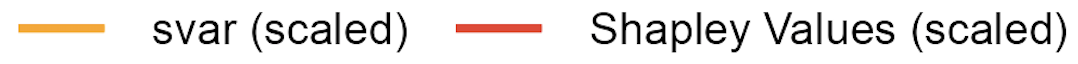} 
      \hspace*{1cm} \includegraphics[width=0.51\textwidth, trim = 0mm 0mm 0mm 0mm, clip]{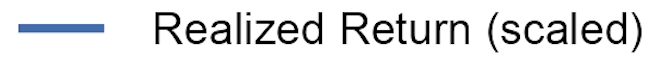}
      \vskip 2pt
      \caption{\footnotesize MACE and $\mathtt{svar}$ over time} \label{fig:svar_Shap_TS_scatter_ts}
  \end{subfigure} \ \\
  \end{minipage}
 \begin{minipage}{\textwidth}
 	\begin{threeparttable}
 \begin{tablenotes}[para,flushleft]
 \setlength{\lineskip}{0.2ex}
	\scriptsize
	\textit{Notes} Panel (a): we plot the Shapley Values for the 8$^\text{th}$ lag of $\mathtt{svar}$ against the observed stock-market volatility, lagged by 8 months ($\mathtt{L8\_svar}$). The points in blue represent the local Shapley Values for the in-sample period of the expanding-window ending in 12/2006. The points in red depict the local out-of-sample Shapley Values of our expanding-window exercise. The orange line is fitted to the joint distribution of in- and out-of-sample points. \ \\
		\textit{Notes} Panel (b): in blue, we show the \textit{scaled} realized return of the MACE portfolio from a buy-\&-hold strategy. The red line shows the \textit{scaled} localized Shapley Values of the grouped $\mathtt{svar}$ on the left, and of the 8$^\text{th}$ lag of $\mathtt{svar}$ on the right.

	\end{tablenotes}
	 \end{threeparttable}\ \\
 \end{minipage}
         \addtocounter{figure}{-1}
  \caption{MACE \& Stock-Market Volatility ($\mathtt{svar}$) around the Great Recession \\ 01/2007 - 12/2010}
  \label{fig:svar_Shap_TS_scatter}
\end{figure}

Lastly,  we investigate \textit{how} $\mathtt{L8\_svar}$ appears to contribute.   Figure \ref{fig:svar_Shap_TS_scatter_sc} shows a scatter plot of the Shapley Values for $\mathtt{L8\_svar}$ (the "local contributions") with $\log\left(\mathtt{L8\_svar} \times 100\right)$ \textcolor{black}{on the} x-axis.\footnote{The dots in blue represent the Shapley Values over the training-set of the expanding window with an OOS start date in 01/2007. The red dots are the OOS Shapley Values collected for the expanding window 01/2007-12/2010. See Appendix \ref{sec:vi} for a detailed description of the collection process.} The yellow line represents a suggestive fit as the mean of all point realizations of $\log\left(\mathtt{L8\_svar} \times 100\right) < -0.25$ and the mean of all realizations of $\log\left(\mathtt{L8\_svar} \times 100\right) \geq -0.25$.  While being inherently imperfect because of $\mathtt{L8\_svar}$'s various interactions with other predictors in RF,  this fit is nonetheless instructive.  First,  the sign is right: more risk commands a higher premium.  At a level of around $\mathtt{L8\_svar} \approx 0.0078$, which translates into a measure of daily stock-market volatility of $\sigma = \sqrt{\mathtt{L8\_svar}} \approx 0.088 \equiv 8.8\%$,  market uncertainty seemingly triggers a regime of higher expected rate of return on the MACE portfolio.  This observation speaks to the findings in \cite{CampbellHentschel1992} that the volatility feedback channel emerges during times of elevated volatility.  Here,  the suggestive fit points to a simple two regimes relationship: a first one where there is basically no risk premium,  and second where there is a constantly higher premium, irrespective of the specific values of $\mathtt{L8\_svar}$ as long as it is above a certain threshold.

Figure \ref{fig:svar_Shap_TS_scatter_ts} shows how this nonlinear relationship plays out in the time space.   We plot the \textit{scaled} versions of the realized stock variance ($\mathtt{svar}$), the local Shapley Values for the grouped $\mathtt{svar}$, and the realized MACE returns.  The positive relationship between $\mathtt{svar}$ and the realized MACE returns is clearly emerging from the midst of the Great Recession in late 2008 onwards.  The delay is visible from the red bump appearing much later than the original $\mathtt{svar}$ impulse.  The nonlinearity is also discernible from the red line following a very different pattern than the orange one -- perfect linearity would imply a mere rightward translation of the orange line.   The red plateau is well timed with high (unconditional) MACE realized returns at the outset of the Great Recession. From that and other observations,  we can conclude that part of MACE's success in that era is uncovering a portfolio with a well dissimulated,  yet stronger reaction to changes in volatility regimes.

\subsection{Explainability by Factor-based Strategies} \label{sec:factor_m}

We now conduct a similar exercise to that of Section \ref{sec:factor_d}.   The first regression uses the five-factor model augmented by the momentum factor constructed based on all the stocks available in each period.  Given the proliferation of other legitimate anomalies in the literature \citep{FengGiglioXiu2020,JensenKellyPedersen2023} -- and their availability at the monthly frequency -- we also consider a wider set of factors  and augment the six-factors mentioned above by the dataset of \cite{JensenKellyPedersen2023}.  As in this exercise the number of factors ($N=159$) approaches the number of out-of-sample observations $T=180$, we first run LASSO to select the relevant set of factors and conduct inference afterwards by simply regressing the MACE portfolio on the factors with a non-zero loading from the first pass regression.\footnote{Note the we pre-distill the zoo by taking out factors that show a correlation with either of the five \cite{FF2015} factors that exceed 0.8 in absolute terms over the monthly out-of-sample period from January 1987 - December 2019. } In Table \ref{tab:Factors_m_post} we report results for the six-factor model on the left-hand side and for the zoo of factors on the right-hand side. We run each exercise for the two subperiods of January 1987 - December 2004 and January 2005 until the end of our sample in December 2019.

There are some recurring themes from the same analysis of Section \ref{sec:factor_d} on daily data.  Trading according to RF -- when predictability does indeed emerge, as it does for the second half of the sample -- neutralizes a lot of factor-based explanatory power,  with factor regression's $R^2$ shrinking by more than half when moving from MACE (PM) to MACE.  Despite loading on four out of six factors in a statistically significant manner, the six-factor workhorse model is only able to explain about 26\% of the MACE portfolio.   Again,  MACE has a moderate market beta of around 0.5, thus offering the investor some protection against market swings.  However,  here,  it is not due to the construction of the portfolio itself ($\boldsymbol{w}\neq0$ is enforced),  but the action of trading with RF.  Indeed,  MACE (PM) highlights that the raw portfolio is in fact very much exposed to the market factor.  Also, we note that MACE loads on the size factor, yet, this time with a positive sign, suggesting that MACE successfully exploits the return pattern of buying small and shorting large stocks. 

The spanning regression  results confirm earlier observations for the monthly exercise : first, pre-2005, MACE does not generate any alpha, but the story changes entirely in the second half of the sample where  MACE generates an alpha that amounts up to 99 basis points on a monthly basis.  This excess return survives when moving to the \textit{factor zoo} on the right-hand side.  In fact, the alpha even increases by more than 20 basis points. Sticking with the post 2005 sample and our benchmark MACE specification, the first-pass LASSO keeps only the market factor from the core six-factor model.  The only anomalies that join the exposure to systematic market swings in a statistically significant manner are \textit{residual momentum over the last six months} (resff3\_6\_1) -- i.e. momentum in that part of stock returns that is unexplained by the \cite{FF1992} three-factor model -- developed in \cite{BlitzHuijMartens2011}, and \textit{asset tangibility} (tangibility) \citep{HahnLee2009}.  In summary, the spanning regressions have unveiled some additional insights, which were not apparent hitherto: in the latter part of the sample, MACE identifies and successfully leverages prominent anomalies in the market, while still finding some patterns that allow it to generate some alpha that is not to be explained by factor strategies ranging from a conservative 6-factors strategy to considering the whole (listed) Factor Zoo.

\section{Concluding Remarks}\label{sec:con}

We introduce the MACE algorithm to construct maximally machine-learnable portfolios.  \textcolor{black}{It} does so by directly optimizing the portfolio weights to make life easier for the prediction function.  As we have discussed,  this does not neglect variance,  quite to the contrary,  as the MMLP problem is intimately linked to traditional mean-variance optimization.   Advantages with respect to the various strands of literature building \textit{linear} mean-reverting portfolio is MACE's flexibility through the use of Random Forest and its scalability.  Peaking into the future,  those qualities are essential to discover increasingly complex patterns of predictability in an era where a flock of humans and machines are constantly on the lookout for those.  With respect to key ML applications in empirical asset pricing,  MACE provides a low maintenance (data- and computations-wise) alternative which can deliver the goods leveraging only basic time series data, or lagged returns themselves.  Our two applications,  daily and monthly trading,  illustrate that by scoring enviable returns and Sharpe Ratios in evaluation periods where gains from using ML methods have often been anticlimactic.

There are quite a few directions for future research, beyond more or less straightforward applications to new assets and information sets,  and changing ridge regularization for any other shrinkage one's heart desires.  First,  MACE could be extended to solely learn buy-sell signals where the cutoff point itself is trainable within the loop.  In that way,  we could potentially construct ``episodic portfolios'' where trading rarely occurs and typically does so when a rarer event is expected with moderate uncertainty.  Second,  some structured form of nonlinearities could be accommodated on the left-hand side of the equation.  While from a statistical standpoint,  \textcolor{black}{nothing is impossible},  from a financial one,  the LHS must remain a tradeable combination of securities. Nonetheless,  some nonlinear transformations of returns can be approximated by appropriately designed options and MACE could learn a maximally predictable combination of financial instruments.  Third  and more ambitiously,  MACE's alternating EM-style algorithm could potentially be replaced by a single hemisphere neural network (à la \cite{HNN}) that minimizes directly the MMLP loss function combined with bagging strategies to deal with the inevitability of overfitting and finding a trivial solution.  As discussed earlier, there are numerous headwinds to such modifications and bagging by itself may not be enough.  But,  keeping in mind deep learning's edge with large and non-traditional data,  the additional efforts could perhaps bring MMLPs to new highs.

 \pagebreak
 
 \setlength\bibsep{5pt}
		
\bibliographystyle{apalike}

\setstretch{0.75}
\bibliography{ref_GCG_MACEP.bib}

\clearpage

\appendix
\newcounter{saveeqn}
\setcounter{saveeqn}{\value{section}}
\renewcommand{\theequation}{\mbox{\Alph{saveeqn}.\arabic{equation}}} \setcounter{saveeqn}{1}
\setcounter{equation}{0}
	
\section{Appendix}
\setstretch{1.25}

\subsection{Transaction Costs}\label{sec:TC}





In the section,  we quantifying the reduction in economic performance due to transactions costs (TC).  To do so, we calculate
\vspace*{-0.25cm}
\begin{equation} \label{equ:GKX}
	\text{TC}_t = \textfrak{c} \; \underbrace{\sum^N_{n=1} \left\vert \varpi_{t,n} - \frac{\varpi_{t-1,n}\left(1+r_{t,n}\right)}{1 + \sum_j \varpi_{t-1,j} r_{t,j}} \right\vert }_{\text{Turnover}_t } \quad ,
\end{equation}
where $\varpi_{t,n}$ is the effective weight, i.e. $\varpi_{t,n} \equiv \omega_{t} \, w_{t,n}$, and where Turnover$_t$ follows \cite{gu2020empirical}.

{\vspace{0.15cm}}
{\noindent \sc \textbf{Daily Results.}}  Given that our strategy utilizes highly liquid stocks listed on the NASDAQ,  we expect transaction costs (TC) to be low.   Thus,  we set $\textfrak{c} \in \{0.01\%, 0.015\%,0.03\% \}$. These reflect the 50$^\text{th}$, 75$^\text{th}$, and 90$^\text{th}$ percentile of the one-half bid-ask spread \citep{DuTepperVerdelhan2018} distribution of the MACE$_{100}$ stocks between 2017-01-01 and 2022-12-31.

\vspace*{0.25cm}
\begin{table}[h!]
\centering
		\vspace*{0.45em}
  \begin{threeparttable}
\centering
\footnotesize
\caption{Daily Stock Returns after Transaction Costs \label{tab:TC_daily} \vspace{-0.3cm}}
\setlength{\tabcolsep}{0.85em}
\begin{tabular}{ l *{10}{c} }
\toprule
\rowcolor{white}
\hspace*{0.075cm} &
\multicolumn{3}{c}{$0.01\%$} &
\multicolumn{3}{c}{$0.015\%$} &
\multicolumn{3}{c}{$0.03\%$}  \\
\cmidrule(lr){2-4} \cmidrule(lr){5-7} \cmidrule(lr){8-10} 
&
\makebox[3em]{$r^A$} &
\makebox[3em]{$SR$} &
\makebox[3em]{$\Omega$} &
\makebox[3em]{$r^A$} &
\makebox[3em]{$SR$} &
\makebox[3em]{$\Omega$} &
\makebox[3em]{$r^A$} &
\makebox[3em]{$SR$} &
\makebox[3em]{$\Omega$} \\
\midrule
 MACE$_{20}$    & 19.72 & 0.84 & 1.14 & 18.07 & 0.77 & 1.12 & 13.14 & 0.56 & 1.07 \tabularnewline \addlinespace[2pt]

MACE$_\text{loose bag}$    & 20.23 & 1.19 & 1.20 & 18.82 & 1.11 & 1.18 & 14.58 & 0.86 & 1.11 \tabularnewline \addlinespace[2pt]

 MACE$_{100}$   & 32.83 & 1.26 & 1.24 & 28.58 & 1.1 & 1.20 & 15.81 & 0.61 & 1.08 \tabularnewline
\bottomrule
\end{tabular}
\begin{tablenotes}[para,flushleft]
\scriptsize \textit{Notes}: This table reports annualized returns ($r^A$), Sharpe Ratio ($SR$) and the Omega Ratio ($\Omega$) for various MACE portfolios after accounting for transaction costs with $\textfrak{c} \in \{0.01\%, 0.015\%, 0.03\% \}$.
  \end{tablenotes}
  \end{threeparttable}
\end{table} 

Table \ref{tab:TC_daily} reports the corresponding summary statistics for various MACE portfolios after subtracting $TC_t$ from the realized returns $r_t$. For the two portfolios with $N=20$ stocks,  annualized returns before TC amounted to 23.16\% for MACE$_{20}$ (Table \ref{tab:summstats_daily} and to 23.10\% for MACE$_\text{loose bag}$ (Table \ref{tab:summstats_daily_refine}). As Table \ref{tab:TC_daily} shows, the fallout due to transaction costs is well contained for these portfolios\textcolor{black}{, assuming $\textfrak{c}$ to range at the between the 50$^\text{th}$, 75$^\text{th}$ percentile of the bid-ask-spread distribution described above.}.  Evidently, the degradation has to be higher for MACE$_{100}$ since it implies trading five times more stocks.  \textcolor{black}{Still, even under the most conservative scenario with $\textfrak{c} = 0.03\%$ it outperforms EW and the S\&P 500 (before accounting for TC) in terms of annualized returns and is competitive in terms of risk-adjusted returns}. \textcolor{black}{Note also that for $\textfrak{c} \in \{0.01\%, 0.015\%\}$,  TC-adjusted $r^A$,  $SR$,  and $\Omega$ for all three MACEs are still  well above what is reported for competing strategies , including passive ones (with TCs $\approx 0$) and more proactive ones.} 



{\vspace{0.15cm}}
{\noindent \sc \textbf{Monthly Results.}} Table \ref{tab:TC_monthly} shows the monthly returns for MACE and its refinements after accounting for several levels of transaction costs.  
\textcolor{black}{We now use $\textfrak{c} \in \{ 0.0005, 0.001, 0.01\}$, of which the latter two are based on \cite{CongEtAl2021}, and $\textfrak{c} = 0.05\%$ is added given the ongoing reduction in TC with the advent of commission-free trading platforms in the 2010s.} 

\textcolor{black}{For $\textfrak{c} \in \{ 0.0005, 0.001\}$, we see that TCs only eat up a minor fraction of average monthly returns, such that also the corresponding risk-metrics, $SR$ and $\Omega$ remain in the neighborhood of those reported in Table \ref{tab:summstats_monthly}. Only the upper bound of $\textfrak{c} = 0.1\%\}$ eats up a significant part of the generated returns. Yet, still in that case, MACE$_{\mu \geq \underline{\mu}}$ remains highly competitive to the non-TC adjusted benchmarks reported in Table \ref{tab:summstats_monthly}. }

\vspace*{0.5cm}
\begin{table}[h!]
\centering
		\vspace*{0.45em}
  \begin{threeparttable}
\centering
\footnotesize
\caption{Monthly Stock Returns after Transaction Costs \label{tab:TC_monthly} \vspace{-0.3cm}}
\setlength{\tabcolsep}{0.85em}
\begin{tabular}{ l *{10}{c} }
\toprule
\rowcolor{white}
\hspace*{0.075cm} &
\multicolumn{3}{c}{$0.05\%$} &
\multicolumn{3}{c}{$0.1\%$} &
\multicolumn{3}{c}{$1.0\%$}  \\
\cmidrule(lr){2-4} \cmidrule(lr){5-7} \cmidrule(lr){8-10} 
&
\makebox[3em]{$r^A$} &
\makebox[3em]{$SR$} &
\makebox[3em]{$\Omega$} &
\makebox[3em]{$r^A$} &
\makebox[3em]{$SR$} &
\makebox[3em]{$\Omega$} &
\makebox[3em]{$r^A$} &
\makebox[3em]{$SR$} &
\makebox[3em]{$\Omega$} \\
\midrule \addlinespace[5pt]

& \multicolumn{9}{c}{01/2005 - 12/2019} \\
\cmidrule(lr){2-10}
 MACE & 18.43  & 1.03  &  1.87  &  17.90  & 1.00  &  1.82  &  8.37  & 0.46  &  1.16 \tabularnewline \addlinespace[2pt]

MACE$_\text{bag}$    & 16.64  & 0.96  &  1.71  &  16.13  & 0.93  &  1.67  &  6.94  & 0.40  &  1.09  \tabularnewline \addlinespace[2pt]

 MACE$_{\mu \geq \underline{\mu}}$   & 19.33  & 1.00  &  1.77  &  19.08  & 0.99  &  1.75  &  14.60  & 0.76  &  1.46  \tabularnewline \addlinespace[2pt]
 
 & \multicolumn{9}{c}{01/1987 - 12/2004} \\
\cmidrule(lr){2-10}
 MACE & 7.57  & 0.34  &  1.10  &  7.02  & 0.32  &  1.07  &  -2.90  & -0.13  &  0.75  \tabularnewline \addlinespace[2pt]

MACE$_\text{bag}$    & 7.72  & 0.37  &  1.11  &  7.22  & 0.35  &  1.08  &  -1.73  & -0.08  &  0.78 \tabularnewline \addlinespace[2pt]

 MACE$_{\mu \geq \underline{\mu}}$   & 11.79  & 0.56  &  1.28  &  11.49  & 0.54  &  1.26  &  6.03  & 0.28  &  1.04  \tabularnewline
 
\bottomrule
\end{tabular}
\begin{tablenotes}[para,flushleft]
\scriptsize \textit{Notes}: This table reports annualized returns ($r^A$), Sharpe Ratio ($SR$) and the Omega Ratio ($\Omega$) for various MACE portfolios after accounting for transaction costs with $c \in \{0.5\%, 1.0\%, 2.0\% \}$. See Equation \eqref{equ:GKX} for the exact calculation.
  \end{tablenotes}
  \end{threeparttable}
\end{table}

\subsection{Variable Importance Calculations}\label{sec:vi}

In our monthly expanding window exercise, both periods are obviously not static but ``evolving''. With $e = 1,...,E$ expanding windows, $T^{ins}_e$ denotes the end of the in-sample period for window $e$. Hence, we collect the corresponding Shapley Values for the OOS period as follows: for each variable $i$, we collect only those Shapley Values that fall into the interval starting with the month following the end of the current window's in-sample period ($T^{ins}_e + 1$) and ending with the end of the in-sample period of the next expanding window ($T^{ins}_{e+1}$). As we expand our in-sample period each quarter by another three months, the period between  $T^{ins}_e + 1$ and $T^{ins}_{e+1}$ amounts to three months. The corresponding OOS variable importance of variable $i$ ($VI^{oos}_i$) is thus calculated as follows:

\begin{align} \label{equ:VI_oos}
	VI^{oos}_i = \sum_{e=1}^E \sum^{T^{ins}_{e+1}}_{t=T^{ins}_e + 1} \, \left\vert \, \phi_{i,t} \, \right\vert \quad .
\end{align}

Taking Figure \ref{fig:svar_Shap_TS_scatter} as an example, where the OOS period runs from 01/2007 through 12/2010: we start in 12/2006 and collect the first three local Shapley Values of the OOS-period (01/2007-03/2007). We then expand our training set until 03/2007. Hence we collect the first three local Shapley Values of the new OOS period (04/2007-06/2007). We proceed until our training set ends in 09/2010.
\newpage
Summarizing indicator $i$'s contribution across all it's lags, we calculate the \textit{grouped} $VI$ for group $g$ ($VI^{oos}_g$) as follows:
\begin{align} \label{equ:VI_oos_grouped}
	VI^{oos}_{g} = \sum_{e=1}^E \sum^{T^{ins}_{e+1}}_{t=T^{ins}_e + 1} \sum_{i \in g} \, \left\vert \, \phi_{i,t} \, \right\vert \quad .
\end{align}
where $g$ includes all lags of with which indicator $i$ is represented in the feature set.


\subsection{Volatility-Adjusted $VI$-Plots}\label{sec:Adj_VI}

In Figure \ref{fig:AdjVI_monthly} we show volatility-adjusted $VI$ plots. That is, we adjust $VI^{oos}_i$ in Equation \eqref{equ:VI_oos} by indicator $i$'s ratio of OOS to in-sample standard deviation:

\begin{align} \label{equ:AdjVI_oos}
	AdjVI^{oos}_z = VI^{oos}_z \times \left( \frac{\sigma^{oos}_z}{ \sigma^{ins}_z}\right)^{-1} \quad \text{for} \; z = i,g \quad,
\end{align}

where $\sigma^{oos}_i$ is the standard deviation over the OOS period (here: 01/2008 - 12/2009) and $\sigma^{ins}_i$ the standard deviation over the in-sample period (here: 03/1957 - 12/2007) respectively.

For the grouped case ($AdjVI^{oos}_g$),  the standard deviation ($\sigma^{oos}_g$) is calculated as the standard deviation of the moving-average of indicator $i$, where the length of the moving average corresponds to the number of lags (here 12) with which $i$ enters the predictor matrix.

\vspace*{0.25cm}
\begin{figure}[h!]
\captionsetup{skip=2mm}
      \begin{subfigure}[b]{0.333\textwidth}
  \centering
  \captionsetup{skip=-0.5mm}
        \includegraphics[width=1\textwidth, trim = 0mm 0mm 0mm 0mm, clip]{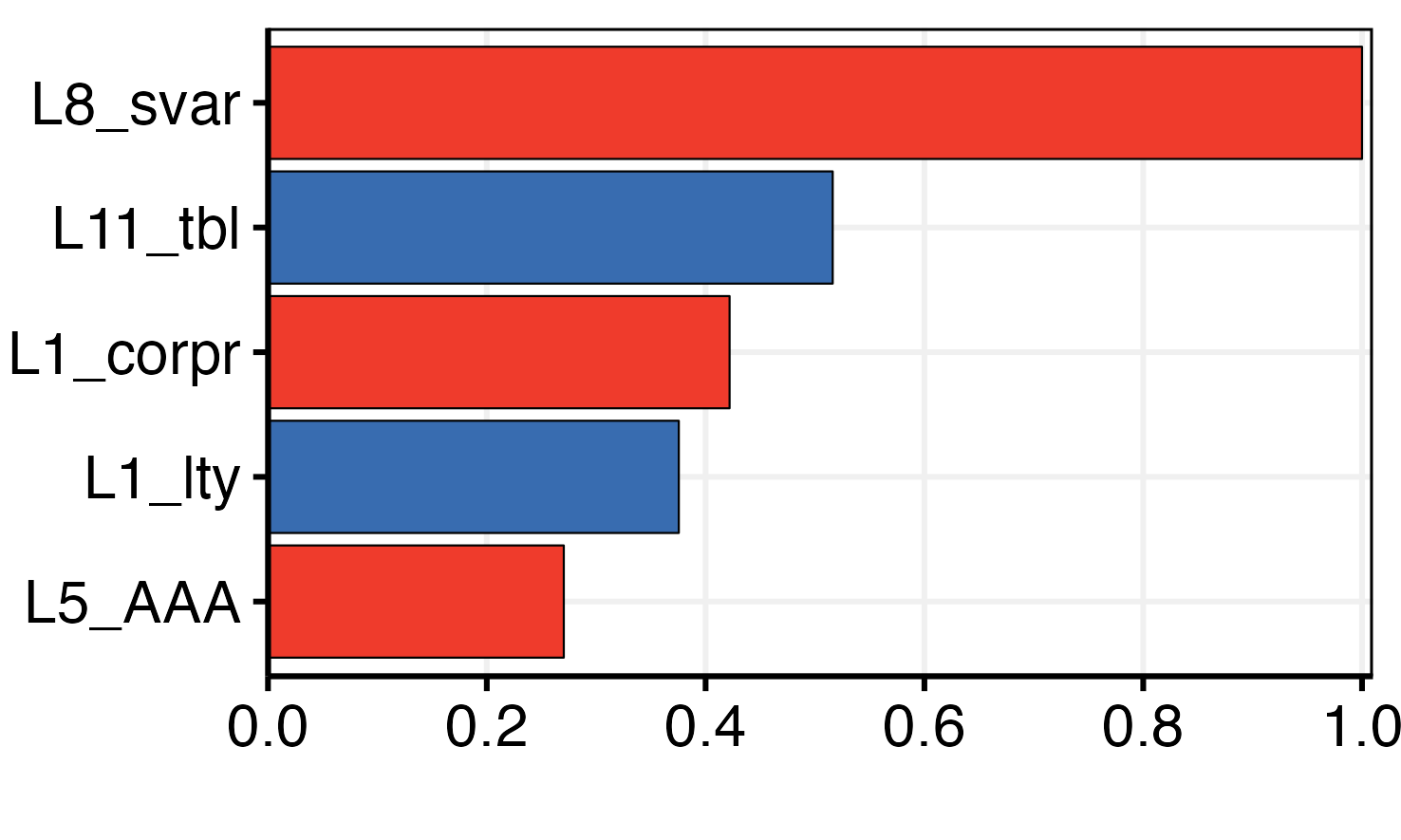}
\caption{MACE:  $AdjVI^{oos}_i$}
      \end{subfigure}
  \begin{subfigure}[b]{0.33\textwidth}
  \centering
  	\captionsetup{skip=-0.5mm}
	\includegraphics[width=\textwidth, trim = 0mm 0mm 0mm 0mm, clip]{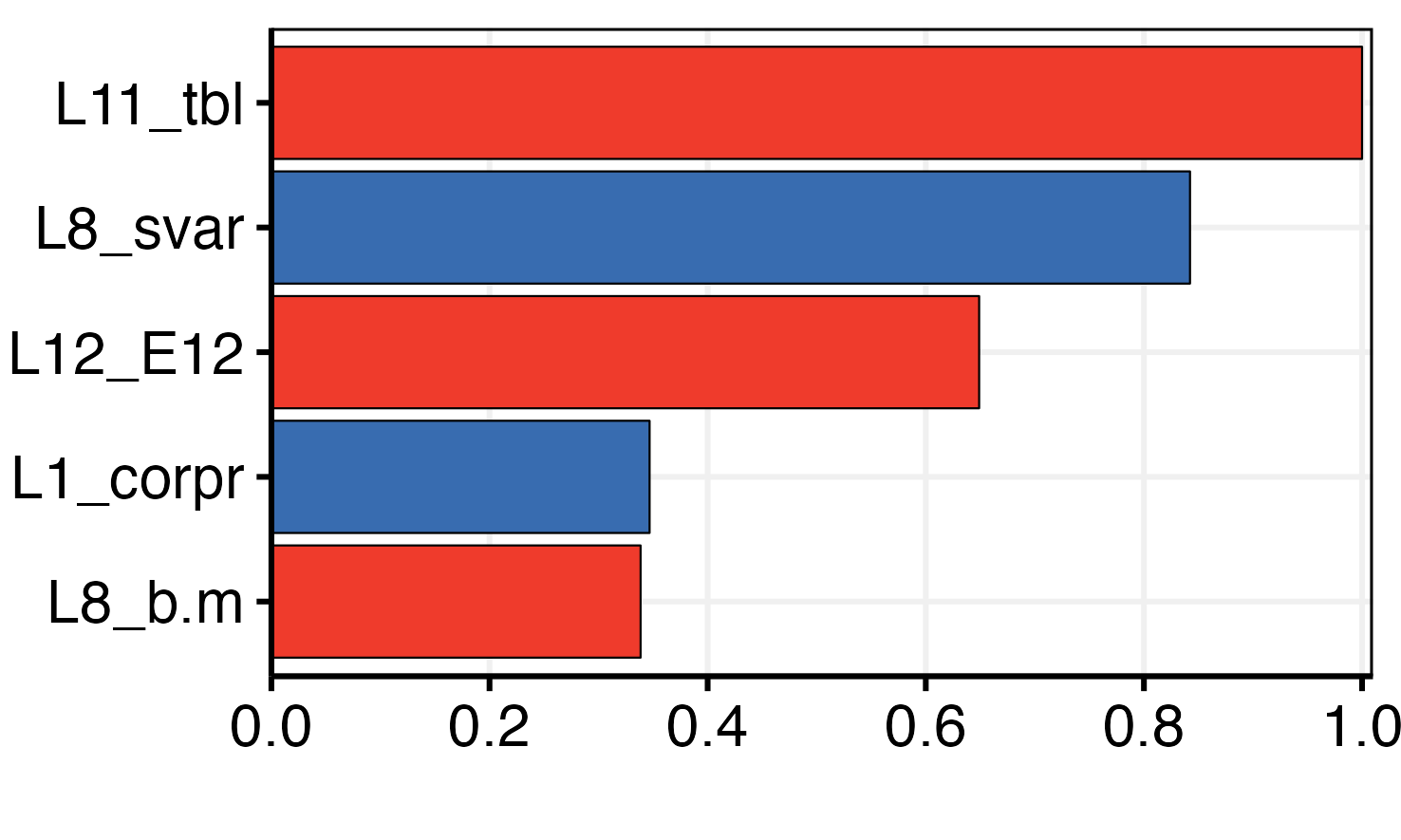}
\caption{EW (RF):  $AdjVI^{oos}_i$}
      \end{subfigure}%
      \begin{subfigure}[b]{0.33\textwidth}
  \centering
  \captionsetup{skip=-0.5mm}
        \includegraphics[width=\textwidth, trim = 0mm 0mm 0mm 0mm, clip]{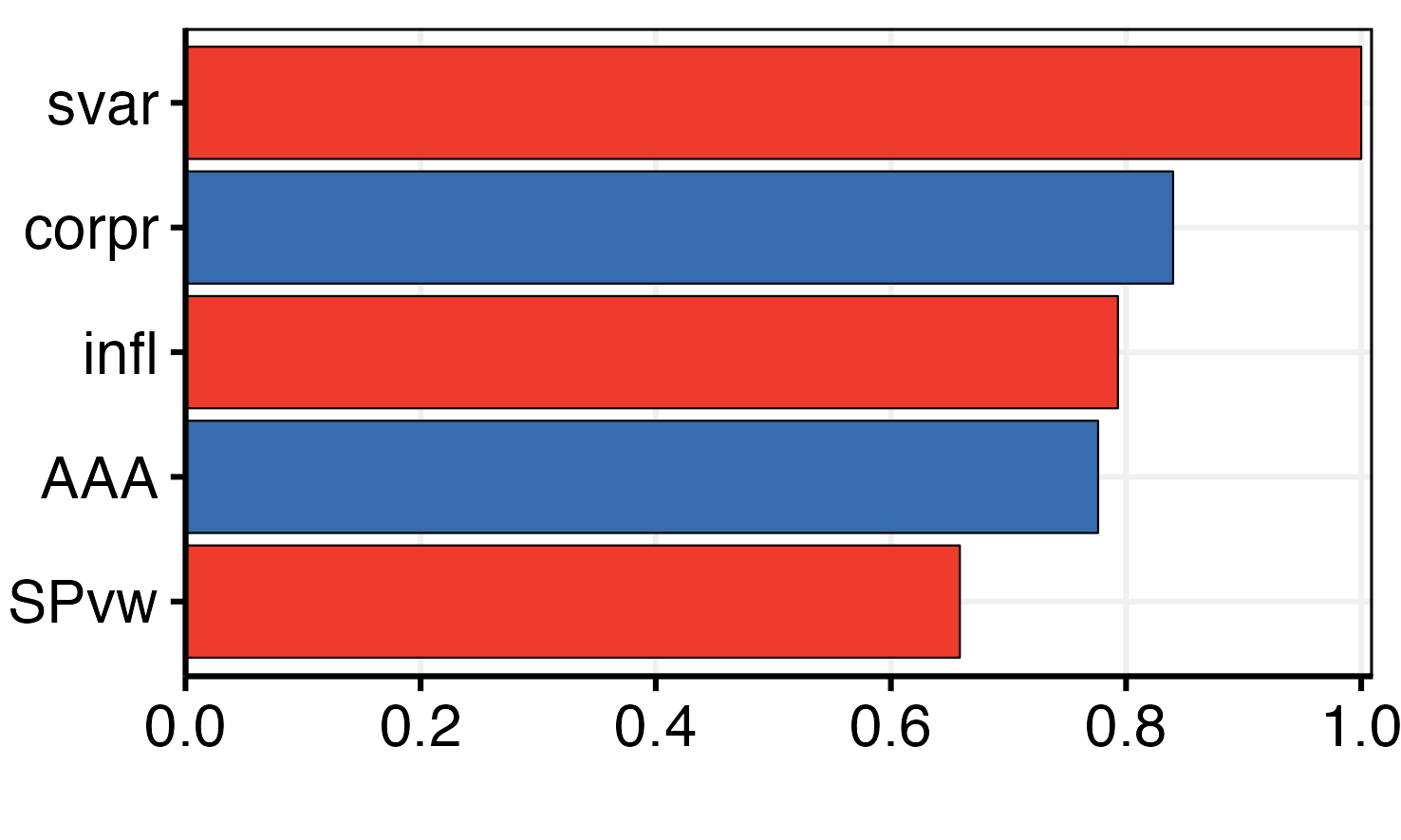}
\caption{MACE:  $AdjVI^{oos}_g$}
      \end{subfigure}%
 \vspace*{-0.1cm}
        \begin{threeparttable}
 \begin{tablenotes}[para,flushleft]
 \setlength{\lineskip}{0.2ex}
	\scriptsize 
		\textit{Notes}: The bars represent the volatility-adjusted $VI$ of predictor $i$ and grouped versions ($AdjVI_{\text{grouped}}$, i.e. ,  summing the Shapley Values across all lags of variable $i$) as in Equation \ref{equ:AdjVI_oos}. We scale indicator $i$'s Shapley Values by the ratio of $i$'s out-of-sample and in-sample standard deviation. The in-sample period runs from 03/1958 to 12/2007. In the \textit{grouped} version, we calculate the standard deviation of the 12-month moving average, as each indicator $i$ enters the predictor matrix with 12 lags. Afterwards, the Shapley Values are further scaled by the maximum Shapley Value.
	\end{tablenotes}
	 \end{threeparttable}\ \\
       \vspace*{0.02cm}
        \addtocounter{figure}{-1}
  \caption{Volatility-Adjusted Shapley Value Importance: 01/2008 - 12/2009}
  \label{fig:AdjVI_monthly}
\end{figure}

\subsection{Additional Tables}\label{sec:tables}

\begin{table}[H]
\caption{Daily Trading Application -- Factor Regressions Results }
\vspace{-0.65cm}
\begin{center}
\begin{footnotesize}
\begin{threeparttable}
\begin{tabular}{l D{.}{.}{2.7} D{.}{.}{2.7} D{.}{.}{2.7} | D{.}{.}{2.7} D{.}{.}{2.7}}
\toprule

& \multicolumn{3}{c}{$N=20$} &  \multicolumn{2}{c}{$N= 100$} \\
\cmidrule(lr){2-4} \cmidrule(lr){5-6}

 & \multicolumn{1}{c}{MACE} & \multicolumn{1}{c}{MACE (PM)} & \multicolumn{1}{c}{MACE$_\text{loose bag}$}  & \multicolumn{1}{c}{MACE} & \multicolumn{1}{c}{MACE (PM)} \\
\midrule

\addlinespace[5pt]

& \multicolumn{5}{c}{Sample Period: 2016/01/01 - 2022/12/07} \\
\cmidrule(lr){2-6}
 
$\alpha$ & 0.05        & 0.01           & 0.05^{**}    & 0.13^{***} & 0.00           \\ \addlinespace[1pt]

MKT         & 0.64^{***}  & 0.59^{***}    & 0.45^{***}    & 0.35^{***} & 0.39^{***}           \\ \addlinespace[1pt]

SMB         & -0.31^{***} & -0.07  & -0.22^{***}    & -0.07      & -0.05           \\ \addlinespace[1pt]

HML         & 0.01        & -0.28^{***}& 0.00    & -0.11      & -0.48^{***}           \\ \addlinespace[1pt]

RMW         & -0.02       & 0.17^{***} & 0.00    & 0.02       & 0.41^{***}          \\ \addlinespace[1pt]

CMA         & 0.55^{***}  & 0.75^{***}  & 0.44^{***}    & 0.15       & 0.24^{***}           \\ \addlinespace[1pt]

MOM         & 0.01        & -0.01   & 0.00     & 0.03       & -0.02^{*}           \\ 

\midrule
$R^2$       & 0.29        & 0.43 & 0.29    & 0.08       & 0.41              \\
Observations  & \multicolumn{1}{D{.}{.}{2.2}}{1241 }       & \multicolumn{1}{D{.}{.}{2.2}}{1241}               & \multicolumn{1}{D{.}{.}{2.2}|}{1241}         & \multicolumn{1}{D{.}{.}{2.2}}{1241}                  & \multicolumn{1}{D{.}{.}{2.2}}{1241}             \\

\addlinespace[5pt]

& \multicolumn{5}{c}{Sample Period: \textit{No Covid}} \\
\cmidrule(lr){2-6}
 
$\alpha$ & 0.04      & 0.01           & 0.04^{*}    & 0.08^{**}  & 0.00           \\ \addlinespace[1pt]

MKT         & 0.46^{***} & 0.86^{***}   & 0.38^{***} & 0.33^{***} & 0.49^{***} \\ \addlinespace[1pt]

SMB         & -0.15^{**} & -0.21^{***} &  -0.14^{***}  & -0.06              & -0.10^{**} \\ \addlinespace[1pt]

HML         & -0.06    & -0.19^{***}      & -0.06        & -0.17^{**} & -0.47^{***}        \\ \addlinespace[1pt]

RMW         & 0.05       & 0.16^{***}           & 0.06       & 0.08                & 0.44^{***}    \\ \addlinespace[1pt]

CMA         & 0.45^{***} & 0.85^{***}  & 0.43^{***} & 0.15  & 0.29^{***} \\ \addlinespace[1pt]

MOM         & 0.00   & -0.01      & -0.00      & -0.00               & -0.02           \\

\midrule
$R^2$       & 0.16     & 0.58                    & 0.22        & 0.08                & 0.45           \\
Observations  & \multicolumn{1}{D{.}{.}{2.2}}{1179 }       & \multicolumn{1}{D{.}{.}{2.2}}{1179}               & \multicolumn{1}{D{.}{.}{2.2}|}{1179}         & \multicolumn{1}{D{.}{.}{2.2}}{1179}                  & \multicolumn{1}{D{.}{.}{2.2}}{1179}             \\

\addlinespace[5pt]

& \multicolumn{5}{c}{Sample Period: \textit{Only Covid}} \\
\cmidrule(lr){2-6}

$\alpha$ & 0.46     & -0.21           & 0.39      & 1.20^{*} & -0.12         \\ \addlinespace[1pt]

MKT         & 0.93^{***} & 0.17^{***}  & 0.50^{***} & 0.36     & 0.22^{***} \\ \addlinespace[1pt]

SMB         & -0.49    & 0.04       & -0.42 & 0.32     & -0.06          \\ \addlinespace[1pt]

HML         & 0.11  & -0.18^{*}           & 0.37   & 0.53     & -0.33^{***}           \\ \addlinespace[1pt]

RMW         & -0.87    & 0.38    & -0.59   & -0.27  & 0.02   \\ \addlinespace[1pt]

CMA         & 1.31 & 0.33     & 0.73   & 1.16   & 0.12   \\ \addlinespace[1pt]

MOM         & 0.17   & -0.00   & 0.10  & 0.54   & -0.07 \\

\midrule
$R^2$       & 0.54   & 0.26       & 0.45     & 0.18  & 0.29   \\

Observations   & \multicolumn{1}{D{.}{.}{2.3}}{62}    & \multicolumn{1}{D{.}{.}{2.3}}{62}      & \multicolumn{1}{D{.}{.}{2.3}|}{62}                  & \multicolumn{1}{D{.}{.}{2.3}}{62}     & \multicolumn{1}{D{.}{.}{2.3}}{62}           \\

\bottomrule
\end{tabular}
\begin{tablenotes}[flushleft]
\scriptsize{\item[] $^{***}p<0.01$; $^{**}p<0.05$; $^{*}p<0.1$.  \textit{No Covid} covers the period 2016/01/01-2020/01/31 \& 2020/05/01-2022/12/07. \textit{Only Covid} captures the three months of the first Covid wave: 2020/02/01-2020/04/30.
 Standard-Errors are adjusted for heteroskedasticity and autocorrelation.}
\end{tablenotes}
\end{threeparttable}
\end{footnotesize}
\label{tab:FF5_d_Trading}
\end{center}
\end{table}


\begin{table}[H]
\caption{Monthly Trading Application -- Factor Regressions Results}
\vspace{-0.65cm}
\begin{center}
\begin{scriptsize}
\begin{threeparttable}
\begin{tabular}{l D{.}{.}{2.7} D{.}{.}{2.7} D{.}{.}{2.7} D{.}{.}{2.7} | D{.}{.}{2.7} D{.}{.}{2.7} D{.}{.}{2.7} D{.}{.}{2.7}}
\toprule

& \multicolumn{4}{c}{Model: \textit{6-Factor}} &  \multicolumn{4}{c}{Model: \textit{Factor Zoo}} \\
\cmidrule(lr){2-5} \cmidrule(lr){6-9} 
\addlinespace[2pt]

& \multicolumn{2}{c}{01/1987-12/2004} &  \multicolumn{2}{c}{01/2005-12/2019} & \multicolumn{2}{c}{01/1987-12/2004} &  \multicolumn{2}{c}{01/2005-12/2019} \\
\cmidrule(lr){2-3} \cmidrule(lr){4-5} \cmidrule(lr){6-7} \cmidrule(lr){8-9}

 & \multicolumn{1}{c}{MACE} & \multicolumn{1}{c}{MACE (PM)} & \multicolumn{1}{c}{MACE} & \multicolumn{1}{c}{MACE (PM)} & \multicolumn{1}{c}{MACE} & \multicolumn{1}{c}{MACE (PM)} & \multicolumn{1}{c}{MACE} & \multicolumn{1}{c}{MACE (PM)} \\
\midrule

\addlinespace[5pt]

$\alpha$ & -0.33    & -0.54^{**} & 0.99^{***} & 0.11 & 0.43      & 0.45^{**}  & 1.20^{***} & 0.27 \\ \addlinespace[1pt]

MKT         & 1.09^{***} & 1.26^{***} & 0.58^{***} & 1.20^{***} & 0.93^{***} & 1.13^{***} & 0.50^{***} & 1.59^{***} \\ \addlinespace[1pt]

SMB         & 0.22       & 0.26^{***} & 0.42^{***} & 0.33^{**} & & &   & 0.76^{***}  \\ \addlinespace[1pt]

HML         & 0.69^{***} & 0.82^{***} & -0.07      & -0.05 & 0.39^{*} & 0.44^{***} & & \\ \addlinespace[1pt]

RMW         & 0.16      & 0.31^{**}  & 0.51^{***} & 0.46^{**} & &  &  &     \\ \addlinespace[1pt]

CMA         & 0.00 & -0.03 & 0.69^{**}  & 0.43^{*} & &  &   &  \\ \addlinespace[1pt]

MOM         & -0.14      & -0.04   & 0.09      & 0.01 & -0.11  &  & & \\ \addlinespace[5pt]

aliq\_mat       & & & & &             & -0.14     &  &  \\ \addlinespace[1pt]

ami\_126d & & & & &             &     & 0.09      &              \\ \addlinespace[1pt]

beta\_dimson\_21d & & & & &  &  &  & 0.20^{*} \\ \addlinespace[1pt]

betadown\_252d & & & & &  &  &  & 0.43^{***}\\ \addlinespace[1pt]

cowc\_gr1a     & & & & & -0.41^{*}  & -0.35^{**} &  & -0.57^{**} \\ \addlinespace[1pt]

debt\_gr3  & & & & &  &  & 0.70 &  \\ \addlinespace[1pt]

debt\_me        & & & & &           & -0.06      &  &  \\ \addlinespace[1pt]

dolvol\_126d & & & & &  &  &  & -0.06  \\ \addlinespace[1pt]

dolvol\_var\_126d & & & & &  &  &  & 0.69^{***}  \\ \addlinespace[1pt]

dsale\_drec & & & & &  &  &  & -0.27  \\ \addlinespace[1pt]

dsale\_dsag & & & & &  &  &  & 0.37 \\ \addlinespace[1pt]

f\_score & & & & &  &  &  & 0.15  \\ \addlinespace[1pt]

iskew\_capm\_21d & & & & &  &  & 0.74 &  \\ \addlinespace[1pt]

iskew\_hxz4\_21d & & & & & 0.46       & &  &  \\ \addlinespace[1pt]

kz\_index & & & & &  &  &  & 0.28^{**}  \\ \addlinespace[1pt]

lti\_gr1a & & & & &  &  &  & 0.45   \\ \addlinespace[1pt]

ncol\_gr1a  & & & & &  &  & 0.42 &  \\ \addlinespace[1pt]

netdebt\_me     & & & & &              & 0.00    &  &  \\ \addlinespace[1pt]

nfna\_gr1a & & & & & -0.18    &  &  &  \\ \addlinespace[1pt]

noa\_at   & & & & & -0.16    & -0.58^{***} &  &  \\ \addlinespace[1pt]

qmj\_safety     & & & & & -0.29     & -0.43^{**}  &  &  \\ \addlinespace[1pt]

resff3\_6\_1  & & & & &  &  & 0.53^{**} &  \\ \addlinespace[1pt]

ret\_60\_12  & & & & &  &  & 0.23&  \\ \addlinespace[1pt]

seas\_11\_15an  & & & & &            & -0.19  &  & -0.20 \\ \addlinespace[1pt]

seas\_6\_10an   & & & & &            & -0.16   &  &  \\ \addlinespace[1pt]

seas\_6\_10na   & & & & &            & 0.36^{***}  &  &  \\ \addlinespace[1pt]

seas\_6\_20an   & & & & &            &      &  & -0.39^{**} \\ \addlinespace[1pt]

seas\_6\_20na   & & & & &            &  &  & 0.46^{**} \\ \addlinespace[1pt]

seas\_2\_5na   & & & & &            &  &  & 0.27^{**} \\ \addlinespace[1pt]

taccruals\_ni     & & & & & 0.38    &          &  &  \\ \addlinespace[1pt]

tangibility  & & & & &  &  & 0.53^{**} &  \\ \addlinespace[1pt]

\midrule 
$R^2$       & 0.50       & 0.72 & 0.26     & 0.67 & 0.52      & 0.80 & 0.33     & 0.83     \\

Observations   & \multicolumn{1}{D{.}{.}{2.2}}{216}    & \multicolumn{1}{D{.}{.}{2.2}}{216}  & \multicolumn{1}{D{.}{.}{2.2}}{180}    & \multicolumn{1}{D{.}{.}{2.2}|}{180}  & \multicolumn{1}{D{.}{.}{2.2}}{216}    & \multicolumn{1}{D{.}{.}{2.2}}{216}  & \multicolumn{1}{D{.}{.}{2.2}}{180}    & \multicolumn{1}{D{.}{.}{2.2}}{180} \\     

\bottomrule
\end{tabular}
\begin{tablenotes}[flushleft]
\scriptsize{\item[] $^{***}p<0.01$; $^{**}p<0.05$; $^{*}p<0.1$. Standard-Errors are adjusted for heteroskedasticity and autocorrelation.\\
 \textit{6-Factor Model}: we run simple OLS on a \cite{FF2015} 5-factor + momentum model.\\
 \textit{Factor Zoo}: we augment the factors in \cite{JensenKellyPedersen2023} with \cite{FF2015} 5-factor + momentum. We first conduct variable-selection via LASSO. We then collect the variables with non-zero coefficients and run simple OLS.}
\end{tablenotes}
\end{threeparttable}
\end{scriptsize}
\label{tab:Factors_m_post}
\end{center}
\end{table}

\end{document}